\documentclass[10pt]{article}
\usepackage{amsfonts}
\usepackage{amsmath}
\usepackage{amsthm}
\usepackage{amssymb}
\usepackage{dsfont}
\usepackage{mathrsfs}
\usepackage{graphicx}
\usepackage[bf,center]{titlesec}
\usepackage{verbatim}
\usepackage[section]{placeins}

\newtheoremstyle{break}
  {}
  {}
  {\itshape}
  {}
  {\bfseries}
  {}
  {\newline}
  {}

\theoremstyle{break}

\numberwithin{equation}{section}

\newtheorem{cor}{Corollary}[section]

\newtheorem{lem}{Lemma}[section]
\newtheorem{prp}{Proposition}[section]
\newtheorem{thm}{Theorem}[section]

\renewcommand{\iint}{\int \!\!\!\! \int}

\long\def\symbolfootnote[#1]#2{\begingroup%
\def\thefootnote{\fnsymbol{footnote}}\footnote[#1]{#2}\endgroup}
\newcommand{\arctanh}{\mathrm{artanh}}
\newcommand{\arccoth}{\mathrm{arcoth}}
\newcommand{\sinc}{\mathrm{sinc}}

\title{Landau Damping in Relativistic Plasmas}
\author{Brent Young \\ Universit\"at zu K\"oln \\ bojy77@gmail.com}

\begin{document}
\maketitle

\begin{abstract}
We examine the phenomenon of Landau Damping in relativistic plasmas via a study of the relativistic Vlasov-Poisson system (rVP) on the torus for initial data sufficiently close to a spatially uniform steady state.  We find that if the steady state is regular enough (essentially in a Gevrey class of degree in a specified range) and that the deviation of the initial data from this steady state is small enough in a certain norm, the evolution of the system is such that its spatial density approaches a uniform constant value sub-exponentially fast (i.e. like $\exp(-C|t|^{\overline{\nu}})$ for $\overline{\nu} \in (0,1)$).  We take as \emph{a priori} assumptions that solutions launched by such initial data exist for all times (by no means guaranteed with rVP, but reasonable since we are close to a spatially uniform state) and that the various norms in question are continuous in time (which should be a consequence of an abstract version of the Cauchy-Kovalevskaya Theorem).  In addition, we must assume a kind of ``reverse Poincar\'e inequality'' on the Fourier transform of the solution.  In spirit, this assumption amounts to the requirement that there exists $0<\varkappa<1$ so that the mass in the annulus $\varkappa \le |v| < 1$ for the solution launched by the initial data is uniformly small for all $t$.  Typical velocity bounds for solutions to rVP launched by small initial data (at least on $\mathbb{R}^6$) imply this bound.  We note that none of our results require spherical symmetry (which is a crucial assumption for most results known about rVP).
\end{abstract}

\tableofcontents

\section{Introduction}

Landau Damping is one of the more striking phenomena to emerge from the field of plasma physics and has received much study since its discovery nearly 70 years ago.  The initial findings centered around plasmas modeled by the Vlasov-Poisson system
\begin{equation}
\textrm{VP}^+: \left\{ \begin{array}{l} \partial_t f + p\cdot \nabla_q f + E\cdot\nabla_p f = 0\\
\nabla\cdot E = 4 \pi (\rho - \overline{\rho})\\
\rho(t,q) = \int f(t,q,p)\;d^3p
 \end{array}\right.,
\end{equation}
which describes a non-relativistic one-component Coulomb plasma with a (uniform) neutralizing background charge density, $\overline{\rho}$, in unbounded space.  As first noted by Landau in 1946 \cite{L46}, solutions to the linearization of VP$^+$ about a uniform, Maxwellian equilibrium can exhibit exponential decay of Fourier modes associated to non-zero wavevectors in their spatial distributions $\rho$ (the electric field will also decay exponentially fast in such situations).  As such, the linearized system seems to exhibit a \emph{time-irreversible} behavior (exponential decay to a constant background).  What makes this result so surprising at first glance is that VP$^+$ itself is \emph{time-reversible} (and so also is its linearization).  By contrast, traditional approach-to-equilibrium results (e.g., plasmas described by the Boltzmann equation tending to the uniform Maxwellian background) all involve irreversible equations describing dissipative systems where some Lyapunov functional is decreasing as $t$ increases (such as the negative entropy functional for the Boltzmann equation).  Due to the form of the Vlasov equation however, any reasonable Lyapunov functional will be preserved under the evolution!  Many different physical mechanisms have been put forward to explain this apparent paradox, but from a purely mathematical perspective,  the decrease in amplitude of $\rho$ is paid for at the expense of increasing derivative norms for $f$.  The increasing filamentation in phase space (accompanied by increasingly higher frequency oscillations) is such that it averages out in the marginal distribution $\rho$.  Hence, from a mathematical point-of-view, Landau Damping is a kind of weak convergence result.  For a nice introduction to these ideas, see \cite{V10}.

Once we move from a mode-by-mode analysis to a full treatment of the linearized system,  things become much trickier.  Since the zero Fourier mode is always preserved in time (as this represents the total charge), it is reasonable to expect that only modes associated to wavevectors well separated from zero can exhibit \emph{uniform} exponential decay in time.  On the 3-dimensional torus of size $L$, this is a given since the smallest non-zero wavevector has magnitude $L^{-1}$.  For plasmas distributed on the entirety of $\mathbb{R}^3$, we can only hope that modes above a certain threshold will decay uniformly exponentially fast.  Indeed,  Glassey and Schaeffer \cite{GS94} have shown that for VP$^+$ in one spatial dimension linearized about a uniform kinetic equilibrium, $f_0(p)$, the best overall decay rate (as measured by $L^2$-norm) one can hope for is $\mathcal{O}(t^{-1})$.  For the Maxwellian equilibrium in one spatial dimension, the decay rate is only like $(\ln t)^{-3/2}$ (in three spatial dimensions, they also show the decay for the Maxwellian is like $(\ln t)^{-5/2}$), and for $f_0$ radially decreasing and compactly supported, there can be no decay at all.  In general, the faster $f_0'$ limits to zero as $|p| \to \infty$, the slower $\rho$ tends to zero in $L^2$-norm.  Hence, confinement of the plasma (as in the torus) seems essential for true exponential decay.

For many years, this damping phenomenon was known rigorously only for the linearized system.  Recently, Mouhot and Villani \cite{MV11} have succeeded in showing that sufficiently regular solutions to both the fully non-linear VP$^{+}$ and VP$^{-}$ systems on the torus do indeed exhibit the damping phenomenon exponentially in time; here, VP$^-$ is the gravitational analog of the Coulombic VP$^+$.  More precisely, they show that for an analytic kinetic equilibrium, $f_0 = f_0(p)$, satisfying certain stability criteria (along with constraints on the sizes of its derivatives and Fourier transform) there is an $\epsilon >0$ so that all initial data, $f_i = f_i(p,q)$, within $\epsilon$ of $f_0$ in an appropriate norm limit exponentially fast in $t$ to a spatially uniform state as $t \to \infty$.

When we inquire about Landau Damping for relativistic plasmas, we find much less information in the literature.  In 1994, Schlickeiser \cite{S94} examined the phenomenon for mono-charged, relativistic plasmas close to the spatially uniform J\"uttner distribution (which is the relativistic version of the Maxwellian profile).  Working with an expression for the plasma conductivity tensor linearized about this relativistic equilibrium (derived earlier by Trubnikov),  he found that there is a temperature-dependent critical magnitude, $k_c$, so that ``superluminal oscillations undergo no Landau damping,'' (see the abstract of \cite{S94}) corresponding to wavevectors of magnitude below the critical value (it is our reading of the paper \cite{S94} that ``Landau damping'' is meant in the strict sense of exponential decay).  This is in sharp contradistinction to the situation for non-relativistic plasmas linearized about the Maxwellian where no such critical $k_c$ is to be found.  As such, mono-charged relativistic plasmas may not exhibit exponential decay to the uniform equilibrium on the torus even for very nice initial data (depending, of course, on the ambient temperature and size of the torus).

In a previous paper \cite{Y14}, the author examined Landau damping on a mode-by-mode basis for the relativistic Vlasov-Poisson system (rVP) linearized around a sufficiently nice, spatially uniform kinetic equilibrium.  One of the major outcomes of this study was that true exponential decay of modes is not possible for sufficiently symmetric initial data (e.g. spherically symmetry) due to the universal speed limit imposed by relativity.\symbolfootnote[2]{It is worth noting that asymmetric data may possibly exhibit true exponential decay for $t>0$ -- so long as the decay for $t<0$ is sub-exponential.  However, all currently known results for rVP hinge on spherical symmetry, and so these cases preclude true exponential decay of modes.}  However, under certain conditions on the uniform background, sub-exponential decay of certain modes (i.e. decay like $\exp(-C |t|^{\overline{\nu}})$ for $\overline{\nu} \in (0,1)$) is possible.  In accordance with Schlickeiser's results, such decay is possible in the plasma physics case only for modes associated to wavevectors larger than a certain size (depending on the background).  Curiously, no such limitation is seen for gravitational plasmas - though this seems to be an artifact of working on a mode-by-mode basis (rather than requiring some uniform decay for \emph{all modes} with wavevectors larger than some critical size).  It is the purpose of the current paper to examine Landau damping for the fully non-linear rVP system - at least in certain regimes.

Incidentally, ``relativistic Vlasov-Poisson'' may sound like a strange mix of relativistic and non-relativistic physics, but in fact it is a special case of the \emph{relativistic Vlasov-Maxwell system} (rVM):
\begin{equation}
\textrm{rVM:} \left\{ \begin{array}{l} \partial_t f + v(p)\cdot \nabla_q f + \sigma\left(E+v(p)\times B\right)\cdot\nabla_p f = 0\\
\partial_t E = \nabla \times B - j\\
\partial_t B = -\nabla\times E,\\
\nabla\cdot E = 4\pi (\rho - \overline{\rho}), \;\; \nabla \cdot B = 0\\
\rho(t,q) = \int f(t,q,p)\;d^3p\\
j(t,q) = \int v(p)f(t,q,p)\;d^3p
 \end{array}\right.,
\end{equation}
which describes the evolution of a mono-charged, dilute (i.e. collisionless) plasma with phase-space distribution function $f >0$ (the magnitude of charge for the particles comprising the plasma is given by $\sigma >0$); these equations are understood to be in the rest frame of the neutralizing background (otherwise, one would need to add a corresponding background current $\overline{j}$).  The relativistic velocity in terms of the momentum is given by
\begin{equation}
v(p) = \frac{p}{\sqrt{1+|p|^2}},
\end{equation}
in units where the speed of light and the mass of the particles in question are both equal to 1.  For an excellent introduction to this system, see \cite{R04}.  \symbolfootnote[1]{In particular, this paper gives a nice review of the existence of global weak solutions to rVM.  As for results on decay rates, there are a few results (at least in certain special cases).  In 2010, Glassey, Pankavich, and Schaeffer \cite{GPS10} showed that there are solutions to rVM in 1.5 dimensions (i.e. one spatial dimension and two momenta dimensions) for which the spatial distribution of charge exhibits no decay in $t$.  In fact, all $L^p$-norms of the distribution for $p\in [1,\infty]$ are bounded below by a constant which is independent of $t$.  They also show that there are no non-trivial, steady-state solutions in 1.5 dimensions which are compactly supported.}  Should we make the ansatz that $B$ is identically zero for all times, we arrive at the \emph{relativistic Vlasov-Poisson system} (rVP):
\begin{equation}
\textrm{rVP:} \left\{ \begin{array}{l} \partial_t f + v(p)\cdot \nabla_q f + \sigma E\cdot\nabla_p f = 0\\
\nabla\cdot E = 4\pi (\rho - \overline{\rho})\\
\rho(t,q) = \int f(t,q,p)\;d^3p
 \end{array}\right. .
\end{equation}
In the case that the initial data, $f_0$, for rVM is spherically symmetric (and there is no stray electromagnetic radiation from sources at infinity), we obtain $B\equiv 0$ for all times without further ado.  Hence, we expect rVP to be significant for spherical, single-specie plasmas.

If we allow the parameter $\sigma$ appearing in rVP to become negative, we obtain a model which formally describes a gas of relativistic particles interacting through Newtonian gravitation.  Such a model might well be assumed to be valid for a sufficiently ``hot'' gas (so that the use of the relativistic velocity is justified) but rarefied enough that gravity is adequately modeled by the Poisson equation.  Currently, the only work along these lines known to the author is \cite{R94} wherein Rendall proves that sufficiently regular, asymptotically flat initial data for the fully covariant Vlasov-Einstein system launches solutions which are well approximated by the \emph{non-relativistic} Vlasov-Poisson system.  Nonetheless, rVP with the Newtonian interaction is still of great interest in kinetic theory since it provides a tractable model where relativistic effects can be studied for self-gravitating systems.  Surprisingly,  the seemingly innocuous replacement of the relativistic formula for velocity (in terms of momentum) drastically alters the behavior of solutions from VP to rVP.  We close out the introduction with a brief overview of results pertaining to rVP which highlight these differences.

One of the earliest papers to appear on rVP$^{\pm}$ is \cite{GS85} wherein Glassey and Schaeffer show that global classical solutions will exist for initial data that are spherically symmetric, compactly supported in momentum space, and vanish on characteristics with vanishing angular momentum which are in addition compactly supported in $\mathbb{R}^6$ and have $L^{\infty}$-norm below a critical constant $\mathcal{C}_{\infty}^{\pm}$, with $\mathcal{C}_{\infty}^{+} = \infty$ and $\mathcal{C}_{\infty}^{-} < \infty.$  More recently,  Kiessling and Tahvildar-Zadeh \cite{KTZ08} have extended the theorem of Glassey and Schaeffer for rVP$^-$ by proving global existence of classical solutions for initial data which satisfy the same basic requirements as above but are in $\mathfrak{P}_1\cap C^1$ \symbolfootnote[2]{ $\mathfrak{P}_n\cap C^k$ is the set of probability measures on $\mathbb{R}^6$ absolutely continuous w.r.t. Lebesgue measure whose first $n$ moments are finite and whose Radon-Nikodym derivative is $C^k$. } and have $L^{\beta}$-norm below a critical constant $\mathcal{C}_{\beta}^{-}$ with  $\mathcal{C}_{\beta}^{-} < \infty$, and $\mathcal{C}_{\beta}^{-}$ identically zero iff $\beta < 3/2.$  The authors explicitly computed $\mathcal{C}_{3/2}^{-}$ but characterized the constant for other values of $\beta > 3/2$ as a variational problem.  The constants for the remaining cases were computed by the author in terms of Lane-Emden functions \cite{Y11-1}.

Glassey and Schaeffer also investigated what may happen when solutions to rVP$^-$ are launched by initial data with $\|f\|_{\infty} > \mathcal{C}_{\infty}^{-}$.  They proved that negative energy data lead to ``blow-up" (i.e. formation of a singularity) in finite time.  This is in sharp contrast with the non-relativistic Vlasov-Poisson system with attractive coupling (VP$^-$) for which there is a robust global existence theory for fairly general initial data (c.f. \cite{H90} and \cite{P92}).  Indeed, the possibility of collapse for solutions to rVP$^-$ is a primary motivation for studying the system - as the collapse is due solely to ``relativistic effects."  In \cite{LMR08b},  Lemou, M\'ehats, and Rapha\"el proved that systems launched by initial data with negative total energy approach a self-similar collapse profile. Around the same time, Kiessling and Tahvildar-Zadeh proved that any spherically symmetric classical solution of rVP$^-$ launched by initial data satisfying $f_0 \in \mathfrak{P}_3\cap C^1$ (along with other technical requirements) and having \emph{zero total energy} and total (scalar) virial less than or equal to $-1/2$ will blow up in finite time (Theorem 6.1 of \cite{KTZ08}).  However, they left open the question whether such initial data existed.  Explicit examples of such data were found by the author and reported in \cite{Y11-2}.

There has also been much work concerning the nonlinear stability of stationary solutions of rVP$^-$ and the dynamical details of the solutions which blow-up in finite time.  Had\v zi\'c and Rein \cite{HR07} showed the non-linear stability of a wide class of steady-state solutions of $\textrm{rVP}^-$ against certain allowable perturbations utilizing energy-Casimir functionals.  Shortly thereafter, Lemou, M\'ehats, and Rapha\"el \cite{LMR08a} investigated non-linear stability versus the formation of singularities in $\textrm{rVP}^-$ through concentration compactness techniques.

As for work on decay rates for the full rVP system in unbounded space, Horst \cite{H90} showed in 1990 that continuously differentiable, spherically symmetric initial data which are compactly supported launch solutions whose spatial matter distributions decay almost like $t^{-3}$ in $L^{\infty}$-norm (there is a logarithmic factor in the decay rate).  In 2009, Glassey, Pankavich, and Schaeffer \cite{GPS09} proved that non-trivial, continuously differentiable initial data for rVP$^{-}$ in 1.5 dimensions with compact support exhibit no decay whatsoever in $L^p$-norm for $p \in [1,\infty]$.  Lastly, we have the mode-by-mode results in \cite{Y14} cited above.  We note that the considerations in this last paper are equally valid on $\mathbb{R}^6$ (i.e. unbounded space) and on $\mathbb{T}^3_L \times \mathbb{R}^3$ (toroidal space).

Based on the numerous results outlined above, our intuition is that the full story of the evolution of solutions to VP and rVP on $\mathbb{R}^6$  is enormously complicated.  We anticipate that modes associated to spatial wavevectors of sufficiently large magnitude will indeed be damped out exponentially (or sub-exponentially) fast.  However, since the 0-mode (corresponding to the total mass or charge) is constant in time, there must be some complicated regime where the damping rate sharply drops.  On longer time scales, we also expect damping due to dispersive effects (especially for the plasma physics case).  Sorting out these effects will likely be a rather difficult piece of analysis indeed!  As in \cite{MV11}, we bypass many of these issues by working with  plasmas on $\mathbb{T}^3_L \times \mathbb{R}^3$ that are reasonably close to a spatially uniform background.  As the plasma is confined in space, there are no dispersive effects to worry about, and there is a clear gap in the spatial wavevectors from the 0-mode.  Of course from a physical point-of-view, this is less than desirable, but our setting is not utterly devoid of merit.  One could imagine that these results are pertinent to the local behavior of very large plasmas (large enough that there are sizable regions of roughly constant spatial density) when they are perturbed by small, roughly periodic disturbances (small in both amplitude and wavelength).  This is especially true for rVP where deviations from boundary effects propagate at a maximum speed (linearly in time) inside the plasma (as opposed to the expected sub-exponential decay we are investigating).

Instead of the original analysis of Mouhot and Villani \cite{MV11}, we choose to follow the analysis of Landau damping for VP on the torus given by Bedrossian, Masmoudi, and Mouhot in \cite{BMM13}.  Our results are largely identical.  Namely, should we take initial data $f_{in}(q,v)$ sufficiently close (in a type of Gevrey norm) to a very regular time-independent background $g_0(v)$, then the solution to rVP launched by this initial data, $f_t(q,v)$, approaches a free-streaming solution (i.e.  a function of the form $f_{\infty}(q-vt,v)$) at a rate like $\exp(-|t|^{\nu})$ (in a related Gevrey norm).   The decay rate in $t$ of the spatial density, $\rho_{\infty}$, for free-streaming solutions is directly controlled by the Fourier Transform of its profile, $f_{\infty}$ --- which will decay rapidly should the profile be regular enough.  As expected from the results in \cite{Y14}, we can only hope for $\nu \in (0,1)$.  The general outline of the analysis for rVP will be directly analogous to the work for VP.  Since certain portions are substantially different in detail, we choose to give a mostly complete account except for the portions where the calculations are precisely the same (in which case we only give an outline of the relevant analysis).  Most interestingly, the results of this study do not require spherical symmetry.  As noted above, almost all studies of rVP have needed spherical symmetry of initial data (and the solutions they launch) to make progress.  Of course, the current study is concerned with the torus where things are much simpler than the full space.  We note that the results for rVP require a number of \emph{a priori} assumptions that seem a bit restrictive in general.  However, given that we are close to a time-independent solution of rVP, these assumptions are presumably not unreasonable.  Most importantly, the typical existence results for spherically symmetric initial data (on the full space, at least) imply the sort of bounds needed to ensure our results on Landau Damping go through with no issue.

\bigskip

\textbf{Acknowledgements:}  The author wishes to thank Michael Kiessling for suggesting a study of Landau Damping in relativistic plasmas.  The author also thanks Markus Kunze for many enlightening conversations and for pointing out the paper by Bedrossian, Masmoudi, and Mouhot \cite{BMM13} on which this paper is heavily dependent.

\section{Basic Setup}

We consider the relativistic Vlasov equation over the $3$-dimensional torus $\mathbb{T}_L^3$ of characteristic size $L$:
\begin{align}
&\partial_t f + v(p)\cdot \nabla_q f + F(t,q) \cdot \nabla_p f = 0,\label{rVP_mom}\\
&F(t,q)  = -\nabla_q W \ast_q \left(\rho_f(t,\cdot) - L^{-3}\int_{\mathbb{T}^3_L}\rho_f(t,y)\;dy \right)(q),\\
&\rho_f(t,q) = \int_{\mathbb{R}^d}f(t,q,p)\;dp,\\
&f(0,q,p)  = f_{in}(q,p), \label{rVP_mom_ic}
\end{align}
where
\begin{equation}
v(p) = \frac{p}{\sqrt{1+|p|^2}}. \label{def_v(p)}
\end{equation}
We will think of $\mathbb{T}_L^3$ as $(-L,L]^3$ with periodic boundary conditions.  The quantity subtracted from $\rho_f$ in the convolution above is the total mass of the system which is expected to be constant for reasonable (and interesting) choices of the kernel $W$.  For repulsive Coulomb interactions, we would have $F = e \nabla_q \triangle_q^{-1}(\rho_f-C)$ ($e>0$ giving the magnitude of the like charges constituting the plasma)while for attractive Newton interactions $F = -mG\nabla_q \triangle_q^{-1}(\rho_f-C)$ ($m>0$ being the mass of the particles making up the gas).  We only make an assumption on the transform of $W$:
\begin{equation}
\left|\widehat{W}(k)\right| \le \frac{C_W}{|k|^{1+\gamma}}, \tag{W} \label{ass_W}
\end{equation}
for constants $C_W>0$ and $\gamma \ge 1$ (note that $\gamma = 1$ for the Coulombic and Newtonian interactions).  We will detail our conventions for the various transforms we will need below.  Often, we will simply write $\mathbb{T}^3$ for the torus (leaving the fixed parameter $L$ implicit).

Since we will be working in a regime where the solutions are expected to be close to free-streaming, it will be advantageous to work in terms of the velocity $v$.  We define the function
\begin{equation}
p(v) = \frac{v}{\sqrt{1-|v|^2}}\\
\end{equation}
for $v$ in the open unit ball, $B_1(0).$  Note that this is simply the inverse of \eqref{def_v(p)}.  We also have the relation
\begin{eqnarray}
\frac{1}{\sqrt{1+|p(v)|^2}} &=& \sqrt{1-|v|^2}.
\end{eqnarray}
If we define
\begin{equation}
g(t,q,v) = f(t,q,p(v)),
\end{equation}
then easy calculations give
\begin{eqnarray}
\partial_t f(t,q,p) &=& \partial_t g(t,q,v),\\
\nabla_q f(t,q,p) &=& \nabla_q g(t,q,v),\\
\nabla_p f(t,q,p) &=& \sqrt{1-|v|^2}\left[\mathds{1} - v \otimes v\right]\nabla_v g(t,q,v),
\end{eqnarray}
where the operator in the final line is given by $$\left[\mathds{1} - v \otimes v\right]\nabla_v g(t,q,v) = \nabla_v g(t,q,v) - \left(v \cdot \nabla_v g(t,q,v)\right)v.  $$  The change of variable from momentum to velocity should not change the spatial density, $\rho$.  Hence, the formula for the spatial density in terms of $g$ is given by
\begin{equation}
\rho_g(t,q) = \rho_f(t,q) = \int_{\mathbb{R}^3}f(t,q,p)\;dp = \int_{B_1(0)} \frac{g(t,q,v)}{\left(1-|v|^2\right)^{5/2}}dv.\\
\end{equation}
Note that should $f(t,q,\cdot)$ be in $\mathcal{S}(\mathbb{R}^3)$ as a function of $p$, then $g(t,q,\cdot)$ and all of its $v$-derivatives will be bounded by $\mathcal{C}(t,q)(1-|v|^2)^N$ on $B_1(0)$ for $N$ as large as we wish (the generic constant can depend on the degree of the derivative of $g$).  So, $g(t,q,\cdot)$ will be in $C^{\infty}_c(\mathbb{R}^3)$ with supp$(g) \subset B_1(0)$ for each $(t,q)$ and as a result cannot be analytic in the variable $v$.

Putting these facts together gives the equivalent system for $g$:
\begin{align}
&\partial_t g + v\cdot \nabla_q g + F(t,q) \cdot \sqrt{1-|v|^2}\left[\mathds{1} - v \otimes v\right]\nabla_v g  = 0, \label{rvp_v1}\\
&F(t,q)  = -\nabla_q W \ast_q \left(\rho_g(t,\cdot) - L^{-3}\int_{\mathbb{T}^3}\rho_g(t,y)\;dy \right)(q),\\
&\rho_g(t,q) = \int_{B_1(0)} \frac{g(t,q,v)}{\left(1-|v|^2\right)^{5/2}}dv,\\
&g(0,q,v)  = g_{in}(q,v) = f_{in}(q,p(v))\label{rvp_v2}.
\end{align}
Despite the increased complexity, the advantage to working with the equations in terms of the variable $v$ becomes apparent when we consider the free-streaming operators associated to these systems:
\begin{eqnarray}
\Phi_t^{\textrm{free}}(q,v) &=& (q-vt,v),\\
\Phi_t^{\textrm{free}}(q,p) &=& (q-v(p)t,p).
\end{eqnarray}
Since we will be considering situations where the long-term behavior of the systems is expected to approach a free-streaming solution, there will be a definite benefit to working with velocity over momentum.

We will be interested in solutions of the form
\begin{eqnarray}
g(t,q,v) &=& g_0(v) + h(t,q,v),
\end{eqnarray}
where $g_0(v)$ is a sufficiently regular background state and $$\int_{\mathbb{T}^3}\int_{B_1(0)} \frac{h(0,q,v)}{(1-|v|^2)^{5/2}}dvdq = 0.  $$  For reasonable choices of $W$ this integral should remain zero for all times.  We refer to this assumption on the initial data for $h$ as the \emph{neutral variation condition} (as the variation, $h,$ from the uniform background, $g_0,$ does not alter the total mass or charge of the system).   This leads to the associated equations for $h$:
\begin{align}
&\partial_t h + v\cdot \nabla_q h + F(t,q) \cdot \sqrt{1-|v|^2}\left[\mathds{1} - v \otimes v\right]\nabla_v (g_0 + h)  = 0,\label{pde_hg}\\
&F(t,q)  = -\nabla_q W \ast_q \rho(t,q),\\
&\rho(t,q) = \int_{B_1(0)} \frac{h(t,q,v)}{\left(1-|v|^2\right)^{5/2}}dv,\\
&h(0,q,v)  = h_{in}(q,v),\; \int_{\mathbb{T}^3}\int_{B_1(0)} \frac{h_{in}(q,v)}{\left(1-|v|^2\right)^{5/2}}dvdq = 0.
\end{align}
We should note that we can make the neutral variation condition on the torus with no loss of generality.  Should the initial data for $h$ not meet this condition, we can absorb its total integral by multiplying $g_0$ by an appropriate factor and subtracting the total integral from the initial data.

We expect that for sufficiently small initial data $h$, the solution will converge to a solution of the free transport equation.  Namely, we anticipate that
\begin{align}
h(t,q,v) &\to h_{\infty}(q-vt,v),
\end{align}
as $t$ goes to positive infinity.  If we compose the solution $h(t,q,v)$ with the inverse free transport operator at $-t$ (i.e. we consider the variable $x = q-vt$) and define
\begin{equation}
\varphi(t,x,v) = h(t,x+vt,v),
\end{equation}
a short calculation gives the equivalent equations satisfied by $\varphi$:
\begin{align}
&\partial_t \varphi +F(t,x+vt) \cdot \sqrt{1-|v|^2}\left[\mathds{1} - v \otimes v\right]\nabla_v g_0 \nonumber\\
&\qquad \qquad +F(t,x+vt) \cdot \sqrt{1-|v|^2}\left[\mathds{1} - v \otimes v\right](\nabla_v -t\nabla_x)\varphi = 0,\label{pde_phig}\\
&F(t,q)  = -\nabla_q W \ast_q \rho(t,q),\\
&\rho(t,q) = \int_{B_1(0)} \frac{\varphi(t,q-vt,v)}{\left(1-|v|^2\right)^{5/2}}dv,\\
&\varphi(0,x,v)  = h_{in}(x,v),\; \int_{\mathbb{T}^3}\int_{B_1(0)} \frac{h_{in}(x,v)}{\left(1-|v|^2\right)^{5/2}}dvdx = 0\label{ic_phig}.
\end{align}
For convenience, we also define the auxiliary function,
\begin{eqnarray}
\Phi(t,x,v) &=& \frac{\varphi(t,x,v)}{(1-|v|^2)^{5/2}}.
\end{eqnarray}
If we rewrite the equations above in terms of $\Phi$, we find:
\begin{align}
&\partial_t \Phi +F(t,x+vt) \cdot \sqrt{1-|v|^2}\left[\mathds{1} - v \otimes v\right]\frac{\nabla_v g_0(v)}{(1-|v|^2)^{5/2}}\nonumber\\
&\qquad +F(t,x+vt) \cdot \sqrt{1-|v|^2}\left[\mathds{1} - v \otimes v\right](\nabla_v -t\nabla_x)\Phi \nonumber\\
&\qquad-5\sqrt{1-|v|^2}F(t,x+vt)\cdot v\Phi= 0,\label{pde_phi}\\
&F(t,q)  = -\nabla_q W \ast_q \rho(t,q)\label{force},\\
&\rho(t,q) = \int_{B_1(0)} \Phi(t,q-vt,v)dv,\\
&\Phi(0,x,v)  = \frac{h_{in}(x,v)}{(1-|v|^2)^{5/2}},\; \int_{\mathbb{T}^3}\int_{B_1(0)} \frac{h_{in}(x,v)}{\left(1-|v|^2\right)^{5/2}}dvdx = 0\label{ic_phi}.
\end{align}
From henceforth, we will write
\begin{equation}
\Phi_{in}(x,v) = \frac{h_{in}(x,v)}{(1-|v|^2)^{5/2}},
\end{equation}
for the initial data.  We will also define an auxiliary function, $G_0,$ determined by the uniform background via
\begin{equation}
G_0(v) = \left[\mathds{1} - v \otimes v\right]\frac{\nabla_v g_0(v)}{(1-|v|^2)^{5/2}}.
\end{equation}
We will at the very least assume that $g_0$ and all of its derivatives vanish rapidly enough at $|v|=1$ so that $G_0$ and all of its derivatives also vanish at the boundary.

We define $\alpha(v)=\sqrt{1-|v|^2}$ (which is simply the multiplicative inverse of the standard Lorentz factor $\gamma(v) = (1-|v|^2)^{-1/2}$).  In later computations, it will be advantageous to commute this function past the derivatives acting on $\Phi$. This yields the following system on $\mathbb{T}_L^3 \times´B_1(0)$:
\begin{align}
&\partial_t \Phi +F(t,x+vt) \cdot \alpha G_0(v)\nonumber\\
&\qquad +F(t,x+vt) \cdot \left[\mathds{1} - v \otimes v\right](\nabla_v -t\nabla_x)\alpha \Phi \nonumber\\
&\qquad-4 F(t,x+vt)\cdot v \alpha \Phi= 0,\label{main_pde_phi}\\
&F(t,q)  = -\nabla_q W \ast_q \rho(t,q)\label{main_force},\\
&\rho(t,q) = \int_{B_1(0)} \Phi(t,q-vt,v)dv,\\
&\Phi(0,x,v)  = \Phi_{in}(x,v),\; \int_{\mathbb{T}^3}\int_{B_1(0)} \Phi_{in}(x,v) dvdx = 0\label{main_ic_phi}.
\end{align}
Later, we will specify the exact assumptions on the regularity of the functions $G_0$ and $\Phi_{in}$.  We will also need the following equation which is a direct consequence of the one above (along with the fact that $\alpha(v)^2 = 1 - |v|^2$):
\begin{align}
&\partial_t (\alpha \Phi) +F(t,x+vt) \cdot (1-|v|^2) G_0(v)\nonumber\\
&\qquad +F(t,x+vt) \cdot \left[\mathds{1} - v \otimes v\right](\nabla_v -t\nabla_x)(1-|v|^2) \Phi \nonumber\\
&\qquad-3 F(t,x+vt)\cdot v (1-|v|^2) \Phi= 0.\label{main_pde_aphi}
\end{align}

Finally, we comment on the torus parameter, $L$.  The reason we insist on keeping the parameter $L$ (instead of scaling it out or setting it equal to one) is that the results of \cite{Y14} imply that we can only expect rapid damping of Fourier modes for tori that are sufficiently small (the exact size is determined by the background data).  However, the size of  the torus only enters the discussion in a significant fashion when considering the merits of assumption \eqref{ass_G_2} below for a specific choice of background, $G_0$.

\section{Fourier Transform Conventions}

Our convention for the Fourier Transform of a function on $\mathbb{T}^3_L \times \mathbb{R}^3$ will be:
\begin{align}
\widehat{f}(t,k,\eta) = \frac{1}{L^3}\int_{\mathbb{T}^3}\int_{\mathbb{R}^3} f(t,x,v)e^{-2\pi i \frac{k}{L}\cdot x}e^{-2\pi i \eta \cdot v} dvdx,
\end{align}
where $k \in \mathbb{Z}^3$ and $\eta\in \mathbb{R}^3$.  For more unwieldy expressions, we will occasionally use $$\mathcal{F}\{g(t,x,v)\}(t,k,\eta) = \widehat{g}(t,k, \eta). $$ For functions only of $x$ (and $t$), we write
\begin{align}
\widehat{f}(t,k) = \frac{1}{L^3}\int_{\mathbb{T}^3} f(t,x)e^{-2\pi i \frac{k}{L}\cdot x}dx.
\end{align}

Let us assume that $\Phi$ and $\rho$ are nice enough that all of the following calculations can be justified. By our assumption on the initial data and the fact that the equations preserve the total mass of the system, we have
\begin{equation}
\widehat{\Phi}(t,0,0) = \widehat{\Phi_{in}}(0,0) = 0,
\end{equation}
for all $t$.  Looking at the spatial profile $\rho$, we see that
\begin{eqnarray}
\widehat{\rho}(t,k) &=& \frac{1}{L^3}\int_{\mathbb{T}^3}\int_{\mathbb{R}^3}\Phi(t,q-vt,v)e^{-2\pi i \frac{k}{L}\cdot q}dvdq\\
&=& \widehat{\Phi}\left(t,k,\frac{kt}{L}\right),\\
\widehat{\rho}(t,0) &=& 0.
\end{eqnarray}
Therefore, we see that the transform of the force term is given by
\begin{equation}
\widehat{F}(t,k) = -2\pi i \frac{k}{L} \widehat{W}(k)\widehat{\rho}\left(t,k\right).
\end{equation}
Since the quantity $F(t,x+vt)$ appears in the equation for $\Phi$, we will need the transform of this quantity in both $x$ and $v$.  We find (in the sense of Tempered Distributions)
\begin{equation}
\mathcal{F}\left\{F(t,x+vt)\right\}(t,k,\eta) =  -2\pi i \frac{k}{L} \widehat{W}(k)\widehat{\rho}\left(t,k\right) \delta\left(\eta - \frac{kt}{L}\right).
\end{equation}

Next, we consider the transform of
\begin{equation}
\alpha(v) = \sqrt{1-|v|^2}\mathds{1}_{[0,1]}(|v|).
\end{equation}
We have the explicit form
\begin{equation}
\widehat{\alpha}(\eta) =  \frac{J_1(2\pi |\eta|)}{2\pi |\eta|^3} - \frac{J_0(2\pi |\eta|)}{2|\eta|^2},
\end{equation}
where $J_{\nu}$ is the Bessel function of the first kind with index $\nu$.  A plot of this transform appears below.
\begin{figure}[ht]\centering
  \includegraphics[bb=24 285 575 506,clip=true,scale=0.65]{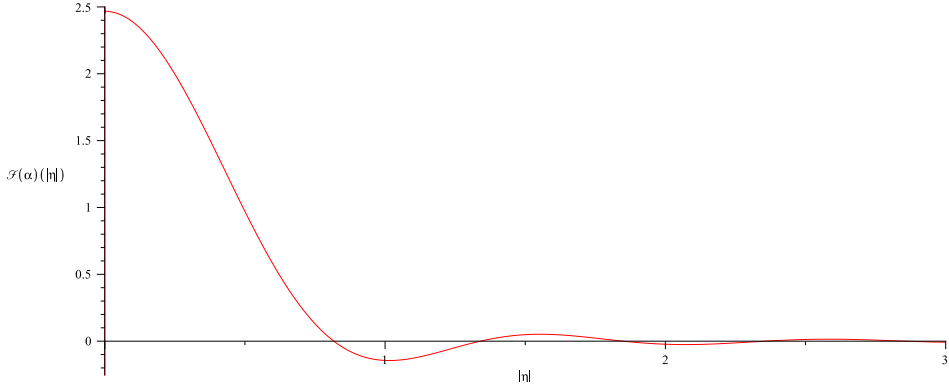}\\
  \caption{Plot of the Fourier Transform of $\alpha$ as a function of $|\eta|$}
  \label{fig1}
\end{figure}
Note that this function is $C^{\infty}$ (by virtue of being the transform of a compactly supported function) and vanishes at infinity (as $\alpha$ is certainly integrable).  The precise decay rate is given by
\begin{equation}
|\widehat{\alpha}(\eta)| \le \frac{4}{(1+|\eta|)^{5/2}},
\end{equation}
($4$ is not quite the optimal constant here, but it will certainly suffice).  Using this, we see that
\begin{equation}
\widehat{\alpha \Phi}(t,k,\eta) = \int \widehat{\alpha}(\eta - \xi) \widehat{\Phi}(t,k,\xi) \; d\xi.
\end{equation}
We will be assuming that $\widehat{\Phi}$ is exponentially decreasing in both of its Fourier arguments (more precisely, decaying like $\exp(-|k|^{\overline{\nu}})\exp(-|\eta|^{\overline{\nu}})$ for $\overline{\nu} \in (0,1)$).  However, due to the fact that $\alpha$ is not analytic on $B_1(0)$ (or in any Gevrey class for that matter), we will also need a similar assumption on $\widehat{\alpha \Phi}$.\symbolfootnote[2]{The convolution with $\widehat{\alpha}$ makes it difficult to estimate rates of decay.  For instance, suppose we have $\widehat{\Phi} \in L^p_kL^{p'}_{\eta}(\mathbb{Z}^3 \times \mathbb{R}^3)$ for any pair $p, p' \in [1,\infty]$ (which is guaranteed by our exponential decay assumptions noted above).  The decay rate quoted for $\widehat{\alpha}$ shows us that it belongs to $L^q(\mathbb{R}^3)$ for $q \in \left(\frac{6}{5},\infty \right]$.  By a variant of Young's Inequality (c.f. \cite{LL01}[p.99]) all we can claim for $\widehat{\alpha \Phi}$ is that it is in $L^p_kL^q_{\eta}(\mathbb{Z}^3 \times \mathbb{R}^3)$ for $p$ and $q$ in the same ranges as above.}  Essentially, this amounts to requiring $\Phi$ to vanish at $|v|=1$ sufficiently rapidly so that both $\Phi$ and $\alpha\Phi$ are in some common Gevrey class.

We want to take the Fourier Transforms of \eqref{main_pde_phi} and \eqref{main_pde_aphi}.  First, the transform of a function of the form $-v \cdot \nabla_v g(v)$ is given by
\begin{eqnarray}
\mathcal{F}\{-v\cdot\nabla_vg(v)\}(\eta) &=& -\frac{i}{2\pi}\nabla_{\eta} \cdot (2\pi i \eta)\widehat{g}(\eta)\nonumber\\
&=& 3\widehat{g}(\eta) + \eta \cdot \nabla_{\eta} \widehat{g}(\eta).
\end{eqnarray}
So, we have
\begin{align}
\mathcal{F}\{ \left[\mathds{1}- v \otimes v\right]&\left(\nabla_v - t\nabla_x\right)g(x,v) - nvg(x,v)\}\nonumber\\
 &= 2\pi i \left(\eta - \frac{kt}{L}\right)\widehat{g}(k,\eta) + \left(\eta - \frac{kt}{L}\right)\cdot \nabla_{\eta} \widehat{vg}(k,\eta)-(n-4)\widehat{vg}(k,\eta),
\end{align}
for any number $n$.  This rather odd way of representing the transform will be useful in our computations.

Looking term by term, we first note the obvious
\begin{eqnarray*}
\mathcal{F}\{\partial_t \Phi(t,x,v)\}(t,k,\eta) &=& \partial_t \widehat{\Phi}(t, k, \eta),\\
\mathcal{F}\{\partial_t \alpha \Phi(t,x,v)\}(t,k,\eta) &=& \partial_t \widehat{\alpha \Phi}(t, k, \eta).
\end{eqnarray*}
Next, we have
\begin{equation}
\mathcal{F}\left\{ F(t,x+vt) \cdot \alpha G_0(v) \right\}(t,k,\eta) = -2\pi i \widehat{W}(k)\widehat{\rho}(t,k)\frac{k}{L}\cdot \widehat{\alpha G_0}\left(\eta - \frac{kt}{L}\right).
\end{equation}
There is, of course, a completely analogous formula for the linear part of \eqref{main_pde_aphi}:
\begin{align}
\mathcal{F}&\left\{ F(t,x+vt) \cdot (1-|v|^2) G_0(v) \right\}(t,k,\eta)\nonumber\\
&\qquad \qquad = -2\pi i \widehat{W}(k)\widehat{\rho}(t,k)\frac{k}{L}\cdot \mathcal{F}\{(1-|v|^2) G_0\}\left(\eta - \frac{kt}{L}\right).
\end{align}
Note that we could rewrite this in terms of $(1+C\triangle_{\eta})\widehat{G_0}$ (where $C$ is just a constant coming from our convention on the transform).  We will leave it as above since this form has decided advantages when we want to compute the sizes of $L^2$-norms.

The remaining (non-linear) terms in \eqref{main_pde_phi} give
\begin{align}
\mathcal{F}&\{\textrm{ non-linear terms \eqref{main_pde_phi}}\}(t,k,\eta)\nonumber\\
&= -2 \pi i\sum_{\ell \in \mathbb{Z}^3\setminus \{(0,0,0)\}} \widehat{\rho}\left(t, \ell\right)\widehat{W}(\ell)\frac{\ell}{L} \cdot \Bigg[2\pi i\left(\eta - \frac{k t}{L} \right) \widehat{\alpha \Phi}\left(t,k-\ell,\eta - \frac{\ell t}{L}\right)\nonumber\\
& \qquad \qquad \qquad \qquad \qquad +\left(\eta - \frac{k t}{L} \right)\cdot \nabla_{\eta}\widehat{v \alpha \Phi}\left(t,k-\ell,\eta - \frac{\ell t}{L}\right)\Bigg].
\end{align}
The sum over $\ell$ excludes the zero mode since $\widehat{\rho}(t,0) = 0$ by our neutral variation condition (in later formulae, we will write $\ell \ne 0$ to emphasize this point). Similarly, we have
\begin{align}
\mathcal{F}&\{\textrm{ non-linear terms \eqref{main_pde_aphi}}\}(t,k,\eta)\nonumber\\
&= -2 \pi i \!\!\sum_{\ell \in \mathbb{Z}^3\setminus \{(0,0,0)\}} \!\!\!\! \widehat{\rho}\left(t, \ell\right)\widehat{W}(\ell)\frac{\ell}{L} \cdot \Bigg[2\pi i\left(\eta - \frac{k t}{L} \right) \mathcal{F}\{(1-|v|^2) \Phi\}\left(t,k-\ell,\eta - \frac{\ell t}{L}\right)\nonumber\\
& \qquad  \qquad \qquad +\left(\eta - \frac{k t}{L} \right)\cdot \nabla_{\eta}\mathcal{F}\{v (1-|v|^2) \Phi\}\left(t,k-\ell,\eta - \frac{\ell t}{L}\right)\nonumber\\
&\qquad \qquad \qquad +\mathcal{F}\{v (1-|v|^2) \Phi\}\left(t,k-\ell,\eta - \frac{\ell t}{L}\right)\Bigg].
\end{align}

Thus, we obtain
\begin{align}
\partial_t \widehat{\Phi}&(t, k, \eta)  - 2 \pi i \widehat{W}(k)\widehat{\rho}\left(t, k\right)\frac{k}{L}\cdot \widehat{\alpha G_0}\left(\eta - \frac{kt}{L}\right)\nonumber\\
&+4 \pi^2 \sum_{\ell \ne 0} \widehat{\rho}\left(t, \ell\right)\widehat{W}(\ell)\frac{\ell}{L} \cdot \left(\eta - \frac{k t}{L} \right) \widehat{\alpha \Phi}\left(t,k-\ell,\eta - \frac{\ell t}{L}\right)\nonumber\\
&-2\pi i \sum_{\ell \ne 0} \widehat{\rho}\left(t, \ell\right)\widehat{W}(\ell)\frac{\ell}{L} \cdot\left[\left(\eta - \frac{k t}{L} \right)\cdot \nabla_{\eta}\widehat{v \alpha \Phi}\left(t,k-\ell,\eta - \frac{\ell t}{L}\right)\right]=0, \label{main_DE_Phi}
\end{align}
and its companion equation
\begin{align}
\partial_t \widehat{\alpha \Phi}&(t, k, \eta)  - 2 \pi i \widehat{W}(k)\widehat{\rho}\left(t, k\right)\frac{k}{L}\cdot \mathcal{F}\{(1-|v|^2)G_0\}\left(\eta - \frac{kt}{L}\right)\nonumber\\
&+4 \pi^2 \sum_{\ell \ne 0} \widehat{\rho}\left(t, \ell\right)\widehat{W}(\ell)\frac{\ell}{L} \cdot \left(\eta - \frac{k t}{L} \right) \mathcal{F}\{(1-|v|^2) \Phi\}\left(t,k-\ell,\eta - \frac{\ell t}{L}\right)\nonumber\\
&-2\pi i \sum_{\ell \ne 0} \widehat{\rho}\left(t, \ell\right)\widehat{W}(\ell)\frac{\ell}{L} \cdot\left[\left(\eta - \frac{k t}{L} \right)\cdot \nabla_{\eta}\mathcal{F}\{v (1-|v|^2) \Phi\}\left(t,k-\ell,\eta - \frac{\ell t}{L}\right)\right]\nonumber\\
&-2\pi i \sum_{\ell \ne 0} \widehat{\rho}\left(t, \ell\right)\widehat{W}(\ell)\frac{\ell}{L} \cdot \mathcal{F}\{v (1-|v|^2) \Phi\}\left(t,k-\ell,\eta - \frac{\ell t}{L}\right)=0. \label{main_DE_aPhi}
\end{align}
These are already a self-consistent system of equations (since we can rewrite $\widehat{\rho}$ in terms of $\widehat{\Phi}$).  Instead of proceeding in such a manner, we get a separate equation for $\widehat{\rho}$ and treat the three equations as a coupled system.  Once we integrate \eqref{main_DE_Phi} in $t$ and evaluate at $\eta = kt/L$, we obtain
\begin{align}
\widehat{\rho}(t,k) &= \widehat{\Phi_{in}}\left(k,\frac{kt}{L}\right)+ \int_0^tL(t-\tau,k)\widehat{\rho}\left(\tau, k\right) d\tau\nonumber\\
& \;\; - 4\pi^2 \int_0^t\sum_{\ell \ne 0}\widehat{\rho}\left(\tau, \ell\right)\widehat{W}(\ell)\frac{k \cdot \ell}{L^2}(t-\tau)\widehat{\alpha \Phi}\left(\tau,k-\ell,\frac{k t - \ell \tau}{L}\right)  d\tau \nonumber\\
&\;\; +2 \pi i \int_0^t\sum_{\ell \ne 0}\widehat{\rho}\left(\tau, \ell\right)\widehat{W}(\ell)\frac{\ell}{L} \cdot (t-\tau)\left[\frac{k}{L} \cdot \nabla_{\eta}\widehat{v \alpha \Phi}\left(\tau,k-\ell,\frac{k t - \ell \tau}{L}\right)\right]  d\tau,\label{main_eq_rho}
\end{align}
where
\begin{equation}
L(t,k) = 2\pi i \widehat{W}(k)\frac{k}{L}\cdot \widehat{\alpha G_0}\left(\frac{kt}{L}\right).
\end{equation}
We refer to the terms in all three equations involving $G_0$ as the \emph{linear terms} (control of which can be guaranteed by appropriate assumptions on the background data).  The other terms will be referred to as the \emph{non-linear terms} of the system.

\section{Gevrey Norms and Paraproduct Decompositions}

\addtocontents{toc}{\setcounter{tocdepth}{2}}

\subsection{Gevrey Norms and Useful Elementary Estimates}

Throughout this paper, we will use norms inspired by the properties of Gevrey Class functions (c.f. \cite{R93}).  For functions $h$ on $\mathbb{T}^3_L \times \mathbb{R}^3$, we define the Gevrey-$\frac{1}{\nu}$ norm ($\nu \in (0,1]$) with Sobolev correction $\sigma \in \mathbb{R}$ as
\begin{equation}
\|h(t)\|^2_{\mathcal{G}^{\lambda, \sigma ; \nu}} = \sum_{k \in \mathbb{Z}^d}\int_{\mathbb{R}^d}\left|\widehat{h}(t,k,\eta)\right|^2\langle k, \eta \rangle^{2\sigma}e^{2\lambda\langle k, \eta \rangle^{\nu}}d\eta.
\end{equation}
The angle brackets represent a modified form of the absolute value function:
\begin{eqnarray}
\langle y \rangle &=& \sqrt{1+y^2} \; \textrm{ for } \; y \in \mathbb{R},\\
\langle k, \eta \rangle &=& \sqrt{1+(|k|+|\eta|)^2} \; \textrm{ for vectors } \; k, \eta \in \mathbb{R}^d,\\
\langle k, y\rangle &=& \sqrt{1+(|k|+|y|)^2} \; \textrm{ for } k\in \mathbb{R}^d, y \in \mathbb{R},
\end{eqnarray}
and similarly for other combinations of vectors and scalars.  For functions independent of the spatial variable $x$ (such as the background data $G_0$), we will also use the norm above but with the understanding that $\widehat{f}$ is meant as $\widehat{f}\delta_{k=0}$ so that only the $k=0$ term survives in the sum.

On the other hand, since the transform of $\rho$ is an evaluation of the transform of $\Phi$ at $\eta = kt/L$, we define a slightly different norm for functions on the torus:
\begin{equation}
\|g(t)\|^2_{\mathcal{F}^{\lambda, \sigma ; \nu}} = \sum_{k \in \mathbb{Z}^d}\left|\widehat{g}(t,k)\right|^2\langle k, kt/L \rangle^{2\sigma}e^{2\lambda\langle k, kt/L \rangle^{\nu}}.
\end{equation}
The parameter $\lambda$ (which will often depend on $t$) can be thought of as a radius of convergence.  The fact that we are forced to have $\lambda(t)$ decreasing in $t$ is an indication of the loss of regularity in $\Phi$ under the evolution \eqref{main_DE_Phi}.

We define an operator $A(t)$ so that
\begin{equation}
\|A(t)h(t)\|_2 = \|h(t)\|_{\mathcal{G}^{\lambda(t), \sigma ; \nu}}.
\end{equation}
In other words, $A(t)$ is the symbolic differential operator satisfying
\begin{equation}
\widehat{Ah}(t,k,\eta) = \widehat{h}(t,k,\eta)\langle k, \eta \rangle^{\sigma}e^{\lambda(t)\langle k, \eta \rangle^{\nu}}.
\end{equation}
By a slight abuse of notation, we write
\begin{equation}
A^{(\lambda, \sigma ; \nu)}(t,k,\eta) = \langle k, \eta \rangle^{\sigma}e^{\lambda(t)\langle k, \eta \rangle^{\nu}}.\label{Adef}
\end{equation}
We will often omit the superscripts on $A$ if the parameters are fixed throughout a computation.  If a particular parameter should change, we will note it explicitly.  For example, we will have occasion to use
\begin{equation}
A^{( \sigma + \sigma')}(t,k,\eta) = \langle k, \eta \rangle^{\sigma+\sigma'}e^{\lambda(t)\langle k, \eta \rangle^{\nu}}.
\end{equation}

There are a number of important identities regarding the terms appearing in these norms which are identified in \cite{BMM13}.  We collect them in the following two lemmas.  All proofs are elementary (and so omitted).
\begin{lem}\label{exp_ineq_lem}
\begin{itemize}
\item[(i)] For $x \ge 0, \alpha > \beta \ge 0, C,\delta >0,$
\begin{equation}
e^{Cx^{\beta}} \le e^{C(C/\delta)^{\frac{\beta}{\alpha - \beta}}}e^{\delta x^{\alpha}}.
\end{equation}

\item[(ii)] For $x \ge 0, \alpha, \sigma,\delta >0,$ there exists a constant $C >0$ so that
\begin{equation}
 e^{-\delta x^{\alpha}} \le \frac{C}{\delta^{\sigma/\alpha}\langle x \rangle^{\sigma}}.
 \end{equation}
\end{itemize}
\end{lem}

\begin{lem}\label{angle_brak_lem}  Let $0 < s < 1$ and $x,y \ge 0.$
\begin{itemize}
\item[(i)]  We have the following versions of the triangle identity for $\langle \cdot \rangle$:
\begin{eqnarray}
\langle x + y\rangle^s & \le & \langle x \rangle^s + \langle y \rangle^s,\\
|\langle x \rangle^s - \langle y \rangle^s| & \le & \langle x - y \rangle^s,\\
C_s\left(\langle x \rangle^s + \langle y \rangle^s\right) &\le& \langle x+y \rangle^s,
\end{eqnarray}
for some $C_s >0$ (in fact, these also hold for $s=1$).

\item[(ii)]  There is some constant $C_s >0$ so that
\begin{equation}
|\langle x \rangle^s - \langle y \rangle^s| \le \frac{C_s}{\langle x \rangle^{1-s}+\langle y \rangle^{1-s}}\langle x-y \rangle.
\end{equation}

\item[(iii)]  If $|x - y| \le x/K$ for some $K>1$, then we have
\begin{equation}
|\langle x \rangle^s - \langle y \rangle^s| \le \frac{s}{(K-1)^{1-s}}\langle x - y \rangle^s.
 \end{equation}

\item[(iv)]  For all $x, y \ge 0$,
\begin{equation}
\langle x + y \rangle^s  \le \left(\frac{\max\{\langle x \rangle, \langle y \rangle\}}{\langle x \rangle +\langle y \rangle}\right)^{1-s}(\langle x \rangle^s+\langle y \rangle^s).
 \end{equation}
\end{itemize}
\end{lem}

On of the principal reasons for using $\langle \cdot \rangle$ as opposed to $|\cdot|$ is the following elementary estimate from \cite{BMM13}:
\begin{lem}\label{est_on_DA}
Let $\lambda(t)$ be decreasing.  Then for all $m \in \mathbb{N}^3$ and all $\sigma \in \mathbb{R}$, there is a constant $C = C(|m|, \lambda(0), \sigma, \nu)$ so that
\begin{equation}
\left|D_{\eta}^{m}A^{(\lambda, \sigma ; \nu)}(t,k,\eta)\right| \le \frac{C}{\langle k,\eta \rangle^{|m|(1-\nu)}}A^{(\lambda, \sigma; \nu)}(t,k,\eta).
\end{equation}
\end{lem}

This rather innocuous looking lemma has a rather profound consequence in our case:
\begin{lem}\label{eta_deriv_lem}
For any multi-index $m \in \mathbb{N}^3$, there is a constant $C_m = C_m(\lambda(0), \sigma, \nu)$ so that for any function $f =f(x,v)$ defined on $\mathbb{T}^3_L \times B_1(0)$
\begin{equation}
\left\| A^{(\lambda, \sigma ; \nu)}D^m_{\eta} \hat{f}\right\|_{L^2_{\eta}L^2_k} \le C_m \left\| A^{(\lambda, \sigma ; \nu)}\hat{f}\right\|_{L^2_{\eta}L^2_k}.
\end{equation}
\end{lem}
Before giving a rigorous proof of this lemma, we give some insight as to why it should be true.  If the norm on the right-hand-side above is finite, then certainly $f$ is in the Gevrey class of order $\nu^{-1}$ (recall that $\nu \in (0,1)$ for our purposes).  As such, $v^mf$ should also be Gevrey of the same degree for any such multi-index (as $v^m$ is analytic).  This alone is not quite enough to give the theorem since requiring the $L^2$-norm of $Af$ to be finite is stronger information than merely belonging to a Gevrey class.  However, the fact that velocities are bounded gives us the extra control needed.

\bigskip

\textbf{Proof:}  We proceed by induction.  The theorem is trivially true for $m=(0,0,0)$, but we give the proof for $|m| = 1$ to see how the induction step should proceed.  Since there are three such multi-indices in this case, we consider $\| A^{(\lambda, \sigma ; \nu)} \partial_{\eta_j} \hat{f}\|$ where $j \in \{1,2,3\}$ refers to the appropriate component of $\eta$.
\begin{eqnarray}
\left\| A^{(\lambda, \sigma ; \nu)} \partial_{\eta_j} \hat{f} \right\|_{L^2_{\eta}L^2_k} &\le & \left\|  \partial_{\eta_j}\left(A^{(\lambda, \sigma ; \nu)} \hat{f}\right)\right\|_{L^2_{\eta}L^2_k}+ \left\| \left( \partial_{\eta_j} A^{(\lambda, \sigma ; \nu)}\right) \hat{f}\right\|_{L^2_{\eta}L^2_k}\nonumber\\
&\le& \left\| \partial_{\eta_j} \left(A^{(\lambda, \sigma ; \nu)} \hat{f}\right)\right\|_{L^2_{\eta}L^2_k}+ C\left\| A^{(\lambda, \sigma ; \nu)} \hat{f}\right\|_{L^2_{\eta}L^2_k}
\end{eqnarray}
where the last inequality follows from Lemma \ref{est_on_DA} and the fact that $\langle k, \eta \rangle \ge 1$ (note that the constant $C$ depends on $ \lambda(0), \sigma,$ and $\nu$).  As for the first term in the last inequality, we appeal to Plancherel's Theorem and the fact that $|v^m| \le 1$ for any multi-index $m$:
\begin{eqnarray}
\left\| \partial_{\eta_j} \left(A^{(\lambda, \sigma ; \nu)} \hat{f}\right)\right\|_{L^2_{\eta}L^2_k} &=& 2\pi \left\| |v_j| A^{(\lambda, \sigma ; \nu)}f \right\|_{L^2_{v}L^2_x}\nonumber\\
&\le& 2\pi\left\| A^{(\lambda, \sigma ; \nu)}f \right\|_{L^2_{v}L^2_x}\nonumber\\
&=& 2\pi\left\|A^{(\lambda, \sigma ; \nu)} \hat{f}\right\|_{L^2_{\eta}L^2_k}
\end{eqnarray}
where on the $(x,v)$ side we interpret $A$ as an operator acting on $f$.  Combining this with the estimate above gives us the lemma for the case of first derivatives.

For the inductive step, we have
\begin{align}
\Big\| & A^{(\lambda, \sigma ; \nu)}  D^m_{\eta} \hat{f}\Big\|_{L^2_{\eta}L^2_k} \nonumber\\
&\le \left\|D^m_{\eta} \left(A^{(\lambda, \sigma ; \nu)}\hat{f}\right)\right\|_{L^2_{\eta}L^2_k} + \sum_{0 < j \le m}\binom{m}{j} \left\| \left(D^j_{\eta} A^{(\lambda, \sigma ; \nu)}\right)D^{m-j}_{\eta} \hat{f}\right\|_{L^2_{\eta}L^2_k}\nonumber\\
&\le 2\pi^{|m|}\left\| |v^m|\left(A^{(\lambda, \sigma ; \nu)}f\right)\right\|_{L^2_{v}L^2_x} + \sum_{0 < j \le m}\binom{m}{j}C(|j|,\ldots) \left\| A^{(\lambda, \sigma ; \nu)}D^{m-j}_{\eta} \hat{f}\right\|_{L^2_{\eta}L^2_k}\nonumber\\
&\le \left( 2\pi^{|m|} +  \sum_{0 < j \le m}\binom{m}{j}C(|j|,\ldots)C_{m-j}\right) \left\|A^{(\lambda, \sigma ; \nu)}\hat{f} \right\|_{L^2_{\eta}L^2_k},
\end{align}
where the constant $C(|j|,\ldots)$ comes from Lemma \ref{est_on_DA} (and also depends on the constants appearing in $A$).  This completes the proof of the lemma. $\blacksquare$

\bigskip

We will have occasion to use Lemma \ref{eta_deriv_lem} frequently throughout the following (often without comment).  The factor, $C_m$, we pick up will usually be referred to as a ``combinatorial factor'' (as it depends on the number of derivatives being taken) despite the fact that it also depends on the various parameters appearing in $A$.  The key observation is that the universal bound on $|v|$ allows us to make these estimates without further assumptions on the differentiability of the transform.

\subsection{Littlewood-Paley Decomposition} \label{LPD}

We fix a function $\psi \in C^{\infty}_c(\mathbb{Z}^d\times \mathbb{R}^d, [0,1])$ which is radially symmetric, $\psi(k,\eta) = 1$ on the set $|k| + |\eta| \le 1/2$, and $\psi(k,\eta) = 0$ on $|k| + |\eta| \ge 3/4$.  We then define the function $\phi(k,\eta) = \psi(k/2,\eta/2) - \psi(k,\eta)$ which has support on the annulus $1/2 \le |k| +|\eta | \le 3/2$.  Finally, we define the scaled functions $\phi_N(k,\eta) = \phi(N^{-1}k,N^{-1}k)$ which have support on $N/2 \le |k| +|\eta | \le 3N/2$.  We then have a partition of unity:
\begin{equation}
1 = \psi(k,\eta) + \sum_{N \in \mathbb{D}}\phi_N(k, \eta),
\end{equation}
where $\mathbb{D} = \{1, 2, 4, 8, \ldots, 2^j, \ldots\}$ is the set of dyadic numbers.  Note that these functions are well-separated in the sense that for any given $(k,\eta)$, at most two terms in the summation are non-zero.

Using these functions, we define the following symbolic differential operators for functions $u \in L^2(\mathbb{T}^d_L \times \mathbb{R}^d)$:
\begin{eqnarray}
S u (x,v) &=& \psi(\nabla_x, \nabla_v) u (x,v) = \mathcal{F}^{-1}\left\{\psi(k,\eta)\widehat{u}(k,\eta)\right\}(x,v),\\
\Delta_N u (x,v) &=& \phi_N(\nabla_x, \nabla_v) u (x,v) = \mathcal{F}^{-1}\left\{\phi_N(k,\eta)\widehat{u}(k,\eta)\right\}(x,v).
\end{eqnarray}
By the fact that these functions give a partition of unity in the Fourier space, we have the Littlewood-Paley decomposition of $u$:
\begin{equation}
u(x,v) = S u (x,v) + \sum_{N \in \mathbb{D}}\Delta_N u (x,v).
\end{equation}
Since each term in the summation above is localized to a compact region of Fourier space, each term is clearly in $C^{\infty}(\mathbb{Z}^d\times \mathbb{R}^d)$.  The decay (or lack thereof) of these terms at infinity depends on the regularity of $u$.

Let $\mathbb{D}_0 = \{0,1, 2, 4, 8, \ldots, 2^j, \ldots\} = \mathbb{D} \cup \{0\}. $ We will abbreviate the terms appearing in the summation above as follows:
\begin{eqnarray}
u_0(x,v) &=& S u (x,v),\\
u_N(x,v) &=& \Delta_N u (x,v),\\
u_{< N}(x,v) &=& \sum_{N' \in \mathbb{D}_0 \; : \; N' < N} u_{N'}(x,v).
\end{eqnarray}
Using this notation, the Littlewood-Paley decomposition of $u$ satisfies an \emph{almost orthogonality property}
\begin{equation}
\sum_{N \in \mathbb{D}_0} \|u_N\|_2^2 \le \|u\|_2^2 \le 2 \sum_{N \in \mathbb{D}_0} \|u_N\|_2^2,
\end{equation}
and an \emph{approximate projection property}
\begin{equation}
\|(u_N)_N\|_2 \le \|u_N\|_2 .
\end{equation}

When we apply the Littlewood-Paley decomposition to the spatial density $\rho(t,x)$, it will be convenient to use the following notation (which is motivated by the fact that $\widehat{\rho}$ is an evaluation of $\widehat{\Phi}$):
\begin{eqnarray}
\widehat{\rho}_0(t,k) &=& \psi(k,kt/L)\widehat{\rho}(t,k),\\
\widehat{\rho}_N(t,k) &=& \phi_N(k,kt/L)\widehat{\rho}(t,k),\\
\widehat{\rho}_{< N}(t,k) &=& \sum_{N' \in \mathbb{D}_0 \; : \; N' < N} \widehat{\rho}_{N'}(t,k).
\end{eqnarray}

\subsection{Paraproduct Decompositions}

Having established our notation for the Littlewood-Paley decomposition of an $L^2$-function $u$, we are now in a position to discuss the decomposition of products of such functions.  Let $u$ and $v$ be two $L^2$-functions.  Expanding each via the Littlewood-Paley decomposition, we can decompose their product into the following three terms (all sums are understood to be over $\mathbb{D}_0$):
\begin{eqnarray}
uv &=& \sum_{N \ge 8} u_{<N/8}v_N + \sum_{N \ge 8}u_{N}v_{<N/8} + \sum_{N}\sum_{N/8 \le N' \le 8N}u_Nv_{N'}\nonumber\\
&=& T_u v + T_v u + \mathcal{R}(u,v).
\end{eqnarray}
Looking at $T_u v$, we see that this term contains only higher derivatives of $v$ while having a useful frequency cut-off on the function $u$.  The same is true of $T_v u$ with the roles of $u$ and $v$ reversed.  The term $\mathcal{R}(u,v)$ contains the contributions from comparable frequencies in $u$ and $v$.

\section{Useful Inequalities}

The following useful inequalities are established in \cite{BMM13} (and so we refer to this reference for all proofs).  The first set are a variant of Young's Inequality (Lemma  3.1 in \cite{BMM13})  while the second set are proven by the paraproduct decomposition above (Lemma 3.3 in \cite{BMM13}).

\begin{lem}\label{young_var_lem}
\begin{itemize}
\item[(i)]  Let $f(k,\eta), g(k, \eta) \in L^2(\mathbb{Z}^d\times \mathbb{R}^d)$ and $\langle k \rangle^{\sigma}h(t,k) \in L^2(\mathbb{Z}^d)$ for all $t$ and some $\sigma > d/2.$  Then for any $t \in \mathbb{R}$, there is a constant $C = C(d, \sigma)$ so that
    \begin{align}
    \left|\sum_{k,\ell}\int f(k,\eta)h(t,\ell)\right.&\left.g(k-\ell,\eta-t\ell)\phantom{\int}\!\!\! d\eta\right| \nonumber\\
    &\le C(d,\sigma) \|f\|_{L^2_{k,\eta}}\|g\|_{L^2_{k,\eta}}\|\langle k \rangle^{\sigma}h(t,\cdot)\|_{L^2_{k}}.
    \end{align}

\item[(ii)]  Let $f(k,\eta), \langle k \rangle^{\sigma}g(k, \eta) \in L^2(\mathbb{Z}^d\times \mathbb{R}^d)$ and $h(t,k) \in L^2(\mathbb{Z}^d)$ for all $t$ and some $\sigma > d/2.$  Then for any $t \in \mathbb{R}$, there is a constant $C = C(d, \sigma)$ so that
    \begin{align}
    \left|\sum_{k,\ell}\int f(k,\eta)h(t,\ell)\right.&\left.g(k-\ell,\eta-t\ell)\phantom{\int}\!\!\! d\eta\right| \nonumber\\
    &\le C(d,\sigma) \|f\|_{L^2_{k,\eta}}\|\langle k \rangle^{\sigma}g\|_{L^2_{k,\eta}}\|h(t,\cdot)\|_{L^2_{k}}.
    \end{align}
\end{itemize}
\end{lem}

Remark:  The restriction $\sigma > d/2$ is essential in the proof.  For instance in the proof of (i), we need the fact that
\begin{eqnarray*}
\|h(t,\cdot)\|_{L^1} &=& \sum_{\ell}|h(t,\ell)|\\
&=& \sum_{\ell}\langle k \rangle^{-\sigma}|\langle k \rangle^{\sigma}h(t,\ell)|\\
&\le& \|\langle k \rangle^{-\sigma}\|_{L^2}\|\langle k \rangle^{\sigma}h(t,\cdot)\|_{L^2},
\end{eqnarray*}
which is only sensible for $\sigma$ large enough.

\begin{lem}\label{product_lem}
For all $0 < s < 1$ and all $\sigma \ge 0$, there is a constant $\widetilde{c} = \widetilde{c}(s,\sigma) \in (0,1)$ so that for all $\lambda >0$ there is a constant $C = C(\lambda, \sigma, s, d) >0$ such that for all $f, g \in \mathcal{G}^{\lambda, \sigma ; s}(\mathbb{T}^d_L \times \mathbb{R}^d)$ and all $h(t) \in \mathcal{F}^{\lambda, \sigma ; s}(\mathbb{T}^d_L)$ we have
\begin{align}
\sum_{k \in \mathbb{Z}^d}&\sum_{\ell \in \mathbb{Z}^d}\int_{\mathbb{R}^d}\langle k, \eta \rangle^{2\sigma}e^{2\lambda\langle k, \eta \rangle^{s}}\left|\widehat{f}(k,\eta)\widehat{h}(t,\ell)\widehat{g}(k-\ell, \eta - \ell t)\right|d\eta\nonumber\\
&\le C \left(\|f\|_{\mathcal{G}^{\lambda, \sigma ; s}}\|g\|_{\mathcal{G}^{\widetilde{c}\lambda, 0 ; s}}\|h(t)\|_{\mathcal{F}^{\lambda, \sigma ; s}}+\|f\|_{\mathcal{G}^{\lambda, \sigma ; s}}\|g\|_{\mathcal{G}^{\lambda, \sigma ; s}}\|h(t)\|_{\mathcal{F}^{\widetilde{c}\lambda, 0 ; s}}\right).
\end{align}
Moreover, there is a constant $C = C(\lambda, \sigma, s, d) >0$ so that we have the following algebra property
\begin{align}
\sum_{k \in \mathbb{Z}^d}\int_{\mathbb{R}^d}\left|\sum_{\ell \in \mathbb{Z}^d}\right.\langle k, \eta \rangle^{\sigma}e^{\lambda\langle k, \eta \rangle^{s}}\widehat{h}(t,\ell)&\left.\widehat{f}(k-\ell, \eta - \ell t)\phantom{\int}\!\!\!\!\right|^2d\eta\nonumber\\
 &\le C \|h(t)\|^2_{\mathcal{F}^{\lambda, \sigma ; s}}\|f\|^2_{\mathcal{G}^{\lambda, \sigma ; s}}.
\end{align}
\end{lem}

We also need the following trace lemma (which follows by induction on the co-dimension from the standard $H^{1/2}$ Trace Lemma).

\begin{lem}\label{trace_lem}
Let $f$ be any smooth function and $\ell_{a,b} = \{at+b : a,b \in \mathbb{R}^3, t\in \mathbb{R}\}$ be an arbitrary straight line in $\mathbb{R}^3$.  Then for any $M > 1$ there is a constant $C = C(M)>0$ so that
\begin{equation}
\|f\|_{L^2(\ell_{a,b})} \le C \|f\|_{H^{M}}.
\end{equation}
\end{lem}

Remark: This trace lemma is useful since (for instance) $\widehat{\rho}(t,k)$ is the restriction of $\widehat{\Phi}(t,k,\eta)$ to the line $\eta = kt/L$ for fixed $k$.

\section{Assumptions on Initial and Background Data and Statement of the Main Theorem}

\subsection{A Priori Assumptions}

We shall assume throughout that the solution $\Phi(t,x,v)$ of \eqref{main_pde_phi} - \eqref{main_ic_phi} launched by initial data $\Phi_{in}(x,v)$ has global existence in the classical sense for all $t \in [0,\infty)$.  Given that negative energy initial data for the rVP with the attractive potential exhibit collapse in finite time (at least for the full space, $\mathbb{R}^6$), this is not a completely unrestrictive assumption.  However, since we will take small initial data (in a sense to be made precise below), we expect that the full solution to rVP will remain close to the uniform background; in particular, the total energy of the system should remain positive.  Hence, global existence is not a completely unreasonable assumption in the cases under study.

We will also assume that for a classical solution of rVP, quantities like $\left\| A \Phi(t) \right\|_{L^2_xL^2_v}$ are continuous in $t$.  Roughly speaking, we need to know that if $\Phi_{in}$ is in some Gevrey class, then there is some time $T$ so that the solution $\Phi(t)$ launched by this initial data remains in this Gevrey class for $0 \le t \le T$.  Presumably, this can be guaranteed by appealing to an abstract form of the Cauchy-Kovalevskaya Theorem (c.f. \cite{N72}). We will leave this question as a loose end to be tied up in future research (with reasonable confidence that it can be resolved positively).

Most importantly, we will need to assume that there is a fixed $\varkappa \in (0,1)$ so that
\begin{equation}
\left\|v A^{(\sigma + m)} \Phi(t) \right\|_{L^2_xL^2_v} \le \varkappa \left\| A^{(\sigma + m )} \Phi(t) \right\|_{L^2_xL^2_v}, \tag{A} \label{apriori_ass}
\end{equation}
holds for all $t$  and all $m \in [-\beta,1]$.\symbolfootnote[2]{This may be an overly strong assumption.  We really only need \eqref{apriori_ass} to be true for $m \in \{-\beta, 0, 1\}$.}  The exact value of $\beta$ is largely arbitrary, but we will at least need $\beta > 2$.  Note that the upper bound $|v| \le 1$ for all velocities makes this inequality true with $\varkappa = 1$ for any classical solution.  We will only need this stronger assumption in one place below - which we will detail as we come to it.  On the Fourier side, this inequality is equivalent to
\begin{equation}
\left\|\nabla_{\eta} A^{(\sigma + m)} \widehat{\Phi}(t) \right\|_{L^2_kL^2_{\eta}} \le 2\pi \varkappa \left\| A^{(\sigma + m )} \widehat{\Phi}(t) \right\|_{L^2_kL^2_{\eta}},
\end{equation}
which is a sort of ``reverse Poincar\'e inequality'' for $A^{(\sigma + m)} \widehat{\Phi}$.  Note that \eqref{apriori_ass} is implied by the slightly stronger assumption
\begin{equation}
V(t) \equiv \sup \left\{|v| : \exists x \in \mathbb{T}^3_L \textrm{ s.t. } (x,v) \in \textrm{supp }\Phi(t) \right\} \le \varkappa < 1,
\end{equation}
for all $t \in [0,\infty)$ (i.e. speeds are uniformly bounded away from the speed of light for all time).  Though this condition is stronger, it is the typical sort of criterion used to establish global existence for solutions of rVP.  See for instance Theorems I and II of \cite{GS85} where choosing initial data small enough (in $L^{\infty}$-norm) implies a global in time upper bound on the momentum (and hence a bound on velocities away from the speed of light).  However, \eqref{apriori_ass} does not seem to hinge on such a strong bound on the velocities.  In fact, it only seems to require that very little of the $L^2$-norm of $A^{(\sigma + m)} \Phi(t)$ be contributed from the region $|v| > \varkappa$ for all $t$.  Further investigation of the merits of this assumption would be an interesting future project.

\subsection{Primary Assumptions}

The linear portion of the equation for $\rho$ is given by
\begin{equation}
\widehat{\rho}(t,k) = \widehat{\Phi_{in}}\left(k,\frac{kt}{L}\right) + \int_0^t L(t-\tau,k)\widehat{\rho}(\tau,k) d\tau + \textrm{ non-linear terms},
\end{equation}
where we have defined
\begin{equation}
L(t,k) = 2\pi i \widehat{W}(k)\frac{k}{L}\cdot \widehat{\alpha G_0}\left(\frac{kt}{L}\right).
\end{equation}
Given the form of this equation, the solution to the linearized equation can be determined by Laplace Transform - provided the transforms of the various quantities involved are reasonably well-behaved.

We will at the very least suppose that $\alpha G_0$ is $C_c(\mathbb{R}^d)$ as a function of $v$ with support in $B_1(0)$.  Since this function cannot be analytic no matter how much regularity we impose, we cannot expect its Fourier Transform to decay exponentially.  However, if we assume $\alpha G_0$ belongs to $G^{1/\nu}_c(\mathbb{R}^d)$ (the set of Gevrey-$\frac{1}{\nu}$ functions of compact support) for some $0 < \nu < 1$, then standard facts about these functions guarantees the existence of constants $C$ and $\epsilon$ so that
\begin{equation}
|\widehat{\alpha G_0}(\eta)| \le Ce^{-\epsilon |\eta|^{\nu}}.
\end{equation}
We will actually assume something slightly more specific:

\bigskip

\noindent \textbf{Primary Assumptions on $G_0$:}

There exist constants $C_0, \overline{\lambda},\kappa >0$, and $\nu \in \left((2+\gamma)^{-1}, 1\right)$ so that
\begin{equation}
\|G_0\|^2_{\mathcal{G}^{\overline{\lambda},0;\nu}} + \| \alpha G_0\|^2_{\mathcal{G}^{\overline{\lambda},0;\nu}}\le C_0,\tag{B1}\label{ass_G_1}
\end{equation}
and for all $z$ with $\Re(z)>-\overline{\lambda}$
\begin{equation}
\inf_{k \in \mathbb{Z}^3\setminus \{(0,0,0)\}}\left|\mathcal{L}[L](|k|z,k) - 1\right| > \kappa,\tag{B2}\label{ass_G_2}
\end{equation}
where $\mathcal{L}[L]$ is the Laplace Transform of the integral kernel $L$ (we give our conventions for this transform below).

\bigskip

Recall that $\gamma \ge 1$ appears in the estimate for $\widehat{W}$ (the Coulombic/Newtonian case is $\gamma = 1$).  Note that in the norms defining the space $\mathcal{G}^{\overline{\lambda},0;\nu}$, the only surviving term in the sum is $k=0$ (as $G_0$ does not depend on $x$).  Also note the critical factor of $|k|$ multiplying $z$ in \eqref{ass_G_2}.  Our convention for the Laplace transform is (the slightly unconventional)
\begin{eqnarray}
\mathcal{L}[f](z,k) &=& \int_0^{\infty}f(t,k)e^{-2 \pi zt}dt.
\end{eqnarray}
The reason for defining things this way is to make the connection with the Fourier Transform more straightforward (which will only play a significant role in the appendix below).  The condition on $\alpha G_0$ is in general not implied by the condition on $G_0$ and vice versa.  Note that $\alpha(v)$ is not in any Gevrey class on the unit ball (as its derivatives blow up at the boundary).  Hence, we need to know that $G_0$ vanishes rapidly enough at the boundary of the unit ball to handle these divergences.

We will comment on assumption \eqref{ass_G_2} for a special class of spherically symmetric initial data in the appendix.  Much of this is similar to results found in \cite{Y14}.  As we require controls that are uniform in the size of the wavevector (as opposed to the simpler mode-by-mode analysis undertaken in the previous study), our results are more modest.

As for Assumption \eqref{ass_G_1}, it tells us immediately that $G_0$ is in the Gevrey class of order $\nu^{-1}$ (which is essentially equivalent to the transform decaying like $\exp(-\langle \eta \rangle^{\nu})$).  Moreover, as
$$\|f\|^2_{\mathcal{G}^{\overline{\lambda},0;\nu}} = \left\| A^{(\overline{\lambda},0;\nu)}(0,0,\cdot)\widehat{f}\right\|_{L^2_{\eta}}, $$
by definition, we know that $A^{(\overline{\lambda},0;\nu)}(0,0,\eta) \widehat{G_0}(\eta)$ decays fast enough at infinity to be in $L^2$ (which is substantially stronger than merely knowing which Gevrey class the function is in).  The preceding statements are equally true for $\widehat{\alpha G_0}(\eta)$ .

As an immediate consequence of this assumption, we have:

\begin{cor}\label{cor_G_1}
For background data satisfying \eqref{ass_G_1} and any $M \in \mathbb{N}$, there is a constant $C_M$ (depending only on $M$) so that
\begin{equation}
\left\|A^{(\overline{\lambda},0;\nu)} \widehat{G_0} \right\|_{H^M_{\eta}} +  \left\|A^{(\overline{\lambda},0;\nu)} \widehat{\alpha G_0} \right\|_{H^M_{\eta}}\le C_MC_0.
\end{equation}
\end{cor}

\textbf{Proof:}  The proof proceeds much like the proof of Lemma \ref{eta_deriv_lem}.  Namely, for any sufficiently nice function $f$ defined on the $B_1(0)$, we have
\begin{eqnarray}
\left\|A^{(\overline{\lambda},0;\nu)} \widehat{f} \right\|^2_{H^M_{\eta}} &=& \sum_{|m| \le M} \left\|D^m_{\eta} A^{(\overline{\lambda},0;\nu)} \widehat{f} \right\|^2_{L^2_{\eta}}\nonumber\\
&=& \sum_{|m| \le M} (2\pi)^{|m|}\left\||v^m| A^{(\overline{\lambda},0;\nu)} f \right\|^2_{L^2_{v}}\nonumber\\
&\le& (2\pi)^{M} M^3 \left\|A^{(\overline{\lambda},0;\nu)} \widehat{f} \right\|^2_{L^2_{\eta}}
\end{eqnarray}
where we have abbreviated $A^{(\overline{\lambda},0;\nu)} \widehat{f}(\eta) = A^{(\overline{\lambda},0;\nu)}(0,0,\eta) \widehat{f}(\eta).$  The factor $M^3$ is a (very) crude upper bound on the number of multi-indices in $\mathbb{N}^3$ of order not more than $M$.  This gives the corollary. $\blacksquare$

\bigskip

Recall that the standard Sobolev embedding theorem guarantees us that $$H^M(\mathbb{R}^3) \hookrightarrow C^{M-2, 1/2}(\mathbb{R}^3).$$  As a result of the corollary above, we see that both
$A^{(\overline{\lambda},0;\nu)} \widehat{G_0}$ and $A^{(\overline{\lambda},0;\nu)} \widehat{\alpha G_0}$ are infinitely differentiable as functions of $\eta$.  Moreover, we have the precise estimate
\begin{equation}
\left\| A^{(\overline{\lambda},0;\nu)} \widehat{G_0}  \right\|_{C^M_{\eta}} + \left\| A^{(\overline{\lambda},0;\nu)} \widehat{\alpha G_0}  \right\|_{C^M_{\eta}} \le C_sC_{M+2}C_0, \label{CM_est_back}
\end{equation}
where $C_s$ is the constant arising from the Sobolev embedding (and $C_{M+2}$ and $C_0$ are as in the corollary).

We make similar assumptions for the initial data:

\bigskip

\noindent \textbf{Primary Assumptions on $\Phi_{in}$:}

Given $\sigma \ge 5, \lambda_0 >0$ and $\overline{\nu} \in \left((2+\gamma)^{-1},\nu\right)$ (where $\nu$ is set by the assumption on the background data), there is an $\epsilon \in (0,\epsilon_0]$ (with the upper bound $\epsilon_0$ to be determined) so that
\begin{equation}
\left\|\Phi_{in}\right\|_{\mathcal{G}^{\lambda_0,\sigma;\overline{\nu}}} + \left\|\alpha \Phi_{in}\right\|_{\mathcal{G}^{\lambda_0,\sigma;\overline{\nu}}}< \epsilon. \tag{I1}\label{ass_Phi_1}
\end{equation}
From our initial setup, we also require
\begin{equation}
\widehat{\Phi_{in}}(0,0) = \frac{1}{L^3}\int_{\mathbb{T}^3_L}\int_{B_1(0)} \Phi_{in}(x,v) dv dx = 0. \tag{I2}\label{ass_Phi_2}
\end{equation}
In addition, we assume that the solution launched by $\Phi_{in}$ satisfies the \emph{a priori} assumptions detailed above (most critically, the existence of $\varkappa \in (0,1)$ given by \eqref{apriori_ass}). The Sobolev factor $\langle k, \eta \rangle^{\sigma}$ is a bit redundant since we can absorb it into the exponential factor appearing in the Fourier multiplier $A$ at the expense of making $\lambda_0$ slightly bigger (`slightly' here means \emph{any} number bigger than $\lambda_0$ will do).  We insert it to be precise since we will need this factor in subsequent calculations.

Just as for the background data, we are forced to assume $\alpha\Phi_{in}(x,v)$ is in an appropriate Gevrey space (in addition to the bound on $\Phi_{in}(x,v)$ itself).  Since the derivatives of $\alpha$ blow up at the boundary of the unit ball, this amounts to assuming $\Phi_{in}$ vanishes rapidly as $|v| \to 1$.  As with the background data, our assumption guarantees a high amount of regularity in the variable $\eta$:

\begin{cor}\label{cor_Phi_1}
For initial data satisfying \eqref{ass_Phi_1} and any $M \in \mathbb{N}$, there is a constant $\widetilde{C_M}$ (depending only on $M$) so that
\begin{equation}
\left\|A^{(\lambda_0,\sigma;\overline{\nu})} \widehat{\Phi_{in}} \right\|_{L^2_k H^M_{\eta}} +  \left\|A^{(\lambda_0,\sigma;\overline{\nu})} \widehat{\alpha \Phi_{in}} \right\|_{L^2_k H^M_{\eta}}\le \widetilde{C_M}\epsilon.
\end{equation}
\end{cor}

The proof is essentially identical to the one above (the summation over $k$ plays no role in the analysis).  Just as above, the standard Sobolev embedding gives us the immediate estimate
\begin{equation}
\left\| A^{(\lambda_0,\sigma;\overline{\nu})} \widehat{\Phi_{in}}  \right\|_{L^2_kC^M_{\eta}} + \left\| A^{(\lambda_0,\sigma;\overline{\nu})} \widehat{\alpha \Phi_{in}}  \right\|_{L^2_kC^M_{\eta}} \le C_s'C_{M+2}\epsilon.\label{CM_est_init}
\end{equation}

\subsection{Main Theorem}

\begin{thm}[Landau Damping for Toroidal, Relativistic Plasmas] \label{main_thm}
Let $G_0$ be a given background datum satisfying Assumptions \eqref{ass_G_1} and \eqref{ass_G_2} for some choice of parameters $C_0, \overline{\lambda},\kappa >0$, and $\nu \in \left((2+\gamma)^{-1}, 1\right)$.  Fix parameters $\sigma \ge 5, \lambda_0 >0$, $\overline{\nu} \in \left((2+\gamma)^{-1},\nu\right),$ and $\varkappa \in (0,1)$.  Then there is an $\epsilon_0 > 0$ depending on these various parameters so that for solutions $\Phi(t)$ to \eqref{main_pde_phi} - \eqref{main_ic_phi} launched by $\Phi_{in}$ satisfying the \emph{a priori} assumptions, \eqref{ass_Phi_1}, and \eqref{ass_Phi_2}, the following holds:  For any $\lambda'\in (0, \lambda_0)$ there exist positive constants $C_1$ and $C_2$ and a $\Phi_{\infty} \in \mathcal{G}^{\lambda',0;\overline{\nu}}(\mathbb{T}^3_L \times B_1(0))$ which is mean zero such that for all $t \ge 0$ we have
\begin{eqnarray}
\left\| \Phi(t) - \Phi_{\infty}\right\|_{\mathcal{G}^{\lambda',0;\overline{\nu}}} &\le& C_1 \epsilon e^{-\frac{1}{2}(\lambda_0-\lambda')t^{\overline{\nu}}} \label{thm_rate_1}\\
\left\| \rho(t) \right\|_{\mathcal{F}^{\lambda',0;\overline{\nu}}} &\le& C_2 \epsilon e^{-\frac{1}{2}(\lambda_0-\lambda')t^{\overline{\nu}}} \label{thm_rate_2}.
\end{eqnarray}
\end{thm}

Tracing back through the various definitions, this is equivalent to the statement that a solution $g(t,q,v)$ of \eqref{rvp_v1} - \eqref{rvp_v2} will converge exponentially fast to some function $g_{\infty}(q-vt,v)$ (i.e. a free-streaming solution) if the initial datum is sufficiently small in an appropriate norm.  We note that the function $\Phi_{\infty}$ in the theorem is completely determined by the choice of background and initial data and is independent of $\lambda'$.

\subsection{Bootstrap Estimates and Proof of Theorem \eqref{main_thm}}

The proof of the theorem above will follow closely the procedure in \cite{BMM13} for the non-relativistic case.  We begin with an examination of damping in the linearized system.  Here we will use Assumptions \eqref{ass_G_1} and \eqref{ass_G_2} to control the contribution from the linear term in the evolution of $\widehat{\rho}$.  After we deal with the linear contribution, we are left with the ostensibly smaller (but more intricate) non-linear terms.  Given the local existence of solutions, the continuity of the following terms, and our assumption on the smallness of the initial data,  we know there is some interval $I=[0,T)$ so that the following estimates hold for $t \in I$:
\begin{align}
\left\|A^{(\sigma+1)}\widehat{ \Phi}(t) \right\|^2_{L^2_kL^2_{\eta}}\!\!- \left\|\mathcal{F}\{v A^{(\sigma+1)} \Phi\}(t) \right\|^2_{L^2_kL^2_{\eta}}\!\!+ \left\|A^{(\sigma+1)}\widehat{\alpha \Phi}(t) \right\|^2_{L^2_kL^2_{\eta}} \!\! & \le \!\!  4K_1\epsilon^2 \langle t \rangle^{6},\label{orig_bstrap_AI}\\
\left\|A^{(\sigma-\beta)}\widehat{\Phi}(t)\right\|_{L^2_kL^2_{\eta}}^2-\left\|\mathcal{F}\{v A^{(\sigma-\beta)} \Phi\}(t)\right\|_{L^2_kL^2_{\eta}}^2+\left\|A^{(\sigma-\beta)}\widehat{\alpha \Phi}(t)\right\|_{L^2_kL^2_{\eta}}^2 &\le  4K_2 \epsilon^2\label{orig_bstrap_AbI},\\
\int_0^t \|A\widehat{\rho}(\tau)\|^2_{L^2_k} d\tau &\le 4K_3\epsilon^2 \label{orig_bstrap_rho},
\end{align}
where we have set $$A(t,k,\eta) = A^{(\lambda, \sigma ; \overline{\nu})}(t,k,\eta) = \langle k, \eta \rangle^{\sigma} e^{\lambda(t)\langle k, \eta \rangle^{\overline{\nu}}}.$$  It turns out that the choice of $\beta$ is rather arbitrary.  We will only need $\beta \in (2, \sigma - 5/2)$.  Note that the differences appearing in both \eqref{orig_bstrap_AI} and \eqref{orig_bstrap_AbI} are positive by Plancherel's Theorem and the fact that $|v| \le 1$.

A few comments on the form of $\lambda(t)$ are in order.  For much of the analysis, all we need to know about $\lambda(t)$ is that it is decreasing and bounded.  Some information about the decay of $\lambda(t)$ is necessary to handle linear damping, but a specific form is only necessary for the proofs of \eqref{reaction_control} and \eqref{reaction_control_2} below (the form of $\lambda(t)$ is largely determined by an analysis of the \emph{plasma echo phenomenon} which is crucial in establishing these estimates).  Moreover, it is in proving these two estimates (and only here) that the lower bound $\nu > \overline{\nu} > (2+\gamma)^{-1}$ mentioned in the assumptions is necessary.  To give a concrete form to these general musings, we can take
\begin{equation}
\lambda(t) = \frac{1}{8}(\lambda_0 - \lambda')(1-t)_+ + \alpha_0 + \frac{1}{4}(\lambda_0-\lambda')\min\left\{1,\frac{1}{t^{a}}\right\},\label{lambda_def}
\end{equation}
where
\begin{eqnarray}
\alpha_0 &=& \frac{1}{2}(\lambda_0 + \lambda'), \label{alpha0_def}\\
a &=& \frac{(2+\gamma)\overline{\nu}-1}{1+\gamma}>0\label{a_def}.
\end{eqnarray}
By $(1-t)_+$, we denote the function that equals $1-t$ for $0\le t \le 1$ and zero otherwise.  Note that
\begin{equation}
\alpha_0 \le \lambda(t) \le \lambda(0) = \frac{7}{8}\lambda_0 + \frac{1}{8}\lambda' < \lambda_0.
\end{equation}
The derivative $\dot{\lambda}(t)$ exists for almost all $t\ge0$ (in fact, it exists for all $t\ne1$), is non-vanishing when it does exist, and satisfies
\begin{equation}
\dot{\lambda}(t) \le -C(\lambda_0-\lambda')\langle t \rangle^{-1-a},
\end{equation}
for some constant $C$.

Given these estimates, we can prove the following result:

\begin{prp}[Bootstrapping]\label{bstrap_prp}
There exist constants $\epsilon_0$ and $K_1, K_2,$ and $K_3$  depending on $\varkappa,\overline{\lambda}, \lambda_0, \lambda', C_0, \overline{\nu}$ and $\kappa$ so that if \eqref{orig_bstrap_AI} -- \eqref{orig_bstrap_rho} hold for $t \in [0,T)$ with $\epsilon < \epsilon_0$, then for $t\in[0,T]$ we have
\begin{align}
\left\|A^{(\sigma+1)}\widehat{ \Phi}(t) \right\|^2_{L^2_kL^2_{\eta}}\!\!-\left\|\mathcal{F}\{v A^{(\sigma+1)} \Phi\}(t) \right\|^2_{L^2_kL^2_{\eta}}\!\!+\left\|A^{(\sigma+1)}\widehat{\alpha \Phi}(t) \right\|^2_{L^2_kL^2_{\eta}} \!\! & \le \!\! 2K_1\epsilon^2 \langle t \rangle^{6}, \label{bstrap_AI} \\
\left\|A^{(\sigma-\beta)}\widehat{\Phi}(t)\right\|_{L^2_kL^2_{\eta}}^2-\left\|\mathcal{F}\{v A^{(\sigma-\beta)} \Phi\}(t)\right\|_{L^2_kL^2_{\eta}}^2+\left\|A^{(\sigma-\beta)}\widehat{\alpha \Phi}(t)\right\|_{L^2_kL^2_{\eta}}^2 & \le \!\! 2K_2 \epsilon^2, \label{bstrap_AbI}\\
\int_0^t \|A\widehat{\rho}(\tau)\|^2_{L^2_k} d\tau\!\! &\le\!\! 2K_3\epsilon^2 \label{bstrap_rho},
\end{align}
from which it follows that $T=+\infty$.
\end{prp}

In proving Theorem \eqref{main_thm}, we actually only need the estimates on $\widehat{\alpha \Phi}$.  Since the time derivative of this quantity involves $\widehat{\Phi} - \widehat{|v|^2\Phi},$ we will need the other terms to propagate the bootstrap estimates for all $T$.  Note the positivity of the difference allows us to use the same upper bound for $\widehat{\alpha \Phi}$ alone.  Our \emph{a priori} assumption \eqref{apriori_ass} allows us to conclude that
\begin{equation}
0 < \left(1-\varkappa^2\right)\left\| A^{(\sigma+m)}\widehat{\Phi}(t) \right\|_2^2 \le \left\| A^{(\sigma+m)}\widehat{\Phi}(t) \right\|_2^2 - \left\|\mathcal{F}\{v A^{(\sigma+m)} \Phi\}(t) \right\|_2^2.
\end{equation}
Hence, our bootstrap estimates provide similar upper-bounds $\widehat{\Phi}$ (at the expense of a factor of $(1-\varkappa^2)^{-1}$).  By Lemma \ref{eta_deriv_lem}, these estimates also provide $L^2_kL^2_{\eta}$-estimates on quantities like $A^{(\sigma+1)}\widehat{v^m\alpha\Phi}(t)$ (at the expense of a constant depending on the multi-index $m$ and the parameters appearing in Fourier multiplier $A$).    We will need to be keep close track on how the constants $K_i$ above depend on one another to ensure they can be chosen appropriately.  Armed with these estimates, the proof of Theorem \ref{main_thm} becomes relatively simple.

As a consequence of these estimates, we note the following immediate corollary.  Define $M[\widehat{\alpha \Phi}]$ as the following symmetric matrix of transforms:
\begin{eqnarray}
M[\widehat{\alpha \Phi}](t,k,\eta) &=& \mathcal{F}\left\{(\mathds{1} - v\otimes v)\alpha \Phi\right\}(t,k,\eta)\nonumber\\
&=& \left[\begin{array}{ccc} \widehat{(1-v_1^2) \alpha \Phi} & -\widehat{v_1v_2\alpha \Phi} & -\widehat{v_1v_3\alpha \Phi}\\ -\widehat{v_1v_2\alpha \Phi} & \widehat{(1-v_2^2) \alpha \Phi} & -\widehat{v_2v_3\alpha \Phi}\\ -\widehat{v_1v_3\alpha \Phi} & -\widehat{v_2v_3\alpha \Phi} &  \widehat{(1-v_3^2) \alpha \Phi}\end{array}\right](t,k,\eta).
\end{eqnarray}
If we take
$$ \left\| A^{(\lambda, \sigma ; \overline{\nu})} M[\widehat{\alpha \Phi}](t) \right\|_2^2 = \sum_{i,j =1}^3 \left\| A^{(\lambda, \sigma ; \overline{\nu})} M[\widehat{\alpha \Phi}](t)_{i,j}\right\|_2^2,  $$
then by commuting $A$ past the derivatives coming from the $v_iv_j$ terms and using \eqref{est_on_DA}, we find that
\begin{equation}
\left\| A^{(\lambda, \sigma ; \overline{\nu})} M[\widehat{\alpha \Phi}](t) \right\|_2 \le C \left\|A^{(\lambda, \sigma ; \overline{\nu})}\widehat{ \alpha \Phi}(t) \right\|_2,
\end{equation}
where $C$ depends only on $\lambda, \sigma,$ and $\overline{\nu}$.  Hence, our bootstrap estimates give rise to analogous estimates for $M[\widehat{\alpha \Phi}]$.

We will present the proof of the bootstrap proposition in the next several sections.  We will also need some preliminary estimates whose proof we defer.  We make the following definition for convenience in subsequent calculations:
\begin{align}
\overline{K}_{k,\ell}(t,\tau) = \frac{2\pi c_1 C_W}{L}\frac{|k|(t-\tau)}{|\ell|^{\gamma}}& e^{(\lambda(t)-\lambda(\tau))\langle k , kt/L \rangle^{\overline{\nu}}} e^{c_2\lambda(\tau)\langle k - \ell, (kt - \ell \tau)/L \rangle^{\overline{\nu}}}\nonumber\\
&\cdot \sum_{|m| \le 2}\left|\widehat{v^m \alpha \Phi}\left(\tau, k-\ell, \frac{kt -\ell \tau}{L} \right)\right|\mathds{1}_{\ell \ne 0},\label{def_Kbar}
\end{align}
where $m$ runs over all multi-indices in $\mathbb{N}^3$ of order not more than two.  The constants $c_1=c_1(\sigma)$ and $c_2 = c_2(\overline{\nu}) \in (0,1)$ will come from estimates made below (through an application of Lemma \ref{angle_brak_lem}[item (iii)]).  In terms of $\overline{K}$, our bootstrap assumptions imply
\begin{align}
\sup_{\tau \ge 0}e^{-(1-c_2)\lambda(0)\langle \tau \rangle^{\overline{\nu}}} \sum_{|m|\le 2}\sum_{k} \sup_{\omega \in \mathbb{Z}^3_{\ne 0} }&\sup_{\zeta \in \mathbb{R}^3}\int_{-\infty}^{\infty}\left|A^{(\sigma + 1)}\widehat{v^m \alpha \Phi}\left(\tau,k, \frac{\omega}{|\omega|}s - \zeta\right)\right|^2ds \nonumber\\
 &\le  C K_1 \epsilon^2 \label{transport_control}\\
\left(\sup_{t\ge 0}\sup_{k}\int_0^t\sum_{\ell \ne 0}\overline{K}_{k,\ell}(t,\tau)d\tau\right)&\left(\sup_{\tau\ge 0}\sup_{\ell \ne 0}\int_{\tau}^{\infty}\sum_{k}\overline{K}_{k,\ell}(t,\tau)dt\right)  \le C K_2 \epsilon^2, \label{reaction_control}\\
\sup_{0\le \tau \le t} \sup_{\ell \ne 0} &\sum_{k \ne 0}\overline{K}_{k,\ell}(t,\tau) \le C \sqrt{K_2}\epsilon \langle t \rangle, \label{reaction_control_2}
\end{align}
where the constant $C$ depends on the various parameters listed in the assumptions.  We will first see how the bootstrap proposition implies our main theorem.  Once that is complete, we will then embark on the lengthy proof that \eqref{orig_bstrap_AI} - \eqref{orig_bstrap_rho} imply the bootstrap proposition.  Finally, we show how our bootstrap assumptions imply the crucial inequalities above.

\bigskip

\textbf{\emph{Proof of Theorem \ref{main_thm}:}}

\bigskip

We assume Proposition \ref{bstrap_prp} and show how it implies the proof of Theorem \ref{main_thm}.  First, note that by the second part of Lemma \ref{product_lem} and our assumption on $\widehat{W}$ we have
\begin{align}
\int_0^t &\left\| F(\tau, x+v\tau) \cdot \left\{\alpha G_0 + \left[\mathds{1}-v\otimes v\right](\nabla_v - \tau \nabla_x)\alpha \Phi(\tau) - 4v \alpha \Phi (\tau)\right\} \right\|_{\mathcal{G}^{\alpha_0,0;\overline{\nu}}} d\tau\nonumber\\
& \le C \int_0^t \left\|\rho(\tau)\right\|_{\mathcal{F}^{\alpha_0,0;\overline{\nu}}}\left\|\alpha G_0 + \left[\mathds{1}-v\otimes v\right](\nabla_v - \tau \nabla_x)\alpha \Phi(\tau) - 4v \alpha \Phi (\tau) \right\|_{\mathcal{G}^{\alpha_0,0;\overline{\nu}}}d\tau\nonumber\\
& \le C \int_0^t \left\|A^{(\alpha_0,0;\overline{\nu})}\widehat{\rho}(\tau)\right\|_{2}\Bigg\|A^{(\alpha_0,0;\overline{\nu})}\widehat{\alpha G_0} \nonumber\\
& \qquad \qquad + A^{(\alpha_0,0;\overline{\nu})}\left[2\pi i\left(\eta - \frac{k\tau}{L}\right)\widehat{\alpha \Phi}(\tau) + \left(\eta - \frac{k\tau}{L}\right)\cdot \nabla_{\eta}\widehat{v\alpha \Phi}(\tau)\right] \Bigg\|_{2}d\tau,
\end{align}
where the constant $C$ depends only on $\alpha_0, \overline{\nu},$ and $C_W$ (the constant appearing in the estimate for $\widehat{W}$).  Since $\alpha_0 = \frac{1}{2}(\lambda_0 + \lambda') \le \lambda(t)$ and $\sigma > \beta + 1$, we certainly have
\begin{align}
A^{(\alpha_0,0;\overline{\nu})}\left(\tau, k, \frac{k\tau}{L}\right) \le C\langle \tau \rangle^{-\sigma} A^{(\lambda,\sigma;\overline{\nu})}\left(\tau, k, \frac{k\tau}{L}\right),\nonumber\\
A^{(\alpha_0,0;\overline{\nu})}\left(\tau, k, \frac{k\tau}{L}\right) \le A^{(\lambda,\sigma - \beta - 1;\overline{\nu})}\left(\tau, k, \frac{k\tau}{L}\right).\nonumber
\end{align}
Noting that the factors of $\eta - k\tau /L$ give us another factor of the form $\langle \tau \rangle \langle k, \eta \rangle$, we see that
\begin{align}
\int_0^t &\left\| F(\tau, x+v\tau) \cdot \left\{\alpha G_0 + \left[\mathds{1}-v\otimes v\right](\nabla_v - \tau \nabla_x)\alpha \Phi(\tau) - 4v \alpha \Phi (\tau)\right\} \right\|_{\mathcal{G}^{\alpha_0,0;\overline{\nu}}} d\tau\nonumber\\
& \le C \int_0^t \langle \tau \rangle^{-\sigma + 1}\left\|A\widehat{\rho}(\tau)\right\|_{2} \left(\left\|A^{(\sigma - \beta)}|\widehat{\alpha G_0}|\right\|_2 +\left\|A^{(\sigma - \beta)}\left|M\left[\widehat{\alpha \Phi}\right](\tau)\right| \right\|_{2}\right)d\tau,
\end{align}
where as before, $C$ depends only on $\alpha_0, C_W,$ and the various parameters appearing in $A$.  H\"older's Inequality then gives
\begin{align}
\int_0^t &\left\| F(\tau, x+v\tau) \cdot \left\{\alpha G_0 + \left[\mathds{1}-v\otimes v\right](\nabla_v - \tau \nabla_x)\alpha \Phi(\tau) - 4v \alpha \Phi (\tau)\right\} \right\|_{\mathcal{G}^{\alpha_0,0;\overline{\nu}}} d\tau\nonumber\\
& \le C \left(\int_0^t \left\|A\widehat{\rho}(\tau)\right\|_{2}^2d\tau\right)^{1/2}\nonumber\\
&\qquad \cdot \left(\int_0^t\langle \tau \rangle^{-2\sigma + 2}\left(\left\|A^{(\sigma - \beta)}\widehat{\alpha G_0}\right\|_2^2 +\left\|A^{(\sigma - \beta)}M\left[\widehat{\alpha \Phi}(\tau)\right]\right\|_2^2\right) d\tau\right)^{1/2}.
\end{align}
Hence, by the Bootstrap Estimates \eqref{bstrap_AI} - \eqref{bstrap_rho} and Assumption \eqref{ass_G_1} (recall that $\overline{\nu} < \nu$), we see that
\begin{align}
\int_0^t &\left\| F(\tau, x+v\tau) \cdot \left\{\alpha G_0 + \left[\mathds{1}-v\otimes v\right](\nabla_v - \tau \nabla_x)\alpha \Phi(\tau) - 4v \alpha \Phi (\tau)\right\} \right\|_{\mathcal{G}^{\alpha_0,0;\overline{\nu}}} d\tau\nonumber\\
&\le C \epsilon
\end{align}
for all $t$.  Hence, we can define $\Phi_{\infty}$ via the absolutely convergent integral
\begin{align}
\Phi_{\infty} = & \Phi_{in}\nonumber\\
&- \int_0^{\infty}  F(\tau, x+v\tau) \cdot \left\{\alpha G_0 + \left[\mathds{1}-v\otimes v\right](\nabla_v - \tau \nabla_x)\alpha \Phi(\tau) - 4v \alpha \Phi (\tau)\right\}  d\tau.
\end{align}

Clearly,
\begin{align}
\Phi(t,x,v) &- \Phi_{\infty}(x,v)\nonumber\\
&= \int_t^{\infty}  F(\tau, x+v\tau) \cdot \left\{\alpha G_0 + \left[\mathds{1}-v\otimes v\right](\nabla_v - \tau \nabla_x)\alpha \Phi(\tau) - 4v \alpha \Phi (\tau)\right\}  d\tau,
\end{align}
and hence
\begin{align}
&\left\| \Phi(t) - \Phi_{\infty}\right\|_{\mathcal{G}^{\lambda',0,\overline{\nu}}}\nonumber\\
&\le \int_t^{\infty} \left\|F(\tau, x+v\tau) \cdot \left\{\alpha G_0 + \left[\mathds{1}-v\otimes v\right](\nabla_v - \tau \nabla_x)\alpha \Phi(\tau) - 4v \alpha \Phi (\tau)\right\}\right\|_{\mathcal{G}^{\lambda',0,\overline{\nu}}}d\tau.
\end{align}
The same estimates as above yield
\begin{align}
&\left\| \Phi(t) - \Phi_{\infty}\right\|_{\mathcal{G}^{\lambda',0,\overline{\nu}}}\nonumber\\
&\le C \int_t^{\infty}e^{-(\alpha_0 - \lambda')\langle \tau \rangle^{\overline{\nu}}} \langle \tau \rangle^{-\sigma + 1} \left\|A\widehat{\rho}(\tau)\right\|_2 \left(\left\|A^{(\sigma - \beta)}\widehat{\alpha G_0}\right\|_2 + \left\|A^{(\sigma - \beta)}M\left[\widehat{\alpha \Phi}\right](\tau)\right\|_{2}\right)d\tau\nonumber\\
&\le C \epsilon e^{-(\alpha_0 - \lambda')\langle t \rangle^{\overline{\nu}}},
\end{align}
which implies the first part of the theorem (as $\alpha_0 = \frac{1}{2}(\lambda_0+\lambda')$).

In order to prove the second part of the theorem, we need the (very) rough estimate provided by Lemma \ref{pwise_rho_lem} below.  This lemma shows us that
\begin{equation}
\|A\rho(t)\|_{L^2_x} \le C\epsilon\langle t \rangle^{1/2},
\end{equation}
where $C$ once again only depends on the various parameters appearing in $A$.  Given this result (and that $\sigma > 1/2$), we have
\begin{eqnarray}
\left\| \rho(t) \right\|_{\mathcal{F}^{\lambda',0,\overline{\nu}}} &=& \left\|A^{(\lambda',0,\overline{\nu})}\widehat{\rho}(t)\right\|_2\nonumber\\
&\le& C e^{-(\alpha_0 - \lambda')\langle t \rangle^{\overline{\nu}}} \langle t \rangle^{-\sigma } \left\|A\widehat{\rho}(t)\right\|_2\nonumber\\
&\le& C \epsilon e^{-(\alpha_0 - \lambda')\langle t \rangle^{\overline{\nu}}},
\end{eqnarray}
which finishes the proof of the theorem. $\blacksquare$

\section{Linear Damping}

Recall that the linear portion of the equation for $\rho$ is given by
\begin{equation}
\widehat{\rho}(t,k) = \widehat{\Phi_{in}}\left(k,\frac{kt}{L}\right) + \int_0^t L(t-\tau,k)\widehat{\rho}(\tau,k) d\tau + \textrm{ nonlinear terms}, \label{linear}
\end{equation}
where
\begin{equation}
L(t,k) = 2\pi i \widehat{W}(k)\frac{k}{L}\cdot \widehat{\alpha G_0}\left(\frac{kt}{L}\right)\label{Ldef}.
\end{equation}
Our goal will be to show that under the assumptions on $G_0$ and the initial datum $\widehat{\Phi_{in}}$, we have finiteness in an appropriate Gevrey norm for $\widehat{\rho}$.  To that end, we formulate the following general lemma:

\begin{lem}[Linear Control]\label{linear_lemma}
Let $G_0$ satisfy assumptions \eqref{ass_G_1} and \eqref{ass_G_2} for appropriate constants $C_0, \overline{\lambda},\kappa >0$ and $\nu \in ((2+\gamma)^{-1}, 1)$.  Let $\overline{\nu} \in ((2+\gamma)^{-1},\nu)$ and $A^{(\lambda, \sigma ; \overline{\nu})}(t,k,\eta)$ be the factor defined in \eqref{Adef} (which appears in the $\mathcal{G}^{\lambda(t), \sigma ; \overline{\nu}}$ and $\mathcal{F}^{\lambda(t), \sigma ; \overline{\nu}}$ norms).  Let $\lambda(t)$ be given by \eqref{lambda_def}.  Specifically,  we need only assume it is a non-negative, decreasing function of $t$, $0 < \alpha_0 \le \lambda(t) \le \alpha_1$, and
\begin{equation}
\dot{\lambda}(t) = \left\{ \begin{array}{ll} -C & t < 1 \\ \frac{-D}{\langle t \rangle^{1+a}} & t \ge 1 \end{array}\right. ,
\end{equation}
where $C,D>0$, $0 < a < \overline{\nu}$, and these constants are such that $\lambda(t)\langle k, kt/L \rangle$ is strictly increasing in $t$ for all $|k| \ge 1$.

Let $F(t,k)$ and the $t$-interval $I = [0,T)$ ($T>0$) be given so that
\begin{equation}
\int_0^T\|F(t)\|_{\mathcal{F}^{\lambda(t), \sigma ; \overline{\nu}}}^2dt = \sum_{k \in \mathbb{Z}^3\setminus \{(0,0,0)\}}\left\|A^{(\lambda, \sigma ; \overline{\nu})}\left(t,k,\frac{kt}{L}\right)F(t,k)\right\|_{L^2_t(I)}^2 < \infty.
\end{equation}
Then there is a constant $C_{LD} = C_{LD}(L,C_0,\overline{\lambda},\kappa,\nu,\overline{\nu},\lambda(0))$ (independent of $T$) so that any solution $\phi(t,k)$ to the system
\begin{equation}
\phi(t,k) = F(t,k) + \int_0^t L(t-\tau, k)\phi(\tau,k) d\tau \label{linearconv}
\end{equation}
(for $t>0$) with $L(t,k)$ given by \eqref{Ldef} satisfies
\begin{equation}
\int_0^T\|\phi(t)\|_{\mathcal{F}^{\lambda(t), \sigma ; \overline{\nu}}}^2dt \le C_{LD}^2 \int_0^T\|F(t)\|_{\mathcal{F}^{\lambda(t), \sigma ; \overline{\nu}}}^2dt.
\end{equation}
\end{lem}

The proof of this lemma is completely analogous to the analysis carried out in \cite{BMM13}[\S 4].  Hence, we will only sketch the proof and refer the reader to this source for most of the details.  Also note that the lower bound $ (2+\gamma)^{-1} < \overline{\nu} < \nu < 1$ is not crucial here (we only include it to make connection with our assumptions above).  All that is really required is that $\overline{\nu} < \nu$ (and both are in the range $(0,1)$).

\bigskip

\textbf{Proof:}  First, since $\lambda(t), \sigma,$ and $\overline{\nu}$ will be fixed throughout the proof, we will drop the superscripts on $A$:
\begin{equation}
A^{(\lambda, \sigma ; \overline{\nu})}\left(t,k,\frac{kt}{L}\right) = A\left(t,k,\frac{kt}{L}\right) = \langle k, kt/L \rangle^{\sigma}e^{\lambda(t)\langle k, kt/L \rangle^{\overline{\nu}}}.
\end{equation}
Note that to prove the lemma, we need only prove
\begin{equation}
\int_0^T A\left(t,k,\frac{kt}{L}\right)^2|\phi(t,k)|^2 dt \le C_{LD}^2 \int_0^T A\left(t,k,\frac{kt}{L}\right)^2|F(t,k)|^2 dt,
\end{equation}
on a mode-by-mode basis (for modes $|k|\ge 1$).  We will accomplish this in a series of intermediate steps.

\bigskip

\emph{Step 1: Well-Posedness of \eqref{linearconv} via Gr\"onwall's Inequality.}  First note that for any $0\le \tau < t < T$ and $|k|\ge1$, we have by Lemma \ref{angle_brak_lem} and the fact that $\lambda(t)$ is decreasing
\begin{align}
&A\left(t,k,\frac{kt}{L}\right) \nonumber\\
&\; \le \left\langle \left(1+\frac{t}{L}\right)|k| \right\rangle^{\sigma}e^{\lambda(0)\left\langle \left(\frac{t-\tau}{L}\right)|k| \right\rangle^{\overline{\nu}}}e^{\lambda(\tau)\left\langle \left(1+\frac{\tau}{L}\right)|k| \right\rangle^{\overline{\nu}}}\nonumber\\
&\; \le \left(\frac{\left\langle \left(1+\frac{t}{L}\right)|k| \right\rangle}{\left\langle \left(\frac{t-\tau}{L}\right)|k| \right\rangle\left\langle \left(1+\frac{\tau}{L}\right)|k| \right\rangle}\right)^{\sigma} \left\langle \left(\frac{t-\tau}{L}\right)|k| \right\rangle^{\sigma} e^{\lambda(0)\left\langle \left(\frac{t-\tau}{L}\right)|k| \right\rangle^{\overline{\nu}}}A\left(\tau,k,\frac{k\tau}{L}\right)\nonumber\\
&\le \left(\sqrt{2}\right)^{\sigma}\left\langle \left(\frac{t-\tau}{L}\right)|k| \right\rangle^{\sigma}e^{\lambda(0)\left\langle \left(\frac{t-\tau}{L}\right)|k| \right\rangle^{\overline{\nu}}}A\left(\tau,k,\frac{k\tau}{L}\right),
\end{align}
Thus, we have
\begin{align}
&A\left(t,k,\frac{kt}{L}\right)|\phi(t,k)| \nonumber\\
&\le A\left(t,k,\frac{kt}{L}\right)|F(t,k)| + \int_0^t A\left(t,k,\frac{kt}{L}\right)|L(t-\tau, k)||\phi(\tau,k)| d\tau\nonumber\\
&\le A\left(t,k,\frac{kt}{L}\right)|F(t,k)|\nonumber\\
&\qquad +(\sqrt{2})^{\sigma}\int_0^t \left\langle \left(\frac{t-\tau}{L}\right)|k| \right\rangle^{\sigma}e^{\lambda(0)\left\langle \left(\frac{t-\tau}{L}\right)|k| \right\rangle^{\overline{\nu}}}|L(t-\tau, k)|A\left(\tau,k,\frac{k\tau}{L}\right)|\phi(\tau,k)| d\tau\nonumber\\
&\le A\left(t,k,\frac{kt}{L}\right)|F(t,k)|\nonumber\\
&\qquad +(\sqrt{2})^{\sigma}\sup_{0\le \tau \le t}\left(\left\langle \left(\frac{t-\tau}{L}\right)|k| \right\rangle^{\sigma}e^{\lambda(0)\left\langle \left(\frac{t-\tau}{L}\right)|k| \right\rangle^{\overline{\nu}}}|L(t-\tau, k)|\right)\nonumber\\
& \qquad \qquad \qquad \qquad \qquad \cdot \int_0^tA\left(\tau,k,\frac{k\tau}{L}\right)|\phi(\tau,k)| d\tau.
\end{align}
Note that
\begin{align}
\Bigg\langle \left(\frac{t-\tau}{L}\right)&|k| \Bigg\rangle^{\sigma}e^{\lambda(0)\left\langle \left(\frac{t-\tau}{L}\right)|k| \right\rangle^{\overline{\nu}}}|L(t-\tau, k)|\nonumber\\
& \le \frac{2\pi C_W}{|k|^{\gamma}L} \left\langle \left(\frac{t-\tau}{L}\right)|k| \right\rangle^{\sigma}e^{\lambda(0)\left\langle \left(\frac{t-\tau}{L}\right)|k| \right\rangle^{\overline{\nu}}}\left|\widehat{\alpha G_0}\left(\frac{k(t-\tau)}{L}\right)\right|\nonumber\\
&\le \frac{2\pi C_W}{L}\sup_{\eta}\left(\left\langle \eta \right\rangle^{\sigma}e^{\lambda(0)\left\langle \eta \right\rangle^{\overline{\nu}}}\left|\widehat{\alpha G_0}\left(\eta\right)\right|\right).\label{sup_est_G0}
\end{align}
Since $\overline{\nu} < \nu$, assumption \eqref{ass_G_1} guarantees us that the derivatives of the function inside the supremum vanish rapidly enough at infinity to be in $L^2(\mathbb{R}^3)$.  Thus, the Sobolev embedding tells us that the supremum is controlled by the $H^{2}(\mathbb{R}^3)$ norm of this quantity.  Thus, we can conclude that
\begin{equation}
\Big\langle \left(\frac{t-\tau}{L}\right)|k| \Big\rangle^{\sigma}e^{\lambda(0)\left\langle \left(\frac{t-\tau}{L}\right)|k| \right\rangle^{\overline{\nu}}}|L(t-\tau, k)| \le C\left\|\langle \cdot \rangle^{\sigma}e^{\lambda(0)\langle \cdot \rangle^{\overline{\nu}}}|\widehat{\alpha G_0}|\right\|_{H^{2}},
\end{equation}
for some constant $C=C(C_W,L)$ and so,
\begin{align}
A\left(t,k,\frac{kt}{L}\right)|\phi(t,k)| &\le A\left(t,k,\frac{kt}{L}\right)|F(t,k)|\nonumber\\
 &\qquad+ C_G \int_0^tA\left(\tau,k,\frac{k\tau}{L}\right)|\phi(\tau,k)| d\tau,
\end{align}
where $C_G = C_G(C_W,L,\sigma,C_0,\nu,\overline{\nu},\lambda(0))$.  An application of Gr\"onwall's Inequality followed by an application of Cauchy-Schwarz on the resulting time integration gives us
\begin{eqnarray}
A\left(t,k,\frac{kt}{L}\right)|\phi(t,k)| &\le& A\left(t,k,\frac{kt}{L}\right)|F(t,k)| \nonumber\\
&&\qquad + C_Ge^{C_G t}\int_0^t A\left(\tau,k,\frac{k\tau}{L}\right)|F(\tau,k)|e^{-C_G\tau}d\tau\nonumber\\
&\le& A\left(t,k,\frac{kt}{L}\right)|F(t,k)| \nonumber\\
&&+ e^{C_G t}\sqrt{C_G/2}\left\|A\left(t,k,\frac{kt}{L}\right)F(t,k)\right\|_{L^2_t(I)},
\end{eqnarray}
which establishes the existence of solutions to \eqref{linearconv} for all times $0\le t \le T$.  Moreover, we clearly have
\begin{equation}
\left\|A\left(t,k,\frac{kt}{L}\right)\phi(t,k)\right\|_{L^2_t(I)} \le \frac{3}{2}e^{C_GT}\left\|A\left(t,k,\frac{kt}{L}\right)F(t,k)\right\|_{L^2_t(I)}.
\end{equation}
It will be important for later calculations to note that the constant $C_G$ is independent of $k$.  Note that this gives us a version of the lemma, but with a constant that depends (badly) on $T$.  Our next job will be to show that we can refine these estimates.

\bigskip

\emph{Step 2: Frequency Localization and Auxiliary Functions.}  Having established that $\phi$ exists and is in $L^2_t(I)$ for each $k$, we would like to take the Laplace Transform of \eqref{linearconv} and use assumption \eqref{ass_G_2}.  The problem to overcome is that \emph{a priori} we have no idea if either $F$ or $\phi$ will be sufficiently regular to justify taking the transform.  To handle this, we decompose $F$ into a number of exponentially decaying components which naturally yields a decomposition of $\phi$.  For these pieces, we can show that the transform exists, and then we can use \eqref{ass_G_2} to show that these components are nicely behaved.  The final steps in the proof of the lemma will justify summing these contributions to reobtain $\phi$.

In this case, we can use a much more primitive decomposition than the Littlewood-Paley method outlined in Section \ref{LPD}.  To that end, fix an $R \ge e$ (which later will be chosen to depend on various constants in our problem) and define
\begin{equation}
F_n(t,k) = F(t,k)\mathds{1}_{Rn\le|kt/L|^{\overline{\nu}}\le R(n+1)}.
\end{equation}
Let $\phi_n$ be the corresponding solution to
\begin{equation}
\phi_n(t,k) = F_n(t,k) + \int_0^t L(t-\tau, k)\phi_n(\tau,k) d\tau. \label{linearconv_n}
\end{equation}
We see that $\phi = \sum_{n=0}^{\infty}\phi_n$ by the linearity of this equation, and $\phi_n$ is supported on $Rn\le|kt/L|^{\overline{\nu}}$.  We also have existence and bounds for each $\phi_n$ coming directly from Step 1 above.

We now define two sets of auxiliary functions:
\begin{eqnarray}
R_n(k,t) &=& e^{2\pi\mu_n|kt/L|}N_{k,n}F_n(k,t),\\
P_n(k,t) &=& e^{2\pi\mu_n|kt/L|}N_{k,n}\phi_n(k,t),
\end{eqnarray}
where the exponents $\mu_n$ and factors $N_{k,n}$ are given by
\begin{eqnarray}
\mu_0\!\!\! &=&\!\!\! \mu_1\\
\mu_n\!\!\! &=&\!\!\! \frac{1}{2\pi (Rn)^{1/\overline{\nu}}}\left[\lambda\left(\frac{L(Rn)^{1/\overline{\nu}}}{|k|}\right)\langle(Rn)^{1/\overline{\nu}}\rangle^{\overline{\nu}} + \sigma\ln \langle(Rn)^{1/\overline{\nu}}\rangle\right], \; n \ge 1,\\
N_{k,0}\!\!\! &=&\!\!\! \langle k \rangle^{\sigma}\exp\left[\lambda\left(\frac{L R^{1/\overline{\nu}}}{|k|}\right)\langle k \rangle^{\sigma}\right],\\
N_{k,n}\!\!\!\! &=&\!\! \!\!\!\!\frac{\langle k,(Rn)^{1/\overline{\nu}}\rangle^{\sigma}}{\langle (Rn)^{1/\overline{\nu}}\rangle^{\sigma}}\exp\!\!\left[\lambda\left(\frac{L(Rn)^{1/\overline{\nu}}}{|k|}\right)\left(\langle k,(Rn)^{1/\overline{\nu}}\rangle^{\overline{\nu}}-\langle (Rn)^{1/\overline{\nu}}\rangle^{\overline{\nu}}\right)\right]\!,\! n \ge 1.
\end{eqnarray}
We will need $\mu_n / L \le \overline{\lambda}$ for all $n\ge 0$ (recall that $\overline{\lambda}$ delimits the region where we are certain the Fourier-Laplace Transform of $L$ is bounded away from $1$).  Since $\overline{\nu} < \nu < 1$, we only need to choose $R$ sufficiently large (recall that $0 < \lambda(t) \le \lambda(0)$ is a decreasing function).  Let us suppose that we take $R$ large enough that $\sup \mu_n \le \overline{\lambda}/2$.  Note that this means we have the dependencies $R = R(L,\overline{\lambda},\nu,\overline{\nu},\lambda(0))$

The reason for these strange factors is that when $|kt/L|^{\overline{\nu}} = Rn$ (i.e. at the boundary of the support for each $\phi_n$) , $$e^{2\pi\mu_n|kt/L|}N_{k,n} =  A\left(t,k,\frac{kt}{L}\right). $$  In general, we have
\begin{eqnarray}
e^{2\pi\mu_0|kt/L|}N_{k,0} &\le& C(L,R)A\left(t,k,\frac{kt}{L}\right) \; \textrm{ on } |kt/L|^{\overline{\nu}} \le R,\label{fac_est1}\\
e^{2\pi\mu_n|kt/L|}N_{k,n} &\le& C(L,R,\lambda(0),\overline{\nu})A\left(t,k,\frac{kt}{L}\right)\label{fac_est2}\\
&& \qquad \qquad \textrm{ on } Rn \le |kt/L|^{\overline{\nu}} \le R(n+1)\nonumber.
\end{eqnarray}
See \cite{BMM13} for more details about these elementary estimates (in particular, these estimates use the decay rate for $\lambda(t)$ quoted in the lemma).

Note that in terms of our new auxiliary functions, we have
\begin{equation}
P_n(t,k) = R_n(t,k) + \int_0^t e^{2\pi\mu_n|k/L|(t-\tau)}L(t-\tau, k)P_n(\tau,k) d\tau. \label{linearconv_n_aux}
\end{equation}

\bigskip

\emph{Step 3: Existence of the Fourier Transform of $P_n$.}  First, let us suppose the Fourier Transform of $P_n$ in $t$ exists.  We extend $P_n, R_n,$ and $L$ to be zero for all negative times (so that the range of integration for the both the Fourier and Fourier-Laplace transforms is $t\in[0,\infty)$)  We will also extend $P_n$ and $R_n$ to be zero outside the interval $I$ under consideration.  We then have by \eqref{linearconv_n_aux}
\begin{equation}
\widehat{P_n}^t(s,k) = \widehat{R_n}^t(s,k) + \mathcal{L}[L]\left(-|k/L|\mu_n+is,k\right)\widehat{P_n}^t(s,k),
\end{equation}
where the superscript `$t$' is to emphasize that the transform is over the time variable.  This in turn would yield
\begin{equation}
\widehat{P_n}^t(s,k) = \frac{\widehat{R_n}^t(s,k)}{1-\mathcal{L}[L]\left(|k|(-\frac{\mu_n}{L}+i\frac{s}{|k|}),k\right)},
\end{equation}
and as a direct consequence of assumption \eqref{ass_G_2}
\begin{eqnarray}
\|P_n(\cdot,k)\|_{L^2_t(I)} &\le& \kappa^{-1}\|R_n(\cdot,k)\|_{L^2_t(I)}\nonumber\\
&\le&\frac{C(L,R,\lambda(0),\overline{\nu})}{\kappa}\left\|AF_n(\cdot,k)\right\|_{L^2_t(I)},\label{P_L2_est}
\end{eqnarray}
where the final inequality follows from the estimates \eqref{fac_est1} and \eqref{fac_est2}.

In order to reach this conclusion, we must first justify taking the Fourier Transform of $P_n$ in the first place.  There are many routes available to us for this purpose.  We will simply appeal to the Paley-Wiener theory for Volterra Equations (c.f. \cite{GLS90}[Chap. 2]).  This theory considers generic fixed point problems of the form $$u(t) = f(t) + k\ast u (t),  $$ for $k \in L^1_t$, $k(t)\le e^{Ct}$ ($C>0$), and $f \in L^1_{loc}(\mathbb{R}_+)$ where we extend $u$ and $f$ to be zero on $\mathbb{R}_-$ (so that the convolution is sensible).  The solution to this problem is $u = f-f \ast r$ where $r\in L^1_{loc}(\mathbb{R}_+)$ is the unique solution to $r = k + k \ast r .$  The major result of this theory is that $r \in L^1(\mathbb{R}_+)$ iff the Fourier-Laplace Transform of $k$ is never equal to 1 on the right half-plane.  Once $r$ is in $L^1$, if we then know $f$ is also in $L^1$, we find that $u \in L^1(\mathbb{R}_+)$.  Once we know a function is $L^1$, we can certainly take its transform.  In our case, we can apply the theory to $f(t) = R_n(t,k), k(t) = e^{2\pi\mu_n|kt/L|}L(t, k),$ and $u(t) = P_n(t,k)$.  This then rigorously justifies estimate \eqref{P_L2_est}.

\bigskip

\emph{Step 4: Summation of the Frequency Blocks.}  Recall that each $\phi_n$ is supported on $|kt/L|^{\overline{\nu}} \ge Rn$.  Thus, we can write
\begin{align}
\Big\|A & \phi(\cdot,k)\Big\|_{L^2_t(I)} = \int_0^T\left|\sum_{n=0}^{\infty}A\left(t,k,\frac{kt}{L}\right)\phi_n(t,k)\right|^2dt\nonumber\\
& \le 2\int_0^T\left|A\left(t,k,\frac{kt}{L}\right)\phi_0(t,k)\right|^2dt + 2\int_0^T\left|\sum_{n=1}^{\infty}A\left(t,k,\frac{kt}{L}\right)\phi_n(t,k)\right|^2dt\nonumber\\
& = 2\int_0^T\left|A\left(t,k,\frac{kt}{L}\right)\phi_0(t,k)\right|^2dt \nonumber\\
&\qquad \qquad + 2\int_0^T\left|\sum_{n=1}^{\infty}A\left(t,k,\frac{kt}{L}\right)\frac{\mathds{1}_{|kt/L|^{\overline{\nu}}\ge Rn}}{N_{k,n}e^{2\pi\mu_n|kt/L|}}P_n(t,k)\right|^2dt\nonumber\\
& = 2\int_0^T\left|A\left(t,k,\frac{kt}{L}\right)\phi_0(t,k)\right|^2dt + 2\int_0^T \sum_{n,n' \ge 1}K_{n,n'}(t,k)P_n(t,k)P_{n'}(t,k)dt,
\end{align}
where the kernel $K$ is given by
\begin{equation}
K_{n,n'}(t,k) = A\left(t,k,\frac{kt}{L}\right)^2\frac{\mathds{1}_{|kt/L|^{\overline{\nu}}\ge Rn}}{N_{k,n}e^{2\pi\mu_{n}|kt/L|}}\frac{\mathds{1}_{|kt/L|^{\overline{\nu}}\ge Rn'}}{N_{k,n'}e^{2\pi\mu_{n'}|kt/L|}}.
\end{equation}
To attack the second integral, we will appeal to Schur's test which says that
\begin{align}
\int_0^T & \sum_{n,n' \ge 1} K_{n,n'}(t,k)P_n(t,k)P_{n'}(t,k)dt\nonumber\\
&\le \left(\sup_{t\in[0,T]}\sup_{n\ge 1}\sum_{n'=1}^{\infty}K_{n,n'}(t,k)\right)^{\frac{1}{2}}\!\!\left(\sup_{t\in[0,T]}\sup_{n'\ge 1}\sum_{n=1}^{\infty}K_{n,n'}(t,k)\right)^{\frac{1}{2}}\!\!\sum_{n=1}^{\infty}\|P_n(k,t)\|^2_{L^2_t(I)},
\end{align}
so long as the two suprema are finite.  Given the form of $K$, it is clear that once we have a bound for one of these suprema, the other will follow by symmetry.  Since this is the same kernel which appears in \cite{BMM13}[Section 4], we will only give an outline of how this is proven.

First, by adapting the arguments leading to \eqref{fac_est1} and \eqref{fac_est2}, we can show that
\begin{equation}
A\left(t,k,\frac{kt}{L}\right)\frac{\mathds{1}_{|kt/L|^{\overline{\nu}}\ge Rn'}}{N_{k,n'}e^{2\pi\mu_{n'}|kt/L|}} \le C(L,R,\lambda(0),\overline{\nu}).
\end{equation}
This leaves us with
\begin{equation}
\sum_{n = 1}^{\infty} K_{n,n'}(t,k) \le C(L,R,\lambda(0),\overline{\nu})\sum_{n = 1}^{\infty}A\left(t,k,\frac{kt}{L}\right)\frac{\mathds{1}_{|kt/L|^{\overline{\nu}}\ge Rn}}{N_{k,n}e^{2\pi\mu_{n}|kt/L|}}.
\end{equation}
We can use the following facts:
\begin{itemize}
\item[1)]  $\lambda(t)$ is decreasing.
\item[2)]  For $e \le (Rn)^{1/\overline{\nu}}\le |kt/L|,$ $$-\langle k, (Rn)^{1/\overline{\nu}}\rangle^{\overline{\nu}} + \langle (Rn)^{1/\overline{\nu}} \rangle^{\overline{\nu}} \le \langle kt/L \rangle^{\overline{\nu}} - \langle k,kt/L \rangle^{\overline{\nu}}.$$
\item[3)]  For $|k| \ge 1$ and $x \ge 0$, $\langle x \rangle \langle k, x \rangle^{-1}$ is increasing in $x$.
\item[4)]  For $x \ge e$, $x/\ln\langle x \rangle$ is an increasing function.
\end{itemize}
to show that
\begin{equation}
\sum_{n = 1}^{\infty} K_{n,n'}(t,k) \le C\sum_{n = 1}^{\infty}\mathds{1}_{|kt/L|^{\overline{\nu}}\ge Rn'}\exp\left[\lambda(t)\langle kt/L \rangle^{\overline{\nu}}\left(1-\frac{|kt/L|\langle (Rn)^{1/\overline{\nu}} \rangle^{\overline{\nu}}}{(Rn)^{1/\overline{\nu}}\langle kt/L\rangle^{\overline{\nu}}}\right)\right],
\end{equation}
where again $C = C(L,R,\lambda(0),\overline{\nu}).$  We now approximate the sum over $n$ with an integral over $y$.  Making the substitution $x = Ry$ and using $\langle x^{1/\overline{\nu}} \rangle^{\overline{\nu}} \ge x $, we find
\begin{eqnarray}
\sum_{n = 1}^{\infty} K_{n,n'}(t,k) &\le& \frac{C}{R} e^{\lambda(t)\langle kt/L\rangle^{\overline{\nu}}}\int_{R}^{|kt/L|^{\overline{\nu}}} e^{-\lambda(t)|kt/L|\frac{\langle x^{1/\overline{\nu}} \rangle^{\overline{\nu}}}{x^{1/\overline{\nu}}}}dx \nonumber\\
&\le&\frac{C}{R} e^{\lambda(t)\langle kt/L\rangle^{\overline{\nu}}}\int_{R}^{|kt/L|^{\overline{\nu}}} e^{-\lambda(t)|kt/L|x^{1-1/\overline{\nu}}} dx\nonumber\\
&\le&\frac{C}{R} e^{\lambda(t)\langle kt/L\rangle^{\overline{\nu}}}\left(\frac{1}{\overline{\nu}}-1\right)\int^{R^{1-1/\overline{\nu}}}_{|kt/L|^{\overline{\nu}-1}}w^{\frac{1}{\overline{\nu}-1}}e^{-\lambda(t)|kt/L|w}dw\nonumber\\
&\le& \tilde{C}(L,R,\lambda(0),\overline{\nu}).
\end{eqnarray}
This shows that the sum above is uniformly bounded in $n'$ and $t$ (in fact, for $t\in[0,\infty)$), which in turn shows that the second supremum appearing in Schur's Test is finite.  The first supremum follows by symmetry, and so Schur's Test gives us
\begin{align}
2\int_0^T\Big|\sum_{n=1}^{\infty}A\left(t,k,\frac{kt}{L}\right)\frac{\mathds{1}_{|kt/L|^{\overline{\nu}}\ge Rn}}{N_{k,n}e^{2\pi\mu_n|kt/L|}}&P_n(t,k)\Big|^2dt \nonumber \\ &\le 2\tilde{C}(L,R,\lambda(0),\overline{\nu}) \sum_{n=1}^{\infty}\|P_n(k,t)\|^2_{L^2_t(I)}.
\end{align}

We are left to consider the $n=0$ term.  For simplicity, we consider the $L^2_t(\mathbb{R}_+)$ norm (from which the control on $L^2_t(I)$ will follow).
\begin{align}
\int_0^{\infty}&\Big|A\left(t,k,\frac{kt}{L}\right)\phi_0(t,k)\Big|^2dt \nonumber\\&
= \int_0^{R^{1/\overline{\nu}}}\left|A\left(t,k,\frac{kt}{L}\right)\phi_0(t,k)\right|^2dt\nonumber +\int_{R^{1/\overline{\nu}}}^{\infty}\left|A\left(t,k,\frac{kt}{L}\right)\phi_0(t,k)\right|^2dt\nonumber\\
&\le Ce^{C_G R^{1/\overline{\nu}}}\left\|A\left(t,k,\frac{kt}{L}\right)F(t,k)\right\|^2_{L^2_t(\mathbb{R}_+)}+\int_{R^{1/\overline{\nu}}}^{\infty}\left|A\left(t,k,\frac{kt}{L}\right)\phi_0(t,k)\right|^2dt,
\end{align}
where in the final inequality we have used our rough Gr\"onwall bound from Step 1 (recall we extended $F(t,k)$ to be zero outside of $I$).  For the second integral, we can show using $|t| \ge R^{\overline{\nu}}$, Lemma \ref{angle_brak_lem}[item (i)], $\lambda(t)$ is decreasing, and Lemma \ref{exp_ineq_lem}[item (i)] that
\begin{equation}
A\left(t,k,\frac{kt}{L}\right) \le C(R) N_{k,0}e^{\mu_0|kt/L|}.
\end{equation}
Combining this with our estimate \eqref{P_L2_est} finally gives
\begin{align}
\int_0^{\infty}\Big|A&\left(t,k,\frac{kt}{L}\right)\phi_0(t,k)\Big|^2dt \nonumber\\
&\le C(R)\left(1+\frac{1}{\kappa^2}\right)\left\|A\left(t,k,\frac{kt}{L}\right)F(t,k)\right\|^2_{L^2_t(\mathbb{R}_+)}.
\end{align}
Finally, putting together all of these various estimates (and recalling that we have extended $F$ to be zero outside of $I$), we arrive at
\begin{equation}
\left\|A\phi(\cdot,k)\right\|^2_{L^2_t(I)} \le C_{LD}^2\left\|AF(\cdot,k)\right\|^2_{L^2_t(I)},
\end{equation}
for some appropriate constant, $C_{LD}$, depending on the parameters $L, C_0, \overline{\lambda}, \kappa, \nu, \overline{\nu},$ and $\lambda(0)$.  This completes the proof of the lemma. $\blacksquare$

\section{Estimates on $\rho$}

\subsection{$L^2_kL^2_t(I)$ Estimate for $\rho$}

We have via the paraproduct decomposition
\begin{equation}
\widehat{\rho}(t,k) = \widehat{\Phi_{in}}\left(k,\frac{kt}{L}\right) + \int_0^tL(t-\tau,k)\widehat{\rho}(\tau,k)d\tau - T_k(t) - R_k(t) - \mathcal{R}_k(t),\label{para_dec_rho}
\end{equation}
where
\begin{align}
&T_k(t) = 4 \pi^2 \int_0^t \sum_{N \ge 8} \sum_{\ell \ne 0} \widehat{\rho}(\tau, \ell)_{<N/8}\widehat{W}(\ell)\frac{k \cdot \ell}{L^2}(t - \tau) \widehat{\alpha \Phi}\left(\tau,k - \ell,\frac{kt - \ell \tau}{L}\right)_{N}d\tau \nonumber\\
& -2\pi i \int_0^t \sum_{N \ge 8} \sum_{\ell \ne 0} \widehat{\rho}(\tau, \ell)_{<N/8}\widehat{W}(\ell)\frac{\ell}{L}(t - \tau) \cdot \left[\frac{k}{L} \cdot \nabla_{2} \widehat{v\alpha \Phi}\left(\tau,k - \ell,\frac{kt - \ell \tau}{L}\right)_{N}\right]d\tau,\\
&R_k(t) =  4 \pi^2 \int_0^t \sum_{N \ge 8} \sum_{\ell \ne 0} \widehat{\rho}(\tau, \ell)_{N}\widehat{W}(\ell)\frac{k \cdot \ell}{L^2}(t - \tau) \widehat{\alpha \Phi}\left(\tau,k - \ell,\frac{kt - \ell \tau}{L}\right)_{<N/8}d\tau \nonumber\\
& -2\pi i \int_0^t \sum_{N \ge 8} \sum_{\ell \ne 0} \widehat{\rho}(\tau, \ell)_{N}\widehat{W}(\ell)\frac{\ell}{L}(t - \tau) \cdot \left[\frac{k}{L} \cdot \nabla_{2} \widehat{v\alpha \Phi}\left(\tau,k - \ell,\frac{kt - \ell \tau}{L}\right)_{<N/8}\right]d\tau,\\
&\mathcal{R}_k(t) =  4 \pi^2 \int_0^t \sum_{N} \sum_{\frac{N}{8} \le N' \le 8N} \sum_{\ell \ne 0} \widehat{\rho}(\tau, \ell)_{N'}\widehat{W}(\ell)\frac{k \cdot \ell}{L^2}(t - \tau) \widehat{\alpha \Phi}\left(\tau,k - \ell,\frac{kt - \ell \tau}{L}\right)_{N}d\tau \nonumber\\
& -2\pi i \!\! \int_0^t \!\! \sum_{N}\!\!\sum_{\frac{N}{8} \le N' \le 8N} \sum_{\ell \ne 0} \widehat{\rho}(\tau, \ell)_{N'}\frac{\widehat{W}(\ell)\ell(t - \tau)}{L} \cdot \!\! \left[\frac{k}{L} \cdot \nabla_{2} \widehat{v\alpha \Phi}\left(\!\tau,k - \ell,\frac{kt - \ell \tau}{L}\!\right)_{N}\right]d\tau,
\end{align}
(recall $N, N'$ are either of the form $2^j$ for $j\ge0$ or $0$ in the decomposition).  The subscript ``2'' on the gradient refers to differentiation with respect to the final slot (which appears as $\nabla_{\eta}$ in previous versions).  We refer to these terms as \emph{transport, reaction,} and \emph{remainder} terms, respectively.  Applying Lemma \ref{linear_lemma}, we have
\begin{align}
\|A\widehat{\rho}(\cdot,k)\|^2_{L^2_t(I)} \le C_{LD}^2&\left\|A\widehat{\Phi_{in}}\left(k,\frac{k\cdot}{L}\right)\right\|^2_{L^2_t(I)} + C_{LD}^2\left\|AT_k\right\|^2_{L^2_t(I)}\nonumber\\
 & + C_{LD}^2\left\|AR_k\right\|^2_{L^2_t(I)} + C_{LD}^2\left\|A\mathcal{R}_k\right\|^2_{L^2_t(I)},\label{main_est_pho}
\end{align}
where $A\left(t, k,\frac{kt}{L}\right)$ is our usual multiplier for transforms independent of $\eta$ (here with parameters $\lambda(t), \sigma,$ and $\overline{\nu}$).  Our goal now is to show that under the bootstrap hypotheses, the norms of $AT_k,  AR_k$ and $A\mathcal{R}_k$ can be controlled by $A\widehat{\rho}$.  First, since we have $\lambda(t) \le \lambda(0) < \lambda_0$, Lemma \ref{trace_lem},  and the fact that the zero-mode is identically zero,
\begin{align}
\sum_{k\ne 0} \left\|A\widehat{\Phi_{in}}\left(k,\frac{k\cdot}{L}\right)\right\|^2_{L^2_t(I)} &\le \sum_{k\ne 0}\int_0^T\left|A\left(0,k,\frac{kt}{L}\right) \widehat{\Phi_{in}}\left(k,\frac{kt}{L}\right)\right|^2dt\nonumber\\
&\le C\left\|A(0)\widehat{\Phi_{in}}\right\|^2_{L^2_kH^2_{\eta}} \le C\epsilon^2 \label{lin_est},
\end{align}
where the constant $C$ is a product of constants coming from the trace lemma and Corollary \ref{cor_Phi_1}.

The estimates for the remaining terms follow very closely the presentation in \cite{BMM13}[\S 5.1].  We present most of the details for the sake of completeness.  However, we will refer to this source for a number of technical estimates which are needed in the course of the proof.

\subsubsection{Reaction Term}

We would like to prove an estimate of the form $$\|AR\|^2_{L^2_kL^2_t(I)} \le K \epsilon^2 \|A\widehat{\rho}\|^2_{L^2_kL^2_t(I)}  ,$$ for some constant $K$.  We can then absorb this term into the left-hand side of \eqref{main_est_pho}.   It will be of utmost importance to keep track of the parameters $K$ depends on!  Inserting our assumption on $\widehat{W}$ yields
\begin{align}
\|AR\|^2_{L^2_kL^2_t(I)} \le \left(\frac{2\pi C_W}{L}\right)^2 \sum_{k\ne 0} \int_0^T \Bigg[A&\left(t,k,\frac{kt}{L}\right)\int_0^t\sum_{\ell \ne 0}\sum_{N \ge 8} \frac{|k|(t-\tau)}{|\ell|^{\gamma}}|\widehat{\rho}\left(\tau, \ell\right)_{N}|\nonumber\\
 & \cdot\Bigg|\widehat{\alpha \Phi}\left(\tau,k-\ell, \frac{kt - \ell \tau}{L}\right)_{< N/8} \Bigg| d\tau \Bigg]^2dt\nonumber\\
 +2\pi \left(\frac{C_W}{L}\right)^2 \sum_{k\ne 0} \int_0^T \Bigg[A&\left(t,k,\frac{kt}{L}\right)\int_0^t\sum_{\ell \ne 0}\sum_{N \ge 8} \frac{(t-\tau)}{|\ell|^{\gamma}}|\widehat{\rho}\left(\tau, \ell\right)_{N}|\nonumber\\
 & \cdot\Bigg|k \cdot \nabla_2 \widehat{v \alpha \Phi}\left(\tau,k-\ell, \frac{kt - \ell \tau}{L}\right)_{< N/8} \Bigg| d\tau \Bigg]^2dt\nonumber\\
 \le \left(\frac{2\pi C_W}{L}\right)^2 \sum_{k\ne 0} \int_0^T \Bigg[A&\left(t,k,\frac{kt}{L}\right)\int_0^t\sum_{\ell \ne 0}\sum_{N \ge 8} \frac{|k|(t-\tau)}{|\ell|^{\gamma}}|\widehat{\rho}\left(\tau, \ell\right)_{N}|\nonumber\\
  & \cdot\sum_{|m|\le 2}\Bigg|\widehat{v^{m}\alpha \Phi}\left(\tau,k-\ell, \frac{kt - \ell \tau}{L}\right)_{< N/8} \Bigg| d\tau \Bigg]^2dt,
\end{align}
where $m$ runs over the set of multi-indices in $\mathbb{N}^3$ of order not more than two.  Our frequency localizations imply the following controls:
\begin{align}
\frac{N}{2} \le |\ell| + |\ell \tau/L| &\le \frac{3N}{2},\\
|k - \ell| + |(k t - \ell \tau)/L| &\le \frac{3N}{32},\\
\frac{13}{16} \le \frac{|k| + |kt/L|}{|\ell| + |\ell \tau /L|} &\le \frac{19}{16}.
\end{align}
Using these controls and Lemma \ref{angle_brak_lem}[item iii], we can make the following estimate:
\begin{align}
A\left(t,k,\frac{kt}{L}\right) &= e^{(\lambda(t)-\lambda(\tau))\langle k , kt/L \rangle^{\overline{\nu}}} A\left(\tau,k,\frac{kt}{L}\right)\nonumber\\
&\le c_1e^{(\lambda(t)-\lambda(\tau))\langle k , kt/L \rangle^{\overline{\nu}}} e^{c_2\lambda(\tau)\langle k - \ell, (kt - \ell \tau)/L \rangle^{\overline{\nu}}}A\left(\tau,\ell,\frac{\ell \tau}{L}\right),
\end{align}
where $c_1=c_1(\sigma)$ and $c_2 = c_2(\overline{\nu}) \in (0,1).$ Substituting this above, dropping the projection on $\widehat{\alpha \Phi}$, and recalling the definition of $\overline{K}_{k,\ell}$ in \eqref{def_Kbar} gives us
\begin{align}
\|AR\|^2_{L^2_kL^2_t(I)} &\le \sum_{k\ne 0} \int_0^T\left[\int_0^t \sum_{\ell \ne 0}\overline{K}_{k,\ell}(t,\tau)A\left(\tau,\ell,\frac{\ell \tau}{L}\right)\sum_{N \ge 8}|\widehat{\rho}(\tau, \ell)_N|d\tau\right]^2dt\nonumber\\
&\le \sum_{k\ne 0} \int_0^T\left[\int_0^t \sum_{\ell \ne 0}\overline{K}_{k,\ell}(t,\tau)A\left(\tau,\ell,\frac{\ell \tau}{L}\right)|\widehat{\rho}(\tau, \ell)|d\tau\right]^2dt,\label{reac_est}
\end{align}
where in the last step we drop the frequency projection acting on $\widehat{\rho}$.  We now apply Schur's test to find
\begin{align}
&\|AR\|^2_{L^2_kL^2_t(I)} \nonumber\\
&\le\left(\sup_{t\ge 0}\sup_{k}\int_0^t\sum_{\ell \ne 0}\overline{K}_{k,\ell}(t,\tau)d\tau\right)\nonumber\\
& \qquad \qquad \cdot\sum_{k\ne 0} \int_0^T\left(\int_0^t\sum_{\ell \ne 0}\overline{K}_{k,\ell}(t,\tau)\left|A\left(\tau,\ell,\frac{\ell \tau}{L}\right)|\widehat{\rho}(\tau, \ell)\right|^2d\tau\right)dt\nonumber\\
&\le\left(\sup_{t\ge 0}\sup_{k}\int_0^t\sum_{\ell \ne 0}\overline{K}_{k,\ell}(t,\tau)d\tau\right)\nonumber\\
&\qquad \qquad \cdot\sum_{\ell \ne 0}\int_0^T\left(\int_{\tau}^T\sum_{k\ne 0} \overline{K}_{k,\ell}(t,\tau)dt \right)\left|A\left(\tau,\ell,\frac{\ell \tau}{L}\right)|\widehat{\rho}(\tau, \ell)\right|^2d\tau\nonumber\\
&\le \left(\sup_{t\ge 0}\sup_{k}\int_0^t\sum_{\ell \ne 0}\overline{K}_{k,\ell}(t,\tau)d\tau\right)\left(\sup_{\tau\ge 0}\sup_{\ell \ne 0}\int_{\tau}^T\sum_{k}\overline{K}_{k,\ell}(t,\tau)dt\right)\|A\widehat{\rho}\|^2_{L^2_kL^2_t(I)}\nonumber\\
&\le C K_2 \epsilon^2 \|A\widehat{\rho}\|^2_{L^2_kL^2_t(I)},
\end{align}
where the final inequality comes from \eqref{reaction_control}.  Recall that the coefficient $C$ appearing in this estimate does not depend on $K_1, K_2,$ or $K_3.$

\subsubsection{Transport Term}

Our goal here is the same as above - estimate $\|AT\|^2_{L^2_kL^2_t(I)}$ in terms of $\|A\widehat{\rho}\|^2_{L^2_kL^2_t(I)}$.  Inserting our assumption on $\widehat{W}$ (and dropping the $|\ell|^{-\gamma}$ altogether) yields
\begin{align}
\|AT\|^2_{L^2_kL^2_t(I)} \le \frac{4\pi^2 C_W^2}{L^2} \sum_{k\ne 0} \int_0^T \Bigg[A&\left(t,k,\frac{kt}{L}\right)\int_0^t\sum_{\ell \ne 0}\sum_{N \ge 8}|k|(t-\tau)\Big|\widehat{\rho}\left(\tau, \ell\right)_{<N/8}\Big|\nonumber\\
 & \cdot \sum_{|m|\le 2}\left|\widehat{v^m \alpha \Phi}\left(\tau,k-\ell, \frac{kt - \ell \tau}{L}\right)_{N} \right| d\tau \Bigg]^2dt
\end{align}
where we have the following controls for each term
\begin{align}
\frac{N}{2} \le |k-\ell| + |(kt -\ell \tau)/L| &\le \frac{3N}{2},\\
|\ell| + |\ell \tau/L| &\le \frac{3N}{32},\\
\frac{13}{16} \le \frac{|k| + |kt/L|}{|k-\ell| + |(kt-\ell \tau) /L|} &\le \frac{19}{16}.
\end{align}
By similar arguments as above (and at the expense of possibly making $c_1$ and $c_2$ larger), we have
\begin{align}
A\left(t,k,\frac{kt}{L}\right) &= e^{(\lambda(t)-\lambda(\tau))\langle k , kt/L \rangle^{\overline{\nu}}} A\left(\tau,k,\frac{kt}{L}\right)\nonumber\\
&\le c_1e^{(\lambda(t)-\lambda(\tau))\langle k , kt/L \rangle^{\overline{\nu}}} e^{c_2\lambda(\tau)\langle \ell, \ell \tau/L \rangle^{\overline{\nu}}}A\left(\tau,k-\ell,\frac{kt-\ell \tau}{L}\right)\nonumber\\
&\le c_1 e^{c_2\lambda(\tau)\langle \ell, \ell \tau/L \rangle^{\overline{\nu}}}A\left(\tau,k-\ell,\frac{kt-\ell \tau}{L}\right),
\end{align}
where we have dropped the first exponential factor since $\lambda(t) - \lambda(\tau) \le 0$ for $\tau \in [0,t]$.  We note that $$|k(t-\tau)| \le \langle \tau \rangle \langle k - \ell, kt - \ell t \rangle.$$ Applying this along with Cauchy-Schwarz (and then dropping all projections) yields
\begin{align}
&\|AT\|^2_{L^2_kL^2_t(I)} \le \frac{4\pi^2 c_1^2 C_W^2}{L^2} \sum_{k\ne 0} \int_0^T\left[\sum_{\ell \ne 0}\int_0^t  \langle \tau \rangle e^{c_2\lambda(\tau)\langle \ell, \ell \tau/L \rangle^{\overline{\nu}}}|\widehat{\rho}(\tau, \ell)|d\tau\right]\nonumber\\
&\cdot \left[\sum_{\ell \ne 0}\int_0^t\langle \tau \rangle \left(\sum_{m \le 2}\left|A^{(\sigma + 1)}\widehat{v^m \alpha \Phi}\left(\tau,k-\ell, \frac{kt - \ell \tau}{L}\right)\right|\right)^2 e^{c_2\lambda(\tau)\langle \ell, \ell \tau/L \rangle^{\overline{\nu}}}|\widehat{\rho}(\tau, \ell)|d\tau\right]\!\! dt.
\end{align}
Applying Cauchy-Schwarz once again yields
\begin{align}
\sum_{\ell \ne 0}\int_0^t &\langle \tau \rangle e^{c_2\lambda(\tau)\langle \ell, \ell \tau/L \rangle^{\overline{\nu}}}|\widehat{\rho}(\tau, \ell)|d\tau \nonumber\\
&\le \|A\widehat{\rho}\|_{L^2_kL^2_t(I)}\left(\sum_{\ell \ne 0}\int_0^t \langle \tau \rangle^2 e^{2(c_2-1)\lambda(\tau)\langle \ell, \ell \tau/L \rangle^{\overline{\nu}}}\langle \ell, \ell \tau/L \rangle^{-2\sigma} d\tau\right)^{1/2}\nonumber.
\end{align}
Recalling that $c_2 \in (0,1)$ and $\sigma > 3/2$,
\begin{align}
\sum_{\ell \ne 0}\int_0^t \langle \tau \rangle^2 e^{2(c_2-1)\lambda(\tau)\langle \ell, \ell \tau/L \rangle^{\overline{\nu}}} & \langle \ell, \ell \tau/L \rangle^{-2\sigma} d\tau \nonumber\\
&\le \sum_{\ell \ne 0} |\ell|^{-2\sigma}\int_0^{\infty} \langle \tau \rangle^2 (1+\tau^2/L^2)^{-\sigma} d\tau\nonumber\\
&\le C(\sigma, L),
\end{align}
which finally yields
\begin{equation}
\sum_{\ell \ne 0}\int_0^t e^{c_2\lambda(\tau)\langle \ell, \ell \tau/L \rangle^{\overline{\nu}}}|\widehat{\rho}(\tau, \ell)|d\tau \le C(\sigma, L) \|A\widehat{\rho}\|_{L^2_kL^2_t(I)}.\label{int_trans_est}
\end{equation}
Thus, we have
\begin{align}
&\|AT\|^2_{L^2_kL^2_t(I)}\nonumber\\
& \le \frac{40\pi^2 c_1^2 C_W^2 C(\sigma, L)}{L^2} \|A\widehat{\rho}\|_{L^2_kL^2_t(I)}\nonumber\\
&\cdot\sum_{k\ne 0} \! \int_0^T \!\!\left[\sum_{\ell \ne 0}\int_0^t\!\langle \tau \rangle \! \sum_{m \le 2}\left|A^{(\sigma + 1)}\widehat{v^m \alpha \Phi}\left(\tau,k-\ell, \frac{kt - \ell \tau}{L}\right)\right|^2 e^{c_2\lambda(\tau)\langle \ell, \ell \tau/L \rangle^{\overline{\nu}}}|\widehat{\rho}(\tau, \ell)|d\tau\!\!\right]\!\!dt\nonumber\\
& \le \frac{40\pi^2 c_1^2 C_W^2 C(\sigma, L)}{L^2} \|A\widehat{\rho}\|_{L^2_kL^2_t(I)}\sum_{\ell \ne 0} \int_0^T \langle \tau \rangle e^{c_2\lambda(\tau)\langle \ell, \ell \tau/L \rangle^{\overline{\nu}}}|\widehat{\rho}(\tau, \ell)|\nonumber\\
&\qquad \qquad \qquad \qquad \cdot\left[\sum_{k \ne 0}\int_\tau^T \sum_{|m| \le 2}\left|A^{(\sigma + 1)}\widehat{v^m \alpha \Phi}\left(\tau,k-\ell, \frac{kt - \ell \tau}{L}\right)\right|^2dt\right] d\tau\nonumber\\
& \le \frac{40\pi^2 c_1^2 C_W^2 C(\sigma, L)}{L^2} \|A\widehat{\rho}\|_{L^2_kL^2_t(I)}\sum_{\ell \ne 0} \int_0^T\langle \tau \rangle e^{\lambda(\tau)\langle \ell, \ell \tau/L \rangle^{\overline{\nu}}}|\widehat{\rho}(\tau, \ell)|d\tau\nonumber\\
& \cdot L\sup_{\tau \ge 0}e^{-(1-c_2)\lambda(0)\langle \tau \rangle^{\overline{\nu}}} \sum_{|m|\le 2}\sum_{k} \sup_{\omega \in \mathbb{Z}^3_{\ne 0} }\sup_{\zeta \in \mathbb{R}^3}\int_{-\infty}^{\infty}\left|A^{(\sigma + 1)}\widehat{v^m \alpha \Phi}\left(\tau,k, \frac{\omega}{|\omega|}s - \zeta\right)\right|^2ds
\end{align}
(the factor of 10 that appears merely comes from taking the summation over $m$ outside of the square).  Making a similar estimate as above and then appealing to \eqref{transport_control}, we arrive at
\begin{equation}
\|AT\|^2_{L^2_kL^2_t(I)} \le CK_1 \epsilon^2 \|A\widehat{\rho}\|_{L^2_kL^2_t(I)}^2.
\end{equation}

\subsubsection{Remainder Term}

Here, we have
\begin{align}
\|A\mathcal{R}\|^2_{L^2_kL^2_t(I)}  \le \frac{4\pi^2 C_W^2}{L^2}\sum_{k\ne 0} \! \int_0^T &   \!\! \Bigg[A\left(t,k,\frac{kt}{L}\right)\int_0^t \! \sum_{N} \sum_{\frac{N}{8} \le N' \le 8N} \sum_{\ell \ne 0} \left|\widehat{\rho}(\tau, \ell)_{N'}\right||k|(t - \tau) \nonumber\\
&\cdot \sum_{|m| \le 2}\left| \widehat{v^m \alpha \Phi}\left(\tau,k - \ell,\frac{kt - \ell \tau}{L}\right)_{N}\right|d\tau\Bigg]^2dt.
\end{align}
Note that for each term we have the controls
\begin{align}
\frac{N}{16} \le \frac{N'}{2} \le |\ell|+|\ell\tau /L| \le \frac{3N'}{2} &\le 12N,\\
\frac{N}{2} \le |k - \ell|+|(kt - \ell\tau)/L| &\le \frac{3N}{2},\\
\frac{1}{24} \le \frac{|k - \ell|+|(kt - \ell\tau)/L|}{|\ell|+|\ell\tau /L|} &\le 24,\\
\langle k, kt/L\rangle^{\overline{\nu}} \le \left(\frac{24}{25}\right)^{1-\overline{\nu}}\Big(\langle k - \ell, (kt-\ell \tau)/L\rangle^{\overline{\nu}}&+\langle \ell, \ell \tau/L\rangle^{\overline{\nu}}\Big),
\end{align}
where the final inequality follows by Lemma \ref{angle_brak_lem}[item iv].  Using Lemma \ref{exp_ineq_lem}[item ii] to handle the Sobolev terms, we have
\begin{align}
A\left(t,k,\frac{kt}{L}\right) &=\langle k, kt/L \rangle^{\sigma} e^{\lambda(t)\langle k, kt/L \rangle^{\overline{\nu}}}\nonumber\\
&\le \langle k, kt/L \rangle^{\sigma}e^{c(\overline{\nu})\lambda(\tau)\langle k - \ell, (kt-\ell \tau)/L\rangle^{\overline{\nu}}}e^{c(\overline{\nu})\lambda(\tau)\langle \ell, \ell \tau/L\rangle^{\overline{\nu}}}\nonumber\\
&\le C(\lambda(0),\sigma,\overline{\nu}) e^{c'(\overline{\nu})\lambda(\tau)\langle k - \ell, (kt-\ell \tau)/L\rangle^{\overline{\nu}}}e^{c'(\overline{\nu})\lambda(\tau)\langle \ell, \ell \tau/L\rangle^{\overline{\nu}}},
\end{align}
where $c'(\overline{\nu}) \in (0,1)$ is only a little larger than $(24/25)^{1-\overline{\nu}}$ (the increase due to the absorbtion of the Sobolev terms).  Applying this estimate, Cauchy-Schwarz, the almost orthogonality of the Littlewood-Paley decomposition, the fact that $\sigma > 3/2$, and our estimate $$|k|(t-\tau) \le \langle \tau \rangle |k-l, kt-\ell \tau|  ,$$ we obtain
\begin{align}
&\|A\mathcal{R}\|^2_{L^2kL^2_t(I)}  \nonumber\\
&\le C(\lambda(0),\sigma,\overline{\nu},C_W,L)\sum_{k\ne 0} \int_0^T \left[\int_0^t\sum_{\ell \ne 0}\left(e^{2\lambda(\tau)\langle \ell, \ell \tau/L\rangle^{\overline{\nu}}}\langle \tau \rangle^2\left|\widehat{\rho}\left(\tau, \ell\right)\right|^2\right)d\tau\right]\nonumber\\
&\qquad \cdot \Bigg[\int_0^t \sum_{\ell \ne 0}\Bigg(\sum_N e^{2c'\lambda(\tau)\langle k - \ell, (kt-\ell \tau)/L\rangle^{\overline{\nu}}}e^{-2(1-c')\lambda(\tau)\langle \ell, \ell \tau/L\rangle^{\overline{\nu}}}\nonumber\\
& \qquad \qquad  \qquad  \cdot |k-l, kt-\ell \tau|^2 \sum_{|m| \le 2}\left|\widehat{v^m \alpha \Phi}\left(\tau,k-\ell, \frac{kt - \ell \tau}{L}\right)_N\right|^2\Bigg)d\tau\Bigg]dt\nonumber\\
&\le C(\lambda(0),\sigma,\overline{\nu},C_W,L) \|A\widehat{\rho}\|_{L^2_kL^2_t(I)}^2\nonumber\\
&\cdot\sum_{k\ne 0} \int_0^T\int_0^t \sum_{\ell \ne 0}\Bigg(\sum_N e^{2c'\lambda(\tau)\langle k - \ell, (kt-\ell \tau)/L\rangle^{\overline{\nu}}}e^{-2(1-c')\lambda(\tau)\langle \ell, \ell \tau/L\rangle^{\overline{\nu}}}\nonumber\\
&\qquad \qquad \cdot |k-l, kt-\ell \tau|^2 \sum_{|m| \le 2}\left|\widehat{v^m \alpha \Phi}\left(\tau,k-\ell, \frac{kt - \ell \tau}{L}\right)_N\right|^2\Bigg)d\tau dt.
\end{align}
Next, we interchange the order of integration and summation and use the Trace Lemma (Lemma \ref{trace_lem}) to deduce:
\begin{align}
&\|A\mathcal{R}\|^2_{L^2kL^2_t(I)} \nonumber\\
&\le C(\lambda(0),\sigma,\overline{\nu},C_W,L) \|A\widehat{\rho}\|_{L^2_kL^2_t(I)}^2\int_0^T \sum_{\ell \ne 0} e^{-2(1-c')\lambda(\tau)\langle \ell, \ell \tau/L\rangle^{\overline{\nu}}}\nonumber\\
&\qquad \cdot\left(\sum_N \sum_{k \ne 0} \int_{\tau}^{T}\sum_{|m| \le 2}\left|A^{(\sigma + 1)}\widehat{v^m \alpha \Phi}\left(\tau,k-\ell, \frac{kt - \ell \tau}{L}\right)_N\right|^2dt\right)d\tau\nonumber\\
&\le C(\lambda(0),\sigma,\overline{\nu},C_W,L) \|A\widehat{\rho}\|_{L^2_kL^2_t(I)}^2 \int_0^T \sum_{\ell \ne 0} e^{-2(1-c')\lambda(\tau)\langle \ell, \ell \tau/L\rangle^{\overline{\nu}}}\nonumber\\
&\qquad \cdot \left(\sum_N \sum_{k \ne 0}\sup_{\omega \in \mathbb{Z}^3\setminus \{0\}} \sup_{\zeta \in \mathbb{R}^3} \int_{-\infty}^{\infty}\sum_{|m| \le 2}\left|A^{(\sigma +1)}\widehat{v^m \alpha \Phi}\left(\tau,k,\frac{\omega}{|\omega|}s-\zeta\right)_N\right|^2 ds\right)d\tau\nonumber\\
&\le C(\lambda(0),\sigma,\overline{\nu},C_W,L,M) \|A\widehat{\rho}\|_{L^2_kL^2_t(I)}^2\nonumber\\
&\qquad \cdot \int_0^T \sum_{\ell \ne 0} e^{-2(1-c')\lambda(\tau)\langle \ell, \ell \tau/L\rangle^{\overline{\nu}}}\left(\sum_N \left\|A^{(\sigma+1)}\widehat{\alpha \Phi}(\tau, k, \cdot)_N\right\|^2_{L^2_kH^{4}_{\eta}}\right)d\tau.
\end{align}
The Littlewood-Paley projections do not commute with the taking of derivatives (and we can not be sure that they preserve the support of our original functions as operators in $(x,v)$), but since the supports of the projectors are well separated, we have
\begin{align}
&\left\|D^{m}_{\eta} \left(A^{(\sigma+1)}\widehat{\alpha \Phi}\right)_N\right\|^2_{L^2_{\eta}L^2_k} \nonumber\\
&\le C(|m|)\Bigg(\left\|\left(D^{m}_{\eta} A^{(\sigma+1)}\widehat{\alpha \Phi}\right)_{N/2}\right\|_{L^2_{\eta}L^2_k}^2 \nonumber\\
&\qquad \qquad \qquad + \left\|\left(D^{m}_{\eta} A^{(\sigma+1)}\widehat{\alpha \Phi}\right)_{N}\right\|_{L^2_{\eta}L^2_k}^2+\left\|\left(D^{m}_{\eta} A^{(\sigma+1)}\widehat{\alpha \Phi}\right)_{2N}\right\|_{L^2_{\eta}L^2_k}^2 \Bigg).
\end{align}
Combining this with the almost projection property of the decomposition gives
\begin{align}
\|A\mathcal{R}\|^2_{L^2_kL^2_t(I)} &\le C(\lambda(0),\sigma,\overline{\nu},C_W,L) \|A\widehat{\rho}\|_{L^2_kL^2_t(I)}^2\nonumber\\
&\qquad \cdot \int_0^{\infty} \sum_{\ell \ne 0} e^{-2(1-c')\lambda(\tau)\langle \ell, \ell \tau/L\rangle^{\overline{\nu}}}\left\|A^{(\sigma+1)}\widehat{\alpha \Phi}(\tau) \right\|^2_{L^2_kH^4_{\eta}} d\tau.
\end{align}
Our bootstrap hypotheses give us
\begin{equation}
\left\|A^{(\sigma+1)}\widehat{\alpha \Phi}(\tau) \right\|^2_{L^2_kH^4_{\eta}} \le 4 C K_1 \epsilon^2\langle \tau \rangle^{6},
\end{equation}
for some constant $C$ which depends only on the fact that we need to estimate four derivatives in $L^2$ (c.f. Corollary \ref{cor_Phi_1}).  Since $c'(\overline{\nu}) \in (0,1)$, we have the estimate
\begin{equation}
\|A\mathcal{R}\|^2_{L^2_kL^2_t(I)} \le C(\lambda(0),\sigma,\overline{\nu},C_W,L) K_1\epsilon^2 \|A\widehat{\rho}\|_{L^2_kL^2_t(I)}^2,
\end{equation}
which allows us to handle the remainder terms.

\subsubsection{Final Estimate on $\rho$}

Putting all of these results together, we find that
\begin{equation}
\left[1-\tilde{C}C_{LD}^2\left(K_1 + K_2\right)\epsilon^2\right]\|A\widehat{\rho}\|^2_{L^2_kL^2_t(I)} \le C_{LD}^2\sum_{k\ne 0}\left\|A\widehat{\Phi_{in}}\left(k,\frac{k\cdot}{L}\right)\right\|^2_{L^2_t(I)},
\end{equation}
where $\tilde{C} = \tilde{C}(\lambda(0),\sigma,\overline{\nu},C_W,L)$.  Thus, if we take $\epsilon^2 < \frac{1}{2}(\tilde{C}C_{LD}^2(K_1 + K_2))^{-1}$, then we find
\begin{eqnarray}
\|A\widehat{\rho}\|^2_{L^2_kL^2_t(I)} &\le& 2C_{LD}^2\sum_{k\ne 0}\left\|A\widehat{\Phi_{in}}\left(k,\frac{k\cdot}{L}\right)\right\|^2_{L^2_t(I)}\nonumber\\
&\le& 2CC_{LD}^2\epsilon^2,
\end{eqnarray}
where the final estimate is from \eqref{lin_est}. The constant $C$ comes from the Sobolev embedding and is independent of the $K_i$.  Thus, if we choose $K_3 = CC_{LD}^2$, we have proven \eqref{bstrap_rho} in Proposition \ref{bstrap_prp}.  Note that we can choose $K_3$ independently of $K_1$ and $K_2$. $\blacksquare$

\subsection{Pointwise in $t$ Estimate on $\rho$}

Next, we prove a (very) rough estimate on the size of the $L^2_x$-norm of $\rho$ as a function of time.  Unlike the global estimate above, the constant appearing below depends on all three $K_i$ (along with the other parameters in the problem).  As in the previous section, the result we detail here is identical to the one found in \cite{BMM13}[\S 5.2].

\begin{lem}[Pointwise $L^2_x$ Estimate for $\rho$] \label{pwise_rho_lem}
For $\epsilon_0$ sufficiently small, the bootstrap assumptions \eqref{orig_bstrap_AI} -- \eqref{orig_bstrap_rho} imply the existence of a constant $K_4 = K_4(C_0, \nu, \overline{\nu}, \kappa, \overline{\lambda}, \lambda_0, \lambda', K_1, K_2, K_3)$ so that for $t\in [0,T]$,
\begin{equation}
\|A\rho(t)\|_{L^2_x}^2 \le K_4\epsilon^2\langle t \rangle.
\end{equation}
\end{lem}

\bigskip

\textbf{Proof:}  Since we are looking for an estimate on the $L^2_x$-norm, it suffices to consider the transform of $\rho$ in $x$.  We look again at the paraproduct decomposition given in \eqref{para_dec_rho}:
\begin{align}
\|A\widehat{\rho}(t)\|_{L^2_k}^2 \le \sum_{k \ne 0}&\left|A\widehat{\Phi_{in}}\left(k,\frac{kt}{L}\right)\right|^2+\sum_{k \ne 0}\left|A(t,k,kt/L)\int_0^tL(t-\tau,k)\widehat{\rho}(\tau,k)d\tau\right|^2\nonumber\\
&+ \sum_{k \ne 0}|AT_k(t)|^2 + \sum_{k \ne 0}|AR_k(t)|^2 + \sum_{k \ne 0}|A\mathcal{R}_k(t)|^2.
\end{align}
The estimate on the first term is immediate from our assumptions on the initial data \eqref{ass_Phi_1}, the embedding $H^{M}(\mathbb{R}^3) \hookrightarrow C^{M-2, 1/2}(\mathbb{R}^3), $ and the fact that $\lambda(t)$ is decreasing:
\begin{eqnarray}
\sum_{k \ne 0}\left|A\widehat{\Phi_{in}}\left(k,\frac{kt}{L}\right)\right|^2 &\le& \sum_{k \ne 0}\sup_{\eta}\left|A\widehat{\Phi_{in}}\left(t, k,\eta\right)\right|^2\nonumber\\
&\le& C_s \left\|A(0)\widehat{\Phi_{in}}\right\|^2_{L^2_kH^2_{\eta}}\nonumber\\
&\le& C\epsilon^2,
\end{eqnarray}
where the final constant $C$ is a product of the $C_s$ coming from the Sobolev embedding and a purely combinatorial factor as in the proof of Corollary \ref{cor_Phi_1}.

\subsubsection{Linear Term}
Employing the same reasoning that led to \eqref{sup_est_G0}, the fact that $\lambda_0 > \lambda(0)$, and the fact that $|k| \ge 1$, we have
\begin{align}
&\sum_{k \ne 0}\Bigg|A(t,k,kt/L)\int_0^tL(t-\tau,k)\widehat{\rho}(\tau,k)d\tau\Bigg|^2\nonumber\\
 &\le 2^{\sigma} \left(\sum_{k \ne 0} \int_0^t \left\langle \left(\frac{t-\tau}{L}\right)|k| \right\rangle^{\sigma}e^{\lambda(0)\left\langle \left(\frac{t-\tau}{L}\right)|k| \right\rangle^{\overline{\nu}}}|L(t-\tau,k)|A(\tau,k,k\tau / L)|\widehat{\rho}(\tau,k)|d\tau\right)^2\nonumber\\
 & \le \frac{4\pi^2 2^{\sigma} C_W^2}{L^2}\left(\sup_{\eta}e^{\lambda_0\langle \eta \rangle^{\overline{\nu}}}\langle \eta \rangle^{\sigma} \left|\widehat{\alpha G_0}(\eta)\right|\right)^2 \left(\sum_{k\ne 0}\int_0^t e^{-(\lambda_0-\lambda(0))\langle (t-\tau)/L \rangle^{\overline{\nu}}} A\widehat{\rho}(\tau, k)d\tau\right)^2\nonumber\\
 &\le C(C_W, L, \sigma, \nu, \overline{\nu}, C_0, \lambda_0) \left(\int_0^{\infty} e^{-2(\lambda_0-\lambda(0))\langle (t-\tau)/L \rangle^{\overline{\nu}}}d\tau\right)\left\|A\widehat{\rho}\right\|_{L^2_kL^2_t(I)}^2\nonumber\\
 &\le C(C_W, L, \sigma, \nu, \overline{\nu}, C_0, \lambda_0, \lambda(0))K_3\epsilon^2
\end{align}
where the final inequality comes from bootstrap estimate \eqref{bstrap_rho}.

\subsubsection{Reaction Term}

For the reaction term, making the same sort of estimates leading to \eqref{reac_est} (and recalling the definition \eqref{def_Kbar}) gives
\begin{equation}
\sum_{k \ne 0}|AR_k(t)|^2 \le \sum_{k \ne 0} \left(\int_0^t \sum_{\ell \ne 0 }\overline{K}_{k,\ell}(t,\tau)\left|A\widehat{\rho}(\tau, \ell)\right| d\tau\right)^2.
\end{equation}
Applying Cauchy-Schwarz and interchanging the order of summation gives
\begin{align}
&\sum_{k \ne 0}|AR_k(t)|^2 \nonumber\\
&\le \sum_{k \ne 0}\left(\int_0^t \sum_{\ell \ne 0 }\overline{K}_{k,\ell}(t,\tau)d\tau\right)\left(\int_0^t \sum_{\ell \ne 0 }\overline{K}_{k,\ell}(t,\tau)\left|A\widehat{\rho}(\tau, \ell)\right|^2 d\tau\right)\nonumber\\
&\le \left(\sup_{t\ge 0} \sup_{k \ne 0} \int_0^t \sum_{\ell \ne 0}\overline{K}_{k,\ell}(t,\tau)d\tau\right) \sum_{\ell \ne 0}\int_0^t\left(\sum_{k \ne 0}\overline{K}_{k,\ell}(t,\tau)\left|A\widehat{\rho}(\tau, \ell)\right|^2 d\tau\right)\nonumber\\
&\le \left(\sup_{0\le t \le T} \sup_{k \ne 0} \int_0^t \sum_{\ell \ne 0}\overline{K}_{k,\ell}(t,\tau)d\tau\right) \left(\sup_{0\le \tau \le t} \sup_{\ell \ne 0} \sum_{k \ne 0}\overline{K}_{k,\ell}(t,\tau) \right)\left\|A\widehat{\rho}\right\|_{L^2_kL^2_t(I)}^2.
\end{align}
The first term appears as a factor in \eqref{reaction_control}.  In the course of proving this estimate, we will see that it is controlled by $C \sqrt{K_2} \epsilon$ (where $C$ will depend on various parameters of the problem but not $K_1, K_2,$ or $K_3$).  The second term is simply \eqref{reaction_control_2} (and is the only source of the $\langle t \rangle$ dependence).  Combining this with the estimate on the $L^2_kL^2_t(I)$-norm from the previous subsection gives
\begin{equation}
\sum_{k \ne 0}|AR_k(t)|^2 \le C K_2 K_3 \epsilon^4 \langle t \rangle,
\end{equation}
where $C$ does not depend on $K_1$, $K_2$, or $K_3$.

\subsubsection{Transport Term}

Proceeding as in the analogous subsection above and applying the estimate for the $L^2_kL^2_t(I)$-norm of $A\widehat{\rho}$ we derived there gives
\begin{align}
\sum_{k \ne 0}&|AT_k(t)|^2\nonumber\\
 &\le C(C_W, \sigma, L)\epsilon \sqrt{K_3} \sum_{k\ne 0}\sum_{\ell \ne 0}\int_0^t \langle \tau \rangle  e^{c_2\lambda(\tau)\langle \ell, \ell \tau/L \rangle^{\overline{\nu}}}|\widehat{\rho}(\tau, \ell)|\nonumber\\
 &\qquad \qquad \qquad \cdot \left(\sum_{m \le 2}\left|A^{(\sigma + 1)}\widehat{v^m \alpha \Phi}\left(\tau,k-\ell, \frac{kt - \ell \tau}{L}\right)\right|\right)^2 d\tau.
\end{align}
Interchanging the order of summation, re-indexing the summation over $k$, and recalling that $c_2 \in (0,1)$ (along with the fact that $|\ell| > 1$) gives
\begin{align}
&\sum_{k \ne 0}|AT_k(t)|^2\nonumber\\
&\le C(C_W, \sigma, L)\epsilon \sqrt{K_3} \left(\sup_{0 \le \tau \le t} e^{-(1-c_2)\lambda(0)\langle \tau / L \rangle^{\overline{\nu}}}\sum_{k \ne 0} \sup_{\eta}\sum_{|m| \le 2}\left|A^{(\sigma+1)}\widehat{v^m \alpha \Phi}\left(\tau,k, \eta\right)\right|^2\right)\nonumber\\
& \qquad \qquad \cdot \sum_{\ell \ne 0}\int_0^t e^{\lambda(\tau)\langle \ell, \ell \tau/L \rangle^{\overline{\nu}}}|\widehat{\rho}(\tau, \ell)|d\tau.
\end{align}
The term involving $A^{(\sigma+1)}\widehat{v^m \alpha \Phi}$, is controlled by the usual Sobolev embedding and bootstrap estimate \eqref{orig_bstrap_AI} up to a constant depending only on the parameters appearing in $A$ (as always, the multiplication by $v^m$ is of no worry whatsoever).  Note that the negative exponential in $\langle t \rangle^{\overline{\nu}}$ kills the time dependence here.  For the integral involving $\widehat{\rho}$, we can proceed as in the derivation of \eqref{int_trans_est} (along with our estimate on the $L^2_kL^2_t(I)$-norm of $A\widehat{\rho}$).  Putting this altogether yields a final estimate
\begin{equation}
\sum_{k \ne 0}|AT_k(t)|^2 \le CK_1K_3 \epsilon^4,
\end{equation}
where $C$ does not depend on $K_1, K_2,$ or $K_3$.

\subsubsection{Remainder Term}

Again proceeding in the same spirit as our calculation for the remainder terms above, we have
\begin{align}
\sum_{k \ne 0}&|A\mathcal{R}_k(t)|^2\nonumber\\
&\le C(\lambda(0),\sigma,\overline{\nu},C_W,L) \|A\widehat{\rho}\|_{L^2_kL^2_t(I)}^2\nonumber\\
&\cdot\sum_{k\ne 0} \int_0^t \sum_{\ell \ne 0}\Bigg(\sum_N e^{2c'\lambda(\tau)\langle k - \ell, (kt-\ell \tau)/L\rangle^{\overline{\nu}}}e^{-2(1-c')\lambda(\tau)\langle \ell, \ell \tau/L\rangle^{\overline{\nu}}}\nonumber\\
&\qquad \qquad \cdot |k-l, kt-\ell \tau|^2 \sum_{|m| \le 2}\left|\widehat{v^m \alpha \Phi}\left(\tau,k-\ell, \frac{kt - \ell \tau}{L}\right)_N\right|^2\Bigg)d\tau,
\end{align}
where $c' = c'(\overline{\nu}) \in (0,1)$ is the same constant as before.  Using Lemma \ref{exp_ineq_lem}[item (ii)], we have
\begin{align}
|k-\ell, kt-\ell \tau| & e^{c'(\overline{\nu})\lambda(\tau)\langle k - \ell, (kt-\ell \tau)/L\rangle^{\overline{\nu}}} \nonumber\\
& \le C(\overline{\nu},\lambda(0),  \beta )A^{(\lambda,\sigma-\beta,\overline{\nu})}\left(\tau, k - \ell, (kt-\ell \tau)/L\right),
\end{align}
where we can take $\beta > 2$.  Hence, our estimate on $\|A\widehat{\rho}\|_{L^2_kL^2_t(I)}$ combined with the observation above yields
\begin{align}
&\sum_{k \ne 0}|A\mathcal{R}_k(t)|^2\nonumber\\
& \le C(\lambda(0),\sigma, \beta, \overline{\nu},C_W,L) K_3 \epsilon^2 \nonumber\\
&\cdot\sum_{k\ne 0} \int_0^t \sum_{\ell \ne 0} \left(\sum_N e^{-2(1-c')\lambda(\tau)\langle \ell, \ell \tau/L\rangle^{\overline{\nu}}} \sum_{|m| \le 2}\left|A^{(\sigma-\beta)}\widehat{v^m \alpha \Phi}\left(\tau,k-\ell, \frac{kt - \ell \tau}{L}\right)_N\right|^2\right)d\tau.
\end{align}
Interchanging the order of summation and using the $H^M(\mathbb{R}^3) \hookrightarrow C^{M-2, 1/2}(\mathbb{R}^3)$ embedding gives
\begin{align}
\sum_{k \ne 0}&|A\mathcal{R}_k(t)|^2\nonumber\\
& \le C(\lambda(0),\sigma, \beta, \overline{\nu},C_W,L) K_3 \epsilon^2 \nonumber\\
&\cdot\sum_{\ell\ne 0} \int_0^t e^{-2(1-c')\lambda(\tau)\langle \ell, \ell \tau/L\rangle^{\overline{\nu}}} \left(\sum_{k \ne 0}\sum_N  \left\|A^{(\sigma-\beta)}\widehat{\alpha \Phi}(\tau)_N\right\|_{H^4_{\eta}}^2\right)d\tau.
\end{align}
As before, we can commute the derivatives with the Paley-Littlewood projectors at the expense of multiplying by a combinatorial factor.  We can then drop the projectors altogether to give
\begin{equation}
\sum_{k \ne 0}|A\mathcal{R}_k(t)|^2 \le  CK_3\sum_{\ell\ne 0} \int_0^t e^{-2(1-c')\lambda(\tau)\langle \ell, \ell \tau/L\rangle^{\overline{\nu}}}   \left\|A^{(\sigma-\beta)}\widehat{\alpha \Phi}(\tau)\right\|_{L^2_kH^4_{\eta}}^2d\tau,
\end{equation}
where $C=C(\lambda(0),\sigma,\beta,\overline{\nu},C_W,L).$ Bootstrap estimate \eqref{orig_bstrap_AbI} then gives us
\begin{align}
\sum_{k \ne 0}|A\mathcal{R}_k(t)|^2 &\le C(\lambda(0),\sigma, \beta, \overline{\nu},C_W,L) K_2K_3 \epsilon^4 \sum_{\ell\ne 0} \int_0^{\infty} e^{-2(1-c')\lambda(\tau)\langle \ell, \ell \tau/L\rangle^{\overline{\nu}}}d\tau\nonumber\\
&\le C(\lambda(0),\sigma, \beta, \overline{\nu},C_W,L) K_2K_3 \epsilon^4,
\end{align}
since the final integration and summation is certainly finite.

\subsubsection{Final Estimate}

Combining all terms and making very gross estimates yields
\begin{align}
\|A\widehat{\rho}(t)\|_{L^2_k}^2 \le \sum_{k \ne 0}&\left|A\widehat{\Phi_{in}}\left(k,\frac{kt}{L}\right)\right|^2+\sum_{k \ne 0}\left|A(t,k,kt/L)\int_0^tL(t-\tau,k)\widehat{\rho}(\tau,k)d\tau\right|^2\nonumber\\
&+ \sum_{k \ne 0}|AT_k(t)|^2 + \sum_{k \ne 0}|AR_k(t)|^2 + \sum_{k \ne 0}|A\mathcal{R}_k(t)|^2\nonumber\\
&\le C\epsilon^2\left(1 + K_3 + K_2K_3 + K_1K_3\right)\langle t \rangle,
\end{align}
where the constant $C$ depends on the various parameters appearing in the problem but not on $K_1, K_2,$ or $K_3$.  Hence, we have the estimate quoted in the lemma.  $\blacksquare$
\section{Bootstrap Estimate \eqref{bstrap_AI}}

\addtocontents{toc}{\setcounter{tocdepth}{1}}

Establishing bootstrap estimates \eqref{bstrap_AI} and \eqref{bstrap_AbI} is similar in outline to the treatment in \cite{BMM13}[\S 5.3 and 5.4].  However, the techniques employed therein require significant alteration in the relativistic case.  In particular, it is here that both the form of the bootstrap estimates and \emph{a priori} Assumption \eqref{apriori_ass} play a crucial role.

We have
\begin{align}
\frac{1}{2}\frac{d}{dt}\Bigg[\Big\|A^{(\sigma+1)} & \widehat{ \Phi}(t) \Big\|^2_2 -\left\|\widehat{v A^{(\sigma+1)} \Phi}(t) \right\|^2_2+\left\|A^{(\sigma+1)}\widehat{\alpha \Phi}(t) \right\|^2_2\Bigg]\nonumber\\
=&\dot{\lambda}(t) \left\|A^{(\sigma + 1 + \overline{\nu}/2)} \widehat{\Phi}(t)\right\|_2^2 + \dot{\lambda}(t) \left\|A^{(\sigma + 1 + \overline{\nu}/2)} \widehat{\alpha \Phi}(t)\right\|_2^2\nonumber\\
&\;\; -\dot{\lambda}(t)\Re \sum_k \int \overline{\frac{i}{2\pi}\nabla_{\eta}A^{(\sigma + 1)}\widehat{\Phi}(t,k,\eta)} \cdot \frac{i}{2\pi}\nabla_{\eta}A^{(\sigma + 1 + \overline{\nu})}\widehat{\Phi}(t,k,\eta) d\eta\nonumber\\
&\;\;+ \Re \sum_k \int A^{(\sigma + 1)}(t,k,\eta)^2 \overline{\widehat{\Phi}(t,k,\eta)}\partial_t \widehat{\Phi}(t,k,\eta) d\eta\nonumber\\
&\;\;+ \Re \sum_k \int A^{(\sigma + 1)}(t,k,\eta)^2 \overline{\widehat{\alpha \Phi}(t,k,\eta)}\partial_t \widehat{\alpha \Phi}(t,k,\eta) d\eta\nonumber\\
&\;\;-\Re \sum_k \int \overline{\frac{i}{2\pi}\nabla_{\eta}A^{(\sigma + 1)}\widehat{\Phi}(t,k,\eta)} \cdot \frac{i}{2\pi}\nabla_{\eta}A^{(\sigma + 1)}\partial_t\widehat{\Phi}(t,k,\eta) d\eta.
\end{align}
Before we insert the time derivatives, we first perform some elementary manipulations on the two terms with gradients.  For the first such term, we would prefer to have an expression where the indices on the multiplier $A$ are symmetric.  To that end, we note that
\begin{align}
\Re \sum_k \int & \overline{\frac{i}{2\pi}\nabla_{\eta} A^{(\sigma + 1)}\widehat{\Phi}(t,k,\eta)} \cdot \frac{i}{2\pi}\nabla_{\eta}A^{(\sigma + 1 + \overline{\nu})}\widehat{\Phi}(t,k,\eta) d\eta\nonumber\\
& = \left\| \widehat{vA^{(\sigma + 1 +\overline{\nu}/2)}\Phi} \right\|_2^2 - \frac{\overline{\nu}^2}{16\pi^2} \sum_k \int \left|A^{(\sigma + \overline{\nu}/2)}\widehat{\Phi}(t,k,\eta)\right|^2\frac{|\eta|^2}{\langle k , \eta \rangle^2} d\eta.
\end{align}
For the final term, we perform an integration-by-parts to get a divergence of the complex conjugate term.  We then expand the resulting Laplacian to obtain
\begin{align}
-\Re \sum_k & \int \overline{\frac{i}{2\pi}\nabla_{\eta}A^{(\sigma + 1)}\widehat{\Phi}(t,k,\eta)} \cdot \frac{i}{2\pi}\nabla_{\eta}A^{(\sigma + 1)}\partial_t\widehat{\Phi}(t,k,\eta) d\eta\nonumber\\
=& \frac{1}{4\pi^2}\Re \sum_k \int A^{(\sigma + 1)}\triangle_{\eta}A^{(\sigma + 1)}(t,k,\eta) \overline{\widehat{\Phi}(t,k,\eta)} \partial_t \widehat{\Phi}(t,k,\eta) d\eta\nonumber\\
& + \frac{1}{2\pi^2}\Re \sum_k \int A^{(\sigma + 1)}\nabla_{\eta}A^{(\sigma + 1)}(t,k,\eta) \cdot \overline{\nabla_{\eta} \widehat{\Phi}(t,k,\eta)} \partial_t \widehat{\Phi}(t,k,\eta) d\eta\nonumber\\
& - \Re \sum_k \int A^{(\sigma + 1)}(t,k,\eta)^2 \overline{\widehat{|v|^2\Phi}(t,k,\eta)}\partial_t \widehat{\Phi}(t,k,\eta) d\eta.
\end{align}

We now input the known derivatives of the transforms (\eqref{main_DE_Phi} and \eqref{main_DE_aPhi}) and obtain a number of terms:
\begin{equation}
\frac{1}{2}\frac{d}{dt}\Bigg[\left\|A^{(\sigma+1)}\widehat{ \Phi}(t) \right\|^2_2-\left\|\widehat{v A^{(\sigma+1)} \Phi}(t) \right\|^2_2+\left\|A^{(\sigma+1)}\widehat{\alpha \Phi}(t) \right\|^2_2\Bigg]= CK(t) + L(t) + NL(t),
\end{equation}
where
\begin{align}
&CK(t) = CK_1(t) + CK_2(t) + CK_3(t)\nonumber\\
&= \dot{\lambda}(t) \left(\left\|A^{(\sigma + 1 + \overline{\nu}/2)} \widehat{\Phi}(t)\right\|_2^2 - \left\| \widehat{vA^{(\sigma + 1 +\overline{\nu}/2)}\Phi} \right\|_2^2\right)\nonumber\\
& \; \; + \dot{\lambda}(t) \left\|A^{(\sigma + 1 + \overline{\nu}/2)} \widehat{\alpha \Phi}(t)\right\|_2^2 + \dot{\lambda}(t)\frac{\overline{\nu}^2}{16\pi^2} \sum_k \int \left|A^{(\sigma + \overline{\nu}/2)}\widehat{\Phi}(t,k,\eta)\right|^2\frac{|\eta|^2}{\langle k , \eta \rangle^2} d\eta,
\end{align}
\begin{align}
L&(t) = L_1(t) + L_2(t)+L_3(t)+L_4(t)\nonumber\\
=& -2\pi \Im \sum_k \int A^{(\sigma + 1)}(t,k,\eta)^2 \overline{\mathcal{F}\{(1-|v|^2)\Phi\}(t,k,\eta)}\widehat{W}(k)\widehat{\rho}(t,k)\frac{k}{L}\cdot \widehat{\alpha G_0}\left(\eta - \frac{kt}{L}\right)d\eta\nonumber\\
&-2\pi \Im \sum_k \int A^{(\sigma + 1)}(t,k,\eta)^2 \overline{\widehat{\alpha\Phi}(t,k,\eta)}\widehat{W}(k)\widehat{\rho}(t,k)\frac{k}{L}\cdot \mathcal{F}\{(1-|v|^2) G_0\}\left(\eta - \frac{kt}{L}\right)d\eta\nonumber\\
&-\frac{1}{2\pi} \Im \sum_k \int A^{(\sigma + 1)}\triangle_{\eta}A^{(\sigma + 1)}(t,k,\eta) \overline{\widehat{\Phi}(t,k,\eta)}\widehat{W}(k)\widehat{\rho}(t,k)\frac{k}{L}\cdot \widehat{\alpha G_0}\left(\eta - \frac{kt}{L}\right)d\eta\nonumber\\
&-\frac{1}{\pi} \Im \sum_k \int A^{(\sigma + 1)}\nabla_{\eta}A^{(\sigma + 1)}(t,k,\eta) \cdot \overline{\nabla_{\eta}\widehat{\Phi}(t,k,\eta)}\widehat{W}(k)\widehat{\rho}(t,k)\frac{k}{L}\cdot \widehat{\alpha G_0}\left(\eta - \frac{kt}{L}\right)d\eta,
\end{align}
and the remaining terms are given by
\begin{equation}
NL(t) = NL_1(t) + NL_2(t) + NL_3(t),
\end{equation}
\begin{align}
NL_1(t) = -4\pi^2 \Re \sum_k \sum_{\ell \ne 0}&\int A^{(\sigma + 1)}(t,k,\eta)^2 \overline{\mathcal{F}\{(1-|v|^2)\Phi\}(t,k,\eta)}\widehat{W}(\ell)\widehat{\rho}(t,\ell)\nonumber\\
& \cdot \frac{\ell}{L}\cdot\left(\eta - \frac{kt}{L}\right)\widehat{\alpha\Phi}\left(t,k-\ell,\eta-\frac{\ell t}{L}\right)d\eta\nonumber\\
 -4\pi^2 \Re \sum_k \sum_{\ell \ne 0}&\int A^{(\sigma + 1)}(t,k,\eta)^2 \overline{\widehat{\alpha \Phi}(t,k,\eta)}\widehat{W}(\ell)\widehat{\rho}(t,\ell)\nonumber\\
&  \cdot \frac{\ell}{L}\cdot\left(\eta - \frac{kt}{L}\right)\mathcal{F}\{(1-|v|^2)\Phi\}\left(t,k-\ell,\eta-\frac{\ell t}{L}\right)d\eta,\\
NL_2(t) = -2\pi \Im \sum_k \sum_{\ell \ne 0}&\int A^{(\sigma + 1)}(t,k,\eta)^2 \overline{\mathcal{F}\{(1-|v|^2)\Phi\}(t,k,\eta)}\widehat{W}(\ell)\widehat{\rho}(t,\ell)\nonumber\\
&  \cdot \frac{\ell}{L}\cdot\left[\left(\eta - \frac{kt}{L}\right)\cdot \nabla_{\eta}\widehat{v\alpha\Phi}\left(t,k-\ell,\eta-\frac{\ell t}{L}\right)\right]d\eta\nonumber\\
 -2\pi \Im \sum_k \sum_{\ell \ne 0}&\int A^{(\sigma + 1)}(t,k,\eta)^2 \overline{\widehat{\alpha \Phi}(t,k,\eta)}\widehat{W}(\ell)\widehat{\rho}(t,\ell)\nonumber\\
&   \cdot \frac{\ell}{L}\cdot\left[\left(\eta - \frac{kt}{L}\right)\cdot \nabla_{\eta}\mathcal{F}\{v(1-|v|^2)\Phi\}\left(t,k-\ell,\eta-\frac{\ell t}{L}\right)\right]d\eta,
\end{align}
\begin{align}
NL_3(t) = -2 \pi \Im \sum_k \sum_{\ell \ne 0} & \int A^{(\sigma + 1)}(t,k,\eta)^2\overline{\widehat{\alpha \Phi}(t,k,\eta)} \widehat{W}(\ell)\widehat{\rho}(t,\ell)\nonumber\\
&\cdot \frac{\ell}{L}\cdot \mathcal{F}\{v(1-|v|^2)\Phi\}\left(t, k-\ell, \eta - \frac{\ell t}{L}\right)d\eta \nonumber\\
 - \Re \sum_k \sum_{\ell \ne 0} & \int A^{(\sigma + 1)}\triangle_{\eta}A^{(\sigma + 1)}(t,k,\eta) \overline{\widehat{\Phi}(t,k,\eta)}\widehat{W}(\ell)\widehat{\rho}(t,\ell)\nonumber\\
& \cdot \frac{\ell}{L}\cdot \left(\eta - \frac{kt}{L}\right)\widehat{\alpha \Phi}\left(t,k-\ell,\eta-\frac{\ell t}{L}\right) d\eta\nonumber\\
 - \frac{1}{2\pi} \Im \sum_k \sum_{\ell \ne 0} & \int A^{(\sigma + 1)}\triangle_{\eta}A^{(\sigma + 1)}(t,k,\eta) \overline{\widehat{\Phi}(t,k,\eta)}\widehat{W}(\ell)\widehat{\rho}(t,\ell)\nonumber\\
&\cdot \frac{\ell}{L}\cdot \left[\left(\eta - \frac{kt}{L}\right)\cdot \nabla_{\eta}\widehat{v \alpha \Phi}\left(t,k-\ell,\eta-\frac{\ell t}{L}\right)\right] d\eta\nonumber\\
- 2 \Re \sum_k \sum_{\ell \ne 0} & \int A^{(\sigma + 1)}\nabla_{\eta}A^{(\sigma + 1)}(t,k,\eta) \widehat{W}(\ell)\widehat{\rho}(t,\ell) \nonumber\\
&\cdot \overline{\nabla_{\eta}\widehat{\Phi}(t,k,\eta)}\frac{\ell}{L}\cdot \left(\eta - \frac{kt}{L}\right)\widehat{\alpha \Phi}\left(t,k-\ell,\eta-\frac{\ell t}{L}\right) d\eta\nonumber\\
 -\frac{1}{\pi} \Im \sum_k \sum_{\ell \ne 0} & \int A^{(\sigma + 1)}\nabla_{\eta}A^{(\sigma + 1)}(t,k,\eta)\widehat{W}(\ell)\widehat{\rho}(t,\ell) \nonumber\\
 &\cdot \overline{\nabla_{\eta}\widehat{\Phi}(t,k,\eta)}\frac{\ell}{L}\cdot \left[\left(\eta - \frac{kt}{L}\right)\cdot \nabla_{\eta}\widehat{v \alpha \Phi}\left(t,k-\ell,\eta-\frac{\ell t}{L}\right)\right] d\eta.
\end{align}

The $CK$-terms (which are manifestly negative) are similar to terms appearing in \emph{Cauchy-Kovalevskaya} type arguments and are used to absorb the highest order contributions from the $NL$-terms.  We note that the $CK$-term involving $A^{(\sigma + \overline{\nu}/2)}$ will play no role in the following estimates and can be dropped altogether.  The $L$-terms are the ones arising from the \emph{linear} portion of the evolution equations for the transforms.  These are fairly easy to estimate since the background, $G_0$, is assumed to have a higher degree of regularity than that measured by $A$ (recall $(2+\gamma)^{-1} < \overline{\nu} < \nu < 1$).  The $NL$-terms are the ones arising from the \emph{non-linear} portions of the evolution equations.

The primary issue in dealing with the non-linear terms is the extra factors of $\eta - kt/L$ which appear thanks to the various time derivatives.  If we try the simple estimate $$\left| \eta  - \frac{kt}{L} \right| \le C \langle t \rangle \langle k, \eta \rangle,  $$ we run into trouble.  This actually works fine in $NL_3$ since a derivative of $A$  gives us an extra $\langle k,\eta \rangle^{\overline{\nu} - 1}$ which will allow us to combine this term with the $CK$-terms.  As for $NL_1$ and $NL_2$, we will decompose these terms via the paraproduct decomposition as above.  It turns out that the reaction and remainder terms can be handled without much ado.  The transport terms require the use of a commutator identity which we will detail after dealing with the linear terms.  Unfortunately, the use of the commutator identities will add even more lower order terms to $NL_3$.

\subsection{Linear Terms}

The first two linear terms have absolute value controlled by an integral of the form
\begin{align}
\left|L_j(t)\right| \le  2\pi \sum_{k \ne 0} \int A^{(\sigma + 1)}(t,k,\eta)^2&\left|\widehat{f_j}(t,k,\eta)\right|\left|\widehat{W}(k)\right|\frac{|k|}{L} \nonumber\\
& \cdot \left|\widehat{\rho}\left(t, k\right)\right|\left|\widehat{\alpha^j G_0}\left(\eta - \frac{kt}{L}\right)\right]d\eta,
\end{align}
where $f_j = (1-|v|^2)\Phi$ or $\alpha \Phi$ for $j= 1$ and $2$, respectively.  Since $G_0$ has no separate $k$ dependence, we can write
\begin{align}
\left|L_j(t)\right| \le  2\pi \sum_{k \ne 0} \sum_{\ell}\int & A^{(\sigma + 1)}(t,k,\eta)^2\Big| \widehat{f_j}(t,k,\eta)\Big|\left|\widehat{W}(\ell)\right|\frac{|\ell|}{L} \nonumber\\
& \cdot \left|\widehat{\rho}\left(t, \ell\right)\right| \mathds{1}_{k=\ell}\left|\widehat{\alpha^j G_0}\left(k-\ell, \eta - \frac{\ell t}{L}\right)\right]d\eta,
\end{align}
in order to bring this term into a form where we can apply Lemma \ref{product_lem}.  Thus, we see that
\begin{align}
\left|L_j(t)\right|\le C \left\|A^{(\sigma + 1)}\widehat{f_j}(t) \right\|_{L^2_kL^2_{\eta}} \Bigg[& \left\|A^{(\tilde{c}\lambda,0)}\widehat{\alpha^j G_0}(t) \right\|_{L^2_{\eta}}\left\|A^{(\sigma +1)}\widehat{W}\frac{k}{L}\widehat{\rho} \right\|_{L^2_k}\nonumber\\
&+ \left\|A^{(\sigma +1)}\widehat{ \alpha^j G_0}(t) \right\|_{L^2_{\eta}}\left\|A^{(\tilde{c}\lambda,0)}\widehat{W}\frac{k}{L}\widehat{\rho} \right\|_{L^2_k}\Bigg],
\end{align}
for some constants $C=C\left(\lambda(0), \sigma, \overline{\nu}\right)$ and $\tilde{c} = \tilde{c}(\sigma, \overline{\nu}) \in (0,1)$.  Note that we have dropped the $L^2_k$ summation on the $G_0$ terms since only the $k=0$ term is non-zero.  Clearly, both terms involving the background data can be controlled by our assumption on the initial data.  More precisely, since $\overline{\nu} < \nu$ we certainly have $$\frac{A^{(\lambda, \sigma'; \overline{\nu})}(t,k,\eta)}{A^{(\overline{\lambda}, 0; \nu)(t,k,\eta)}} \le \frac{\langle k, \eta \rangle^{\sigma'}e^{\lambda(0)\langle k, \eta \rangle^{\overline{\nu}}}}{e^{\overline{\lambda}\langle k, \eta \rangle^{\nu}}} \le C(\lambda(0), \overline{\lambda}, \overline{\nu}, \nu, \sigma' ),  $$  for all $t, k,$ and $\eta$.  Thus, we have
\begin{equation}
\left\|A^{(\tilde{c}\lambda,0)}\widehat{\alpha^j G_0}(t) \right\|_{L^2_{\eta}}, \; \left\|A^{(\sigma +1)}\widehat{ \alpha^j G_0}(t) \right\|_{L^2_{\eta}} \le C(\lambda(0), \overline{\lambda}, \overline{\nu}, \nu, \sigma, C_0 ).
\end{equation}
We also have (thanks to $\gamma \ge 1$)
\begin{align}
A^{(\sigma +1)}\left(t,k,\frac{kt}{L}\right)|\widehat{W}(k)|\frac{|k|}{L} &= \langle k, kt/L \rangle|\widehat{W}(k)|\frac{|k|}{L} A\left(t,k,\frac{kt}{L}\right) \nonumber\\
&\le C(C_W,L) \langle t \rangle  A\left(t,k,\frac{kt}{L}\right),\\
A^{(\tilde{c}\lambda,0)}\left(t,k,\frac{kt}{L}\right)|\widehat{W}(k)|\frac{|k|}{L} &\le C_W e^{-(1-\tilde{c})\alpha_0\langle t \rangle^{\overline{\nu}}} A\left(t,k,\frac{kt}{L}\right)\nonumber\\
&\le C(C_W,\alpha_0,\sigma, \overline{\nu})  A\left(t,k,\frac{kt}{L}\right),
\end{align}
where $\alpha_0$ is our lower bound for $\lambda(t)$.  Putting these facts together gives
\begin{align}
\left|L_1(t)\right| + & \left|L_2(t)\right|  \nonumber\\
 & \le C \langle t \rangle \left\|A\widehat{\rho} \right\|_{L^2_k} \left(\left\|A^{(\sigma + 1)}\widehat{\Phi}(t) \right\|_{L^2_kL^2_{\eta}}+\left\|A^{(\sigma + 1)}\widehat{\alpha \Phi}(t) \right\|_{L^2_kL^2_{\eta}}\right)
\end{align}
where $C=C(C_W,\alpha_0, \lambda(0), \overline{\lambda}, \sigma, \overline{\nu},\nu,C_0)$ is some constant.  As we have done previously, the $1-|v|^2$ term has been dropped in the estimates (which we can do at the expense of a combinatorial factor).

For $L_3$ and $L_4$, we use \eqref{est_on_DA} to deduce that
\begin{eqnarray}
\left| A^{(\sigma + 1)}\triangle_{\eta}A^{(\sigma + 1)}(t,k,\eta) \right| &\le& C A^{(\sigma + \overline{\nu})}(t,k,\eta)^2,\\
\left|A^{(\sigma + 1)}\nabla_{\eta}A^{(\sigma + 1)}(t,k,\eta) \right| &\le& C A^{(\sigma + 1/2 + \overline{\nu}/2)}(t,k,\eta)^2.
\end{eqnarray}
Repeating the same arguments as above (noting $\sigma + \overline{\nu} < \sigma + 1/2 \overline{\nu}/2 < \sigma + 1$) yields
\begin{eqnarray}
|L_3(t)| &\le& C \langle t \rangle^{\overline{\nu}} \left\|A\widehat{\rho} \right\|_{L^2_k} \left\|A^{(\sigma + 1)}\widehat{\Phi}(t) \right\|_{L^2_kL^2_{\eta}},\\
|L_4(t)| &\le& C \langle t \rangle^{1/2+\overline{\nu}/2} \left\|A\widehat{\rho} \right\|_{L^2_k} \left\|A^{(\sigma + 1)}\widehat{\Phi}(t) \right\|_{L^2_kL^2_{\eta}}.
\end{eqnarray}
Putting these altogether, we arrive at the final bound for the linear terms:
\begin{equation}
|L(t)| \le C \langle t \rangle \left\|A\widehat{\rho} \right\|_{L^2_k} \left(\left\|A^{(\sigma + 1)}\widehat{\Phi}(t) \right\|_{L^2_kL^2_{\eta}}+\left\|A^{(\sigma + 1)}\widehat{\alpha \Phi}(t) \right\|_{L^2_kL^2_{\eta}}\right)
\end{equation}
where $C=C(C_W,\alpha_0, \lambda(0), \overline{\lambda}, \sigma, \overline{\nu},\nu,C_0)$ is some constant.

Counting powers of $t$, these terms seem acceptable.  However, leaving them as is would give us an extra factor of $K_1$ at the $\epsilon^2$-level of approximation.  Since we want to find a bound for $K_1$, this is not an optimal state of affairs.  To remedy this, we use the standard identity $$ab \le \frac{1}{2}(a^2 + b^2),$$ and so for a small parameter $b>0$ we have
\begin{align}
|L(t)| \le &\frac{C}{b} \langle t \rangle^4 \left\|A\widehat{\rho} \right\|_{L^2_k}^2 +Cb\langle t \rangle^{-2} \left\|A^{(\sigma + 1+ \overline{\nu}/2)}\widehat{\Phi}(t) \right\|_{L^2_kL^2_{\eta}}^2\nonumber\\
&+\frac{C}{b} \langle t \rangle^4 \left\|A\widehat{\rho} \right\|_{L^2_k}^2 +Cb\langle t \rangle^{-2}\left\|A^{(\sigma + 1 + \overline{\nu}/2)}\widehat{\alpha \Phi}(t) \right\|_{L^2_kL^2_{\eta}}^2.
\end{align}
Replacing $\sigma + 1$ with $\sigma + 1+ \overline{\nu}/2$ (which can only make these terms larger) allows us to combine these with the $CK$ terms.

\subsection{Integral Identities and the Decomposition of the Non-Linear Terms}

We now turn to the non-linear terms.  We will tackle $NL_1$ and $NL_2$ via the paraproduct decomposition and various integral identities (control of $NL_3$ will not require such tools).  In fact, it is the use of the integral identities below which necessitates the form of the bootstrap estimates, i.e. the subtraction of the $vA\Phi$ terms.  This in turn necessitates the \emph{a priori} assumption \eqref{apriori_ass} so that we can close the estimates.  Moreover, these integral identities are only necessary to handle the transport terms in the decomposition (which we will define below).  Essentially, we are forced to assume \eqref{apriori_ass} only to control one part of the paraproduct decomposition.  Whether there is a less restrictive (and less cumbersome) way to ensure that the transport terms are small is an important question.

We begin by noting the following identities:
\begin{align}
\iint_{\mathbb{T}^3_L\times B_1(0)}F(t,x+vt)\cdot  & \left(\nabla_v - t\nabla_x\right)G(x,v) dxdv \equiv 0,\label{int_id_1}\\
\iint_{\mathbb{T}^3_L\times B_1(0)}\Big\{F(t,x+vt)\cdot  \left[v\otimes v\right] & \left(\nabla_v - t\nabla_x\right)G(x,v)\nonumber\\
 & +4 F(t,x+vt)\cdot v G(x,v)\Big\} dxdv \equiv 0 \label{int_id_2},
\end{align}
which hold for any sufficiently nice function $G$ on $\mathbb{T}^3_L\times B_1(0)$ (in particular, it must vanish at the boundary $|v|=1$ with suitable rapidity).  This is, of course, equivalent to the $(0,0)$-mode of the transform of the integrands above being identically zero.

To make use of these identities, we substitute $G(x,v) = f(x,v)g(x,v)$ (for $f$ and $g$ real) and use the product rule to expand the derivatives.  Taking the transform gives us complicated convolution terms which must be identically zero when evaluated at $(0,0)$.  For \eqref{int_id_1}, this process gives us the following identity:
\begin{align}
\sum_{k} \sum_{\ell \ne 0} \int_{\eta} & \overline{\widehat{f}(k,\eta)}\widehat{W}(\ell) \widehat{\rho}(t,\ell) \frac{\ell}{L}\cdot \left(\eta - \frac{kt}{L}\right)\widehat{g}\left(k-\ell, \eta - \frac{\ell t}{L}\right) d\eta \nonumber\\
& +  \sum_{k} \sum_{\ell \ne 0} \int_{\eta} \overline{\widehat{g}(k,\eta)}\widehat{W}(\ell) \widehat{\rho}(t,\ell) \frac{\ell}{L}\cdot \left(\eta - \frac{kt}{L}\right)\widehat{f}\left(k-\ell, \eta - \frac{\ell t}{L}\right)   d\eta \equiv 0.\label{conv_id_1}
\end{align}
From \eqref{int_id_2}, we obtain the following:
\begin{align}
\sum_{k} & \sum_{\ell \ne 0} \int_{\eta}  \overline{\widehat{f}(k,\eta)}\widehat{W}(\ell) \widehat{\rho}(t,\ell) \frac{\ell}{L}\cdot \left[\left(\eta - \frac{kt}{L}\right)\cdot \nabla_{\eta}\widehat{vg}\left(k-\ell, \eta - \frac{\ell t}{L}\right)\right] d\eta \nonumber\\
& +  \sum_{k} \sum_{\ell \ne 0} \int_{\eta} \overline{\widehat{g}(k,\eta)}\widehat{W}(\ell) \widehat{\rho}(t,\ell) \frac{\ell}{L}\cdot \left[\left(\eta - \frac{kt}{L}\right)\cdot \nabla_{\eta}\widehat{vf}\left(k-\ell, \eta - \frac{\ell t}{L}\right) \right]  d\eta \nonumber\\
&+4\sum_{k} \sum_{\ell \ne 0} \int_{\eta} \overline{\widehat{g}(k,\eta)}\widehat{W}(\ell) \widehat{\rho}(t,\ell) \frac{\ell}{L} \cdot \widehat{vf}\left(k-\ell, \eta - \frac{\ell t}{L}\right)d\eta\equiv 0.\label{conv_id_2}
\end{align}
Note that in the final term above, the roles of $f$ and $g$ are interchangeable (as is clear from the form of the analogous term in the original integral).  In both expressions above, we will want to make substitutions like $f(t,x,v) = A(1-|v|^2)\Phi(t,x,v)$ and $g(t,x,v) = A\alpha\Phi(t,x,v)$ (adding a time dependence to $f$ and $g$ does not alter the integral identity whatsoever).  This presents no problem for \eqref{conv_id_1} since the factors of $A$ are just multipliers on the Fourier side.  The story is different for \eqref{conv_id_2} as one of the factors of $A$ is essentially buried behind a second derivative (specifically, the directional derivative of a gradient).  To that end, we insert $Af(t,x,v)$ and $Ag(t,x,v)$ into the second integral identity and expand the derivatives involving the factors of $A$.  This yields a particularly lengthy identity:
\begin{align}
&\sum_{k} \sum_{\ell \ne 0} \int_{\eta}A(t,k,\eta) \overline{\widehat{f}(t,k,\eta)}\widehat{W}(\ell) \widehat{\rho}(t,\ell) \nonumber\\
& \cdot A\left(t,k-\ell, \eta - \frac{\ell t}{L}\right)\frac{\ell}{L}\cdot \left[\left(\eta - \frac{kt}{L}\right)\cdot \nabla_{\eta}\widehat{vg}\left(t,k-\ell, \eta - \frac{\ell t}{L}\right)\right] d\eta \nonumber\\
& +  \sum_{k} \sum_{\ell \ne 0} \int_{\eta} A(t,k,\eta) \overline{\widehat{g}(t,k,\eta)}\widehat{W}(\ell) \widehat{\rho}(t,\ell) \nonumber\\
& \cdot A\left(t,k-\ell, \eta - \frac{\ell t}{L}\right)\frac{\ell}{L}\cdot \left[\left(\eta - \frac{kt}{L}\right)\cdot \nabla_{\eta}\widehat{vf}\left(t,k-\ell, \eta - \frac{\ell t}{L}\right) \right]  d\eta \nonumber\\
& + \sum_{k} \sum_{\ell \ne 0} \int_{\eta} A(t,k,\eta) \overline{\widehat{f}(t,k,\eta)}\widehat{W}(\ell) \widehat{\rho}(t,\ell) \nonumber\\
&  \cdot \frac{i \ell}{2\pi L}\cdot\left[\left(\eta-\frac{kt}{L}\right)\cdot \nabla_{\eta} \nabla_{\eta}A\left(t,k-\ell, \eta - \frac{\ell t}{L}\right) \right] \widehat{g}\left(t,k-\ell, \eta - \frac{\ell t}{L}\right) d\eta \nonumber\\
& + \sum_{k} \sum_{\ell \ne 0} \int_{\eta} A(t,k,\eta) \overline{\widehat{g}(t,k,\eta)}\widehat{W}(\ell) \widehat{\rho}(t,\ell) \nonumber\\
& \cdot \frac{i \ell}{2\pi L}\cdot\left[\left(\eta-\frac{kt}{L}\right)\cdot \nabla_{\eta} \nabla_{\eta}A\left(t,k-\ell, \eta - \frac{\ell t}{L}\right) \right]\widehat{f}\left(t,k-\ell, \eta - \frac{\ell t}{L}\right) d\eta \nonumber\\
& + \sum_{k} \sum_{\ell \ne 0} \int_{\eta} A(t,k,\eta) \overline{\widehat{f}(t,k,\eta)}\widehat{W}(\ell) \widehat{\rho}(t,\ell) \nonumber\\
&  \cdot \left[\frac{\ell}{L}\cdot \nabla_{\eta}A\left(t,k-\ell, \eta - \frac{\ell t}{L}\right) \right] \left(\eta-\frac{kt}{L}\right)\cdot \widehat{vg}\left(t,k-\ell, \eta - \frac{\ell t}{L}\right) d\eta \nonumber\\
& + \sum_{k} \sum_{\ell \ne 0} \int_{\eta} A(t,k,\eta) \overline{\widehat{g}(t,k,\eta)}\widehat{W}(\ell) \widehat{\rho}(t,\ell) \nonumber\\
&  \cdot \left[\frac{\ell}{L}\cdot \nabla_{\eta} A\left(t,k-\ell, \eta - \frac{\ell t}{L}\right) \right]\left(\eta-\frac{kt}{L}\right)\cdot \widehat{vf}\left(t,k-\ell, \eta - \frac{\ell t}{L}\right) d\eta \nonumber\\
& + \sum_{k} \sum_{\ell \ne 0} \int_{\eta} A(t,k,\eta) \overline{\widehat{f}(t,k,\eta)}\widehat{W}(\ell) \widehat{\rho}(t,\ell) \nonumber\\
& \qquad \cdot \left[\left(\eta-\frac{kt}{L}\right)\cdot \nabla_{\eta}A\left(t,k-\ell, \eta - \frac{\ell t}{L}\right) \right] \frac{\ell}{L}\cdot \widehat{vg}\left(t,k-\ell, \eta - \frac{\ell t}{L}\right) d\eta \nonumber\\
& + \sum_{k} \sum_{\ell \ne 0} \int_{\eta} A(t,k,\eta) \overline{\widehat{g}(t,k,\eta)}\widehat{W}(\ell) \widehat{\rho}(t,\ell) \nonumber\\
& \qquad \cdot \left[\left(\eta-\frac{kt}{L}\right)\cdot \nabla_{\eta} A\left(t,k-\ell, \eta - \frac{\ell t}{L}\right) \right]\frac{\ell}{L}\cdot \widehat{vf}\left(t,k-\ell, \eta - \frac{\ell t}{L}\right) d\eta \nonumber
\end{align}
\begin{align}
&+ 4\sum_{k} \sum_{\ell \ne 0} \int_{\eta} A(t,k,\eta)\overline{\widehat{g}(k,\eta)}\widehat{W}(\ell) \widehat{\rho}(t,\ell) \nonumber\\
& \qquad \qquad \cdot \frac{i\ell}{2\pi L} \cdot \nabla_{\eta}A\left(t,k-\ell, \eta - \frac{\ell t}{L}\right) \widehat{f}\left(k-\ell, \eta - \frac{\ell t}{L}\right)d\eta\nonumber\\
&+4\sum_{k} \sum_{\ell \ne 0} \int_{\eta} A(t,k,\eta)\overline{\widehat{g}(k,\eta)}\widehat{W}(\ell) \widehat{\rho}(t,\ell) \nonumber\\
& \qquad \qquad \cdot A\left(t,k-\ell, \eta - \frac{\ell t}{L}\right)\frac{\ell}{L} \cdot \widehat{vf}\left(k-\ell, \eta - \frac{\ell t}{L}\right)d\eta\equiv 0. \label{conv_id_2a}
\end{align}

Via \eqref{conv_id_1}, we can subtract terms from $NL_1$ to get a factor of $$ A(t,k,\eta) - A(t,k-\ell, \eta- \ell t/L).$$  This difference will be crucial for reducing the impact from the extra factor of $\eta - kt/L$. To handle $NL_2$, we see that \eqref{conv_id_2a} allows us to make a similar subtraction at the expense of adding a host of lower order terms.  These extra, lower order terms we will bundle with $NL_3$ since they are either missing the extra factor of $\eta - kt/L$ or have derivatives of the factor $A$ which compensate.  We summarize all of these manipulations by regrouping the non-linear terms as
\begin{equation}
NL_1(t)+  NL_2(t)+NL_3(t)= NL_1'(t) + NL_2'(t) + NL_3'(t),
\end{equation}
where
\begin{align}
&NL_1'(t)= NL_1(t) \nonumber\\
&= -4\pi^2 \Re \sum_k \sum_{\ell \ne 0} \int \overline{A^{(\sigma + 1)}\mathcal{F}\{(1-|v|^2)\Phi\}(t,k,\eta)}\widehat{\rho}\left(t, \ell\right)\widehat{W}(\ell)\frac{\ell}{L} \cdot \left(\eta - \frac{k t}{L} \right)\nonumber\\
&\qquad \cdot \left(A^{(\sigma + 1)}(t,k,\eta) - A^{(\sigma + 1)}\left(t,k-\ell, \eta - \frac{\ell t}{L}\right)\right) \widehat{\alpha \Phi}\left(t,k-\ell,\eta - \frac{\ell t}{L}\right)\; d\eta\nonumber\\
&-4\pi^2 \Re \sum_k \sum_{\ell \ne 0}\! \int \overline{A^{(\sigma + 1)}\widehat{\alpha\Phi}(t,k,\eta)}\!\left(\!A^{(\sigma + 1)}(t,k,\eta) \!-\! A^{(\sigma + 1)}\!\!\left(t,k-\ell, \eta - \frac{\ell t}{L}\right)\!\!\right)\nonumber\\
& \qquad \cdot \widehat{\rho}\left(t, \ell\right)\widehat{W}(\ell)\frac{\ell}{L} \cdot \left(\eta - \frac{k t}{L} \right)    \mathcal{F}\{(1-|v|^2)\Phi\}\left(t,k-\ell,\eta - \frac{\ell t}{L}\right)\; d\eta,
\end{align}
\begin{align}
&NL_2'(t) = \nonumber\\
& -2\pi \Im \sum_k \sum_{\ell \ne 0} \int \overline{A^{(\sigma + 1)}\mathcal{F}\{(1-|v|^2)\Phi\}(t,k,\eta)}\nonumber\\
& \qquad\qquad \cdot \left(A^{(\sigma + 1)}(t,k,\eta) - A^{(\sigma + 1)}\left(t,k-\ell, \eta - \frac{\ell t}{L}\right)\right)\nonumber\\
&\qquad \qquad \cdot  \widehat{\rho}\left(t, \ell\right)\widehat{W}(\ell)\frac{\ell}{L} \cdot\left[ \left(\eta - \frac{k t}{L} \right) \cdot \nabla_{\eta} \widehat{v\alpha \Phi}\left(t,k-\ell,\eta - \frac{\ell t}{L}\right)\right] d\eta\nonumber\\
&-2\pi \Im \sum_k \sum_{\ell \ne 0} \int \overline{A^{(\sigma + 1)}\widehat{\alpha\Phi}(t,k,\eta)}\left(A^{(\sigma + 1)}(t,k,\eta) - A^{(\sigma + 1)}\left(t,k-\ell, \eta - \frac{\ell t}{L}\right)\right)\nonumber\\
&\qquad \qquad \cdot   \widehat{\rho}\left(t, \ell\right)\widehat{W}(\ell)\frac{\ell}{L} \cdot \left[\left(\eta - \frac{k t}{L} \right) \cdot \nabla_{\eta} \mathcal{F}\{v(1-|v|^2)\Phi\}\left(t,k-\ell,\eta - \frac{\ell t}{L}\right)\right] d\eta,
\end{align}
\begin{align}
NL_3'(t) &= \sum_{i=1}^{13} NL_{3,i}'(t)\nonumber\\
& = -2 \pi \Im \sum_k \sum_{\ell \ne 0} \int A^{(\sigma + 1)}(t,k,\eta)^2\overline{\widehat{\alpha \Phi}(t,k,\eta)}\nonumber\\
& \qquad \qquad \cdot \widehat{W}(\ell)\widehat{\rho}(t,\ell)\frac{\ell}{L}\cdot \mathcal{F}\{v(1-|v|^2)\Phi\}\left(t, k-\ell, \eta - \frac{\ell t}{L}\right)d\eta \nonumber\\
& - \Re \sum_k \sum_{\ell \ne 0} \int A^{(\sigma + 1)}\triangle_{\eta}A^{(\sigma + 1)}(t,k,\eta) \overline{\widehat{\Phi}(t,k,\eta)}\nonumber\\
&\qquad \qquad \cdot\widehat{W}(\ell)\widehat{\rho}(t,\ell) \frac{\ell}{L}\cdot \left(\eta - \frac{kt}{L}\right)\widehat{\alpha \Phi}\left(t,k-\ell,\eta-\frac{\ell t}{L}\right) d\eta\nonumber\\
& - \frac{1}{2\pi} \Im \sum_k \sum_{\ell \ne 0} \int A^{(\sigma + 1)}\triangle_{\eta}A^{(\sigma + 1)}(t,k,\eta) \overline{\widehat{\Phi}(t,k,\eta)}\nonumber\\
& \qquad \qquad \cdot \widehat{W}(\ell)\widehat{\rho}(t,\ell)\frac{\ell}{L}\cdot \left[\left(\eta - \frac{kt}{L}\right)\cdot \nabla_{\eta}\widehat{v \alpha \Phi}\left(t,k-\ell,\eta-\frac{\ell t}{L}\right)\right] d\eta\nonumber\\
&- 2 \Re \sum_k \sum_{\ell \ne 0}  \int A^{(\sigma + 1)}\nabla_{\eta}A^{(\sigma + 1)}(t,k,\eta) \cdot \overline{\nabla_{\eta}\widehat{\Phi}(t,k,\eta)}\nonumber\\
& \qquad \qquad \cdot \widehat{W}(\ell)\widehat{\rho}(t,\ell)\frac{\ell}{L}\cdot \left(\eta - \frac{kt}{L}\right)\widehat{\alpha \Phi}\left(t,k-\ell,\eta-\frac{\ell t}{L}\right) d\eta\nonumber\\
& -\frac{1}{\pi} \Im \sum_k \sum_{\ell \ne 0}  \int A^{(\sigma + 1)}\nabla_{\eta}A^{(\sigma + 1)}(t,k,\eta) \cdot \overline{\nabla_{\eta}\widehat{\Phi}(t,k,\eta)}\nonumber\\
&\qquad \qquad \cdot \widehat{W}(\ell)\widehat{\rho}(t,\ell)\frac{\ell}{L}\cdot \left[\left(\eta - \frac{kt}{L}\right)\cdot \nabla_{\eta}\widehat{v \alpha \Phi}\left(t,k-\ell,\eta-\frac{\ell t}{L}\right)\right] d\eta\nonumber
\end{align}
\begin{align}
+2\pi \Im \sum_{k} \sum_{\ell \ne 0} & \int_{\eta} A^{(\sigma + 1)}(t,k,\eta) \overline{\mathcal{F}\{(1-|v|^2)\Phi\}(t,k,\eta)}\widehat{W}(\ell) \widehat{\rho}(t,\ell) \nonumber\\
& \cdot \frac{i \ell}{2\pi L}\cdot\left[\left(\eta-\frac{kt}{L}\right)\cdot \nabla_{\eta} \nabla_{\eta}A^{(\sigma + 1)}\left(t,k-\ell, \eta - \frac{\ell t}{L}\right) \right]\nonumber\\
& \qquad \cdot \widehat{\alpha \Phi}\left(t,k-\ell, \eta - \frac{\ell t}{L}\right) d\eta \nonumber\\
+2\pi \Im \sum_{k} \sum_{\ell \ne 0} & \int_{\eta} A^{(\sigma + 1)}(t,k,\eta) \overline{\widehat{\alpha \Phi}(t,k,\eta)}\widehat{W}(\ell) \widehat{\rho}(t,\ell) \nonumber\\
& \cdot \frac{i \ell}{2\pi L}\cdot\left[\left(\eta-\frac{kt}{L}\right)\cdot \nabla_{\eta} \nabla_{\eta}A^{(\sigma + 1)}\left(t,k-\ell, \eta - \frac{\ell t}{L}\right) \right]\nonumber\\
&\qquad \cdot \mathcal{F}\{(1-|v|^2)\Phi\}\left(t,k-\ell, \eta - \frac{\ell t}{L}\right) d\eta \nonumber\\
+2\pi \Im \sum_{k} \sum_{\ell \ne 0} & \int_{\eta} A^{(\sigma + 1)}(t,k,\eta) \overline{\mathcal{F}\{(1-|v|^2)\Phi\}(t,k,\eta)}\widehat{W}(\ell) \widehat{\rho}(t,\ell) \nonumber\\
& \cdot \left[\frac{\ell}{L}\cdot \nabla_{\eta}A^{(\sigma + 1)}\left(t,k-\ell, \eta - \frac{\ell t}{L}\right) \right] \nonumber\\
&\qquad \cdot \left(\eta-\frac{kt}{L}\right)\cdot \widehat{v\alpha \Phi}\left(t,k-\ell, \eta - \frac{\ell t}{L}\right) d\eta \nonumber\\
+2\pi \Im \sum_{k} \sum_{\ell \ne 0} & \int_{\eta} A^{(\sigma + 1)}(t,k,\eta) \overline{\widehat{\alpha \Phi}(t,k,\eta)}\widehat{W}(\ell) \widehat{\rho}(t,\ell) \nonumber\\
&\cdot \left[\frac{\ell}{L}\cdot \nabla_{\eta} A^{(\sigma + 1)}\left(t,k-\ell, \eta - \frac{\ell t}{L}\right) \right]\nonumber\\
& \qquad \cdot \left(\eta-\frac{kt}{L}\right)\cdot \mathcal{F}\{v(1-|v|^2)\Phi\}\left(t,k-\ell, \eta - \frac{\ell t}{L}\right) d\eta \nonumber\\
+2\pi \Im \sum_{k} \sum_{\ell \ne 0} &\int_{\eta} A^{(\sigma + 1)}(t,k,\eta) \overline{\mathcal{F}\{(1-|v|^2)\Phi\}(t,k,\eta)}\widehat{W}(\ell) \widehat{\rho}(t,\ell) \nonumber\\
& \qquad \cdot \left[\left(\eta-\frac{kt}{L}\right)\cdot \nabla_{\eta}A^{(\sigma + 1)}\left(t,k-\ell, \eta - \frac{\ell t}{L}\right) \right] \nonumber\\
& \qquad\qquad\qquad \cdot \frac{\ell}{L}\cdot \widehat{v\alpha \Phi}\left(t,k-\ell, \eta - \frac{\ell t}{L}\right) d\eta \nonumber\\
+2\pi \Im \sum_{k} \sum_{\ell \ne 0} & \int_{\eta} A^{(\sigma + 1)}(t,k,\eta) \overline{\widehat{\alpha \Phi}(t,k,\eta)}\widehat{W}(\ell) \widehat{\rho}(t,\ell) \nonumber\\
&  \qquad \cdot \left[\left(\eta-\frac{kt}{L}\right)\cdot \nabla_{\eta} A^{(\sigma + 1)}\left(t,k-\ell, \eta - \frac{\ell t}{L}\right) \right]\nonumber\\
& \qquad\qquad\qquad \cdot \frac{\ell}{L}\cdot \mathcal{F}\{v(1-|v|^2)\Phi\}\left(t,k-\ell, \eta - \frac{\ell t}{L}\right) d\eta \nonumber
\end{align}
\begin{align}
+8\pi \Im \sum_{k} \sum_{\ell \ne 0} & \!\int_{\eta} A^{(\sigma + 1)}(t,k,\eta)\overline{\widehat{\alpha \Phi}(k,\eta)}\widehat{W}(\ell) \widehat{\rho}(t,\ell) \nonumber\\
& \cdot \frac{i \ell}{2 \pi L} \cdot \nabla_{\eta}A^{(\sigma + 1)}\left(t,k-\ell, \eta - \frac{\ell t}{L}\right) \mathcal{F}\{(1-|v|^2)\Phi\}\left(t,k-\ell, \eta - \frac{\ell t}{L}\right)d\eta\nonumber\\
+8\pi \Im \sum_{k} \sum_{\ell \ne 0} & \int_{\eta} A^{(\sigma + 1)}(t,k,\eta)\overline{\widehat{\alpha \Phi}(k,\eta)}\widehat{W}(\ell) \widehat{\rho}(t,\ell) \nonumber\\
& \cdot A^{(\sigma + 1)}\left(t,k-\ell, \eta - \frac{\ell t}{L}\right)\frac{\ell}{L} \cdot \mathcal{F}\{v(1-|v|^2)\Phi\}\left(t,k-\ell, \eta - \frac{\ell t}{L}\right)d\eta.
\end{align}

$NL_1'$ and $NL_2'$ will be further decomposed via the usual paraproduct decomposition.  The \emph{transport terms} will be those with the low frequency projection, `$<N/8$', on $\widehat{\rho}(t,\ell)$ while the high frequency projection will be on $\widehat{\alpha \Phi}(t, k-\ell, \eta - t\ell/L)$ (or analogous terms depending on the product) ; the \emph{reaction terms} will swap these projections.  The \emph{remainder terms} will contain the remaining terms from the paraproduct decomposition. $NL_3$ can be dealt with directly.

\subsection{Transport Terms}

We will denote the $N$-th term in the transport portions of the non-linear terms as $(TNL_1')_N$ and $(TNL_2')_N$, respectively.  For both, we have the following localizations for the transport terms:
\begin{align}
\frac{N}{2} \le \left| k-\ell, \eta - \frac{\ell t}{L} \right| \le & \frac{3N}{2},\\
\left|\ell, \frac{\ell t}{L}\right| \le & \frac{3N}{32},\\
\frac{13}{16} \le \frac{\left|k, \eta \right|}{ \left| k-\ell, \eta - \frac{\ell t}{L} \right|} \le & \frac{19}{16},
\end{align}
(recall that $N$ runs over powers of two).  Following estimates in \cite{BMM13}[\textsection 5.3.3], we have
\begin{align}
\Big| & A^{(\sigma + 1)}(t,k,\eta)  - A^{(\sigma + 1)}\left(t,k-\ell, \eta - \frac{\ell t}{L}\right)\Big| \nonumber\\
& \le C \frac{\langle \ell, \ell t/L\rangle}{\langle k, \eta \rangle^{1-\overline{\nu}} + \langle k-\ell, \eta - \ell t/L \rangle^{1-\overline{\nu}}}e^{c\lambda(t)\langle \ell, \ell t / L \rangle^{\overline{\nu}}}A^{(\sigma + 1)}\left(t,k-\ell, \eta - \frac{\ell t}{L}\right)
\end{align}
where $C$ depends on the various parameters appearing in $A$ and $c = c(\overline{\nu}) \in (0,1)$ (but neither depend on $N$).  Using this together with the obvious equality $$\left|\eta - \frac{kt}{L}\right| = \left| \eta - \frac{\ell t}{L} - \frac{(k-\ell)t}{L} \right|, $$ we find that
\begin{align}
&\left| (TNL_1')_N(t)\right| \le \nonumber\\
& C\langle t \rangle^2  \sum_k \sum_{\ell \ne 0} \int \left|A^{(\sigma + 1)}\mathcal{F}\{(1-|v|^2)\Phi\}(t,k,\eta)\right| e^{c\lambda(t) \langle \ell, \ell t/L\rangle^{\overline{\nu}}}\left|\widehat{\rho}(t,\ell)_{<N/8}\right|\nonumber\\
&\cdot \langle k,\eta \rangle^{\overline{\nu}/2}\left\langle k-\ell,\eta-\frac{\ell t}{L} \right\rangle^{\overline{\nu}/2}\left|A^{(\sigma + 1)}\widehat{\alpha\Phi}\left(t,k-\ell,\eta-\frac{\ell t}{L}\right)_N\right| d\eta \nonumber\\
&+ C\langle t \rangle^2 \sum_k  \sum_{\ell \ne 0} \int \left|A^{(\sigma + 1)}\widehat{\alpha\Phi}(t,k,\eta)\right| e^{c\lambda(t) \langle \ell, \ell t/L\rangle^{\overline{\nu}}}\left|\widehat{\rho}(t,\ell)_{<N/8}\right|\nonumber\\
& \cdot \langle k,\eta \rangle^{\overline{\nu}/2}\left\langle k-\ell,\eta-\frac{\ell t}{L} \right\rangle^{\overline{\nu}/2}\left|A^{(\sigma + 1)}\mathcal{F}\{(1-|v|^2)\Phi\}\left(t,k-\ell,\eta-\frac{\ell t}{L}\right)_N\right| d\eta.
\end{align}
Using Lemma \ref{young_var_lem}[(i)] and dropping the projection on $\widehat{\rho}$, we find that
\begin{align}
\left| (TNL_1')_N(t)\right| \le  C \langle t \rangle^2 & \left\| A^{(c\lambda, 5/2)}\widehat{\rho}(t) \right\|_2\nonumber\\
 \cdot \Bigg\{&\left\| A^{(\sigma + 1 +\overline{\nu}/2)}\mathcal{F}\{(1-|v|^2)\Phi\}(t) \right\|_2\left\| A^{(\sigma + 1 +\overline{\nu}/2)}\widehat{\alpha\Phi}_N(t) \right\|_2\nonumber\\
& + \left\| A^{(\sigma + 1 +\overline{\nu}/2)}\widehat{\alpha\Phi}(t) \right\|_2\left\| A^{(\sigma + 1 +\overline{\nu}/2)}\mathcal{F}\{(1-|v|^2)\Phi\}_N(t) \right\|_2\Bigg\},
\end{align}
where $C$ depends on the various parameters in $A$.  Since $c \in (0,1)$ and $\sigma > 5/2$, we get the following estimate for $TNL_1'$ by summing over the $N$:
\begin{align}
\left| TNL_1'(t)\right| \le C  e^{-\frac{1}{2}(1-c)\alpha_0\langle t \rangle^{\overline{\nu}}} & \left\| A\widehat{\rho}(t) \right\|_2\nonumber\\
& \cdot \left\| A^{(\sigma + 1 +\overline{\nu}/2)}\mathcal{F}\{(1-|v|^2)\Phi\}(t) \right\|_2\left\| A^{(\sigma + 1 +\overline{\nu}/2)}\widehat{\alpha\Phi}(t) \right\|_2,
\end{align}
where we have absorbed the $\langle t \rangle^2$ into the exponential factor (which explains the factor of $1/2$ that enters).  As usual, we can drop the $(1-|v|^2)$ term at the expense of a combinatorial factor.  Thus, we have a final estimate:
\begin{align}
\left| TNL_1'(t)\right| \le C e^{-\frac{1}{2}(1-c)\alpha_0\langle t \rangle^{\overline{\nu}}} & \left\| A\widehat{\rho}(t) \right\|_2\nonumber\\
& \cdot \left\| A^{(\sigma + 1 +\overline{\nu}/2)}\widehat{\Phi}(t) \right\|_2\left\| A^{(\sigma + 1 +\overline{\nu}/2)}\widehat{\alpha\Phi}(t) \right\|_2,
\end{align}
where $C$ depends on the various parameters in $A$.  Using again $$ ab \le \frac{a^2+b^2}{2}, $$ we clearly see that
\begin{align}
\left| TNL_1'(t)\right| \le C & e^{-\frac{1}{2}(1-c)\alpha_0\langle t \rangle^{\overline{\nu}}} \left\| A\widehat{\rho}(t) \right\|_2\nonumber\\
& \cdot \left(\left\| A^{(\sigma + 1 +\overline{\nu}/2)}\widehat{\Phi}(t) \right\|_2^2+ \left\| A^{(\sigma + 1 +\overline{\nu}/2)}\widehat{\alpha\Phi}(t) \right\|_2^2\right).
\end{align}

Looking at the form of $NL_2'$, we see that the only essential change is that there is a gradient in $\eta$ and an extra $v$ on certain terms.  Since this amounts (up to a constant) to some second derivative object in $\eta$, we can apply Lemma \ref{eta_deriv_lem} and simply ignore this difference (again, up to a combinatorial factor).  Hence, we get the same sort of bound for this term as well:
\begin{align}
\left| TNL_2'(t)\right| \le C  & e^{-\frac{1}{2}(1-c)\alpha_0\langle t \rangle^{\overline{\nu}}}  \left\| A\widehat{\rho}(t) \right\|_2\nonumber\\
& \cdot \left(\left\| A^{(\sigma + 1 +\overline{\nu}/2)}\widehat{\Phi}(t) \right\|_2^2+ \left\| A^{(\sigma + 1 +\overline{\nu}/2)}\widehat{\alpha\Phi}(t) \right\|_2^2\right).
\end{align}

\subsection{Reaction Terms}

For these terms, we have the following localizations:
\begin{align}
\frac{N}{2} \le \left| \ell, \frac{\ell t}{L} \right| \le & \frac{3N}{2},\\
\left|k-\ell, \eta - \frac{\ell t}{L}\right| \le & \frac{3N}{32},\\
\frac{13}{16} \le \frac{\left|k, \eta \right|}{ \left| \ell, \frac{\ell t}{L} \right|} \le & \frac{19}{16},
\end{align}
along with the easy estimate $|\eta - kt/L| \le \langle t \rangle \left|k-\ell, \eta - \ell t/L\right|$.  From these together with Lemma \ref{angle_brak_lem}[iii], we can conclude that there is some $c = c(\overline{\nu}) \in (0,1)$ so that
\begin{equation}
A^{(\sigma + 1)}(t, k, \eta) \le C A^{(\sigma+1)}(t,\ell, \ell t/L)e^{c\lambda(t)\langle k-\ell, \eta -  \ell t /L\rangle^{\overline{\nu}}},
\end{equation}
where $C$ depends on the various parameters appearing in $A$ (but not on $N$).  Furthermore, by our assumption on $\widehat{W}$ (namely that $\gamma \ge 1$), we have that
\begin{equation}
A^{(\sigma+1)}\left(t,\ell, \frac{\ell t}{L}\right) \left|\widehat{W}(\ell)\frac{\ell}{L}\right| \le \frac{C_W}{L} \frac{\langle \ell, \ell t /L\rangle}{|\ell|}A^{(\sigma)}\left(t,\ell, \frac{\ell t}{L}\right) \le C\langle t \rangle A\left(t,\ell, \frac{\ell t}{L}\right).
\end{equation}
Putting this altogether gives us the following useful estimate for the reaction terms:
\begin{align}
A^{(\sigma + 1)}(t, k, \eta) \left|\widehat{W}(\ell)\frac{\ell}{L}\right|\left|\eta - \frac{kt}{L}\right| & \le C \langle t \rangle^2 A\left(t,\ell, \frac{\ell t}{L}\right)e^{c\lambda(t)\langle k-\ell, \eta -  \ell t /L\rangle^{\overline{\nu}}}\left|k-\ell, \eta - \ell t/L\right| \nonumber\\
&\le C \langle t \rangle^2 A\left(t,\ell, \frac{\ell t}{L}\right)e^{\lambda(t)\langle k-\ell, \eta -  \ell t /L\rangle^{\overline{\nu}}},
\end{align}
where in the final inequality we have absorbed the extra factor of $\left|k-\ell, \eta - \ell t/L\right|$ into the exponential at the expense of replacing $c \in (0,1)$ with a one (and a constant independent of $N$).

For these terms, the difference $$A^{(\sigma + 1)}(t,k,\eta)  - A^{(\sigma + 1)}\left(t,k-\ell, \eta - \frac{\ell t}{L}\right)  $$ is not useful as it was for the transport terms.  So, we simply break up the difference into two pieces which we label as $(RNL_{1,1}')_N$ and $(RNL_{1,2}')_N$.  For the first term, we have
\begin{align}
\left|(RNL_{1,1}'(t))_N\right| \le & C \sum_k \sum_{\ell \ne 0} \int \left|A^{(\sigma + 1)}\mathcal{F}\{(1-|v|^2)\Phi\}(t,k,\eta)\right|A^{(\sigma + 1)}(t,k,\eta) \left|\widehat{W}(\ell)\frac{\ell}{L}\right| \nonumber\\
& \qquad \qquad  \cdot \left| \widehat{\rho}(t,\ell)_N\right|\left|\eta - \frac{kt}{L}\right|\left|\widehat{\alpha \Phi}\left(t,k-\ell,\eta - \frac{\ell t}{L}\right)_{<N/8} \right| d\eta\nonumber\\
&+ C \sum_k \sum_{\ell \ne 0} \int \left|A^{(\sigma + 1)}\widehat{\alpha\Phi}(t,k,\eta)\right|A^{(\sigma + 1)}(t,k,\eta) \left|\widehat{W}(\ell)\frac{\ell}{L}\right| \nonumber\\
&  \qquad \cdot \left| \widehat{\rho}(t,\ell)_N\right|\left|\eta - \frac{kt}{L}\right|\left|\mathcal{F}\{(1-|v|^2) \Phi\}\left(t,k-\ell,\eta - \frac{\ell t}{L}\right)_{<N/8} \right| d\eta\nonumber\\
&\le C \langle t \rangle^2 \sum_k \sum_{\ell \ne 0} \int \left|A^{(\sigma + 1)}\mathcal{F}\{(1-|v|^2)\Phi\}(t,k,\eta)\right| \nonumber\\
& \qquad \qquad \cdot\left|A\widehat{\rho}(t,\ell)_N\right|\left|A^{(0)}\widehat{\alpha \Phi}\left(t,k-\ell,\eta - \frac{\ell t}{L}\right)_{<N/8} \right| d\eta\nonumber\\
& + C \langle t \rangle^2 \sum_k \sum_{\ell \ne 0} \int \left|A^{(\sigma + 1)}\widehat{\alpha\Phi}(t,k,\eta)\right| \nonumber\\
&  \qquad\cdot\left|A\widehat{\rho}(t,\ell)_N\right|\left|A^{(0)}\mathcal{F}\{(1-|v|^2) \Phi\}\left(t,k-\ell,\eta - \frac{\ell t}{L}\right)_{<N/8} \right| d\eta.
\end{align}
We now apply Lemma \ref{young_var_lem}[(ii)] along with the fact that $\sigma - \beta > 5/2$ to get
\begin{align}
\left|\left(RNL_{1,1}'(t)\right)_N\right| \le C \langle t \rangle^2  & \left\| A\widehat{\rho}(t)_N \right\|_2\Bigg\{  \left\|A^{(\sigma + 1)}\mathcal{F}\{(1-|v|^2)\Phi\}(t) \right\|_2 \left\|A^{(\sigma - \beta)}\widehat{\alpha \Phi}(t)_{<N/8} \right\|_2\nonumber\\
 & + \left\|A^{(\sigma + 1)}\widehat{\alpha\Phi}(t) \right\|_2 \left\|A^{(\sigma - \beta)}\mathcal{F}\{(1-|v|^2) \Phi\}(t)_{<N/8} \right\|_2\Bigg\}.
\end{align}
Dropping the `$<N/8$' projections and summing over $N$  (along with our usual device of ignoring factors of $v$) gives
\begin{align}
\left|\left(RNL_{1,1}'(t)\right)\right| \le C \langle t \rangle^2 \left\| A\widehat{\rho}(t) \right\|_2\Bigg\{ & \left\|A^{(\sigma + 1)}\widehat{\Phi}(t) \right\|_2 \left\|A^{(\sigma - \beta)}\widehat{\alpha \Phi}(t) \right\|_2\nonumber\\
 & + \left\|A^{(\sigma + 1)}\widehat{\alpha\Phi}(t) \right\|_2 \left\|A^{(\sigma - \beta)}\widehat{\Phi}(t) \right\|_2\Bigg\},
\end{align}
where $C$ depends only on the parameters appearing in $A$.  Writing $\langle t \rangle^2 = \langle t \rangle^3 \langle t \rangle^{-1}$ and using the inequality controlling a product by the sum of squares gives
\begin{align}
\left|(RNL_{1,1}'(t))\right| \le C & \left\|A^{(\sigma - \beta)}\widehat{\Phi}(t) \right\|_2\nonumber\\
&\cdot\left\{\langle t \rangle^{-2} \left\|A^{(\sigma + 1)}\widehat{\Phi}(t) \right\|_2^2 + \langle t \rangle^6 \left\| A\widehat{\rho}(t) \right\|_2^2\right\}\nonumber\\
+ C & \left\|A^{(\sigma - \beta)}\widehat{\alpha \Phi}(t) \right\|_2\nonumber\\
&\cdot\left\{\langle t \rangle^{-2} \left\|A^{(\sigma + 1)}\widehat{\alpha\Phi}(t) \right\|_2^2 + \langle t \rangle^6 \left\| A\widehat{\rho}(t) \right\|_2^2\right\}.
\end{align}

For the other terms, we note that
\begin{align}
\big|(RNL_{1,2}'(t))_N & \big| \le \nonumber\\
& C \sum_k \sum_{\ell \ne 0} \int \left|A^{(\sigma + 1)}\mathcal{F}\{(1-|v|^2)\Phi\}(t,k,\eta)\right|\left| \widehat{\rho}(t,\ell)_N\right|\nonumber\\
&\qquad \qquad \left| \eta - \frac{kt}{L} \right|\left| A^{(\sigma + 1)}\widehat{\alpha\Phi}\left(t,k-\ell,\eta - \frac{\ell t}{L}\right)_{<N/8}\right|d\eta\nonumber\\
& +  C \sum_k \sum_{\ell \ne 0} \int \left|A^{(\sigma + 1)}\widehat{\alpha \Phi}(t,k,\eta)\right|\left| \widehat{\rho}(t,\ell)_N\right|\nonumber\\
&\qquad \qquad \left| \eta - \frac{kt}{L} \right|\left| A^{(\sigma + 1)}\mathcal{F}\{(1-|v|^2)\Phi\}\left(t,k-\ell,\eta - \frac{\ell t}{L}\right)_{<N/8}\right|d\eta.
\end{align}
By the frequency localizations for this regime, we have $$ \left| \eta - \frac{kt}{L} \right| \le \langle t \rangle \left\langle k-\ell, \eta - \frac{\ell t}{L} \right\rangle \le C\langle t \rangle \left\langle \ell, \frac{\ell t}{L} \right\rangle.   $$  Inserting this estimate, using Lemma \ref{young_var_lem}[(i)], and dropping all projections except for the one on $\widehat{\rho}$ gives
\begin{align}
\left|(RNL_{1,2}'(t))_N\right| \le  C \langle t \rangle \left\|A^{(\sigma + 1)}\mathcal{F}\{(1-|v|^2)\Phi\}(t)\right\|_2\left\| A^{(0,7/2)} \widehat{\rho}(t)_N\right\|_2 \left\| A^{(\sigma + 1)}\widehat{\alpha\Phi}(t)\right\|_2.
\end{align}
To make progress, we first note that we can replace $A^{(0,7/2)}$ by $e^{-\alpha_0\langle t \rangle^{\overline{\nu}}}A^{(\lambda,7/2)}$ by at most increasing the constant $C$.  Also note that in this regime $$ \frac{\langle \ell, \ell t / L\rangle}{N} \approx 1. $$  Finally, noting that $\sigma > 9/2$, we arrive at
\begin{align}
\left|(RNL_{1,2}'(t))_N\right| \le  \frac{C}{N} e^{-\frac{1}{2}\alpha_0\langle t \rangle^{\overline{\nu}}} \left\|A^{(\sigma + 1)}\widehat{\Phi}(t)\right\|_2\left\| A \widehat{\rho}(t)\right\|_2 \left\| A^{(\sigma + 1)}\widehat{\alpha\Phi}(t)\right\|_2.
\end{align}
The factor of $1/2$ that appears in the exponential comes from absorbing the factor of $\langle t \rangle$.  Recall that $N$ runs over the powers of $2$, and so the terms above are summable.  Again employing the standard estimate for a product in terms of squares, we see
\begin{align}
\left|RNL_{1,2}'(t)\right| \le  C e^{-\frac{1}{2}\alpha_0\langle t \rangle^{\overline{\nu}}}&\left\| A \widehat{\rho}(t)\right\|_2 \left\|A^{(\sigma + 1)}\widehat{\Phi}(t)\right\|_2^2 \nonumber\\
 & + C e^{-\frac{1}{2}\alpha_0\langle t \rangle^{\overline{\nu}}}\left\| A \widehat{\rho}(t)\right\|_2\left\| A^{(\sigma + 1)}\widehat{\alpha\Phi}(t)\right\|_2^2,
\end{align}
where once again $C$ depends on the various parameters appearing in $A$.

Following our same logic as above for $NL_2'$ yields two terms which are exactly the same as above (up to a change in the constant $C$):
\begin{align}
\left|RNL_{2,1}'(t)\right| \le C & \left\|A^{(\sigma - \beta)}\widehat{\Phi}(t) \right\|_2\nonumber\\
&\cdot\left\{\langle t \rangle^{-2} \left\|A^{(\sigma + 1)}\widehat{\Phi}(t) \right\|_2^2 + \langle t \rangle^6 \left\| A\widehat{\rho}(t) \right\|_2^2\right\}\nonumber\\
+ C & \left\|A^{(\sigma - \beta)}\widehat{\alpha \Phi}(t) \right\|_2\nonumber\\
&\cdot\left\{\langle t \rangle^{-2} \left\|A^{(\sigma + 1)}\widehat{\alpha\Phi}(t) \right\|_2^2 + \langle t \rangle^6 \left\| A\widehat{\rho}(t) \right\|_2^2\right\},
\end{align}
\begin{align}
\left|RNL_{2,2}'(t)\right| \le  &C e^{-\frac{1}{2}\alpha_0\langle t \rangle^{\overline{\nu}}}\left\| A \widehat{\rho}(t)\right\|_2 \left\|A^{(\sigma + 1)}\widehat{\Phi}(t)\right\|_2^2 \nonumber\\
 & + C e^{-\frac{1}{2}\alpha_0\langle t \rangle^{\overline{\nu}}}\left\| A \widehat{\rho}(t)\right\|_2\left\| A^{(\sigma + 1)}\widehat{\alpha\Phi}(t)\right\|_2^2.
\end{align}

\subsection{Remainder Terms}

For the remainder terms of the paraproduct decomposition, we have the following localizations
\begin{align}
\frac{N'}{2} \le \left| \ell, \frac{\ell t}{L} \right| \le & \frac{3N'}{2},\\
\frac{N}{2} \le \left|k-\ell, \eta - \frac{\ell t}{L}\right| \le & \frac{3N}{2},\\
\frac{1}{24} \le \frac{\left|k - \ell, \eta -\frac{\ell t}{L} \right|}{ \left| \ell, \frac{\ell t}{L} \right|} \le & 24,
\end{align}
where $N/8 \le N' \le 8N$ (recall that both $N$ and $N'$ are powers of 2).  Following the arguments in \cite{BMM13}[\S 5.3.5], we can find $c=c(\overline{\nu}) \in (0,1)$ so that
\begin{equation}
A^{(\sigma + 1)}(t,k,\eta) \le Ce^{c\lambda(t)\langle \ell, \ell t/L \rangle^{\overline{\nu}}}e^{c\lambda(t)\langle k - \ell, \eta - \ell t/L \rangle^{\overline{\nu}}},
\end{equation}
where the constant $C$ only depends on the parameters appearing in $A$.  As before, $$|\eta - kt/L| \le C\langle t \rangle \left\langle k-\ell, \eta - \ell t/L\right\rangle \le C \langle t \rangle \langle \ell, \ell t /L\rangle.$$

Exactly as for the Reaction Terms, we split the Remainder Terms into two pieces: $\mathcal{R}NL_{1,1}'$ and $\mathcal{R}NL_{1,2}'$.  For the first of these, we find
\begin{align}
&\left|\mathcal{R}NL_{1,1}'(t) \right| \nonumber\\
&\le C \sum_N \sum_{N' \approx N} \sum_k \sum_{\ell \ne 0} \int \left|A^{(\sigma + 1)}\mathcal{F}\{(1-|v|^2)\Phi\}(t,k,\eta)\right|e^{c\lambda(t)\langle \ell, \ell t/L \rangle^{\overline{\nu}}} \left|\widehat{W}(\ell)\frac{\ell}{L}\right|\left| \widehat{\rho}(t,\ell)_{N'}\right| \nonumber\\
& \qquad \qquad  \cdot e^{c\lambda(t)\langle k - \ell, \eta - \ell t/L \rangle^{\overline{\nu}}}\left|\eta - \frac{kt}{L}\right|\left|\widehat{\alpha \Phi}\left(t,k-\ell,\eta - \frac{\ell t}{L}\right)_{N} \right| d\eta\nonumber\\
&+C \sum_N \sum_{N' \approx N} \sum_k \sum_{\ell \ne 0} \int \left|A^{(\sigma + 1)}\widehat{\alpha \Phi}(t,k,\eta)\right|e^{c\lambda(t)\langle \ell, \ell t/L \rangle^{\overline{\nu}}} \left|\widehat{W}(\ell)\frac{\ell}{L}\right|\left| \widehat{\rho}(t,\ell)_{N'}\right| \nonumber\\
& \qquad \qquad  \cdot e^{c\lambda(t)\langle k - \ell, \eta - \ell t/L \rangle^{\overline{\nu}}}\left|\eta - \frac{kt}{L}\right|\left|\mathcal{F}\{(1-|v|^2) \Phi\}\left(t,k-\ell,\eta - \frac{\ell t}{L}\right)_{N} \right| d\eta\\
& \le C\langle t \rangle e^{-\frac{1}{2}(1-c)\langle t \rangle^{\overline{\nu}}} \sum_N \sum_{N' \approx N} \frac{1}{N'}\left\|A^{(\sigma + 1)}\mathcal{F}\{(1-|v|^2)\Phi\}(t)\right\|_2 \!\!  \left\| A \widehat{\rho}(t)_{N'}\right\|_2 \!  \left\|A^{(c\lambda, 7/2)}\widehat{\alpha \Phi}(t)_{N} \right\|_2 \nonumber\\
&+C\langle t \rangle e^{-\frac{1}{2}(1-c)\langle t \rangle^{\overline{\nu}}} \sum_N \sum_{N' \approx N} \frac{1}{N'}\left\|A^{(\sigma + 1)}\widehat{\alpha \Phi}(t)\right\|_2 \!\!  \left\| A\widehat{\rho}(t)_{N'}\right\|_2 \!  \left\|A^{(c\lambda, 7/2)}\mathcal{F}\{(1-|v|^2) \Phi\}(t)_{N} \right\|_2\! ,
\end{align}
where the extra factors of $(N')^{-1}$ are justified by the fact that we have included an extra $\langle \ell, \ell t / L \rangle^{\sigma}$ in the norm for $\widehat{\rho}$ (and $\sigma > 1$).  Since $\sigma > 7/2$, $c \in (0,1)$, and $N' \approx N$, we can drop the projections and end with an estimate of the form
\begin{equation}
\left|\mathcal{R}NL_{1,1}'(t) \right| \le C\langle t \rangle^{-1} e^{-\frac{1}{4}(1-c)\langle t \rangle^{\overline{\nu}}} \left\|A^{(\sigma + 1)}\widehat{\Phi}(t)\right\|_2\left\| A \widehat{\rho}(t)\right\|_2 \left\|A^{(\sigma + 1) }\widehat{\alpha \Phi}(t) \right\|_2.
\end{equation}
This yields the following estimate
\begin{align}
\left|\mathcal{R}NL_{1,1}'(t) \right| \le & C\langle t \rangle^{-1} e^{-\frac{1}{4}(1-c)\langle t \rangle^{\overline{\nu}}} \left\| A \widehat{\rho}(t)\right\|_2 \left\|A^{(\sigma + 1) }\widehat{\alpha \Phi}(t) \right\|_2^2\nonumber\\
&+C\langle t \rangle^{-1} e^{-\frac{1}{4}(1-c)\langle t \rangle^{\overline{\nu}}} \left\| A \widehat{\rho}(t)\right\|_2 \left\|A^{(\sigma + 1) }\widehat{\Phi}(t) \right\|_2^2.
\end{align}

For the second portion, we use $\langle k-\ell, \eta - \ell t/L \rangle \approx \langle \ell, \ell t /L \rangle$ to deduce
\begin{align}
&\left|\mathcal{R}NL_{1,2}'(t) \right| \nonumber\\
&\le C \langle t \rangle \sum_N \sum_{N' \approx N} \sum_k \sum_{\ell \ne 0} \int \left|A^{(\sigma + 1)}\mathcal{F}\{(1-|v|^2)\Phi\}(t,k,\eta)\right|e^{\frac{1}{2}\lambda(t)\langle \ell, \ell t/L \rangle^{\overline{\nu}}} \left| \widehat{\rho}(t,\ell)_{N'}\right| \nonumber\\
& \qquad \qquad  \cdot \left|A^{(\sigma + 1)}\widehat{\alpha \Phi}\left(t,k-\ell,\eta - \frac{\ell t}{L}\right)_{N} \right| d\eta\nonumber\\
&+C \langle t \rangle \sum_N \sum_{N' \approx N} \sum_k \sum_{\ell \ne 0} \int \left|A^{(\sigma + 1)}\widehat{\alpha \Phi}(t,k,\eta)\right|e^{\frac{1}{2}\lambda(t)\langle \ell, \ell t/L \rangle^{\overline{\nu}}} \left| \widehat{\rho}(t,\ell)_{N'}\right| \nonumber\\
& \qquad \qquad  \cdot \left|A^{(\sigma +1)}\mathcal{F}\{(1-|v|^2) \Phi\}\left(t,k-\ell,\eta - \frac{\ell t}{L}\right)_{N} \right| d\eta.
\end{align}
Appealing to the same logic as above (especially in regards to the extra factor of $(N')^{-1}$), we find
\begin{align}
&\left|\mathcal{R}NL_{1,2}'(t) \right| \nonumber\\
&\le C \langle t \rangle e^{-\frac{1}{2}\alpha_0 \langle t \rangle^{\overline{\nu}}} \sum_N \sum_{N' \approx N} \frac{1}{N'} \left\|A^{(\sigma + 1)}\mathcal{F}\{(1-|v|^2)\Phi\}(t)\right\|_2 \left\| A\widehat{\rho}(t)_{N'}\right\|_2 \left\|A^{(\sigma + 1)}\widehat{\alpha \Phi}(t)_{N} \right\|_2\nonumber\\
&+C \langle t \rangle e^{-\frac{1}{2}\alpha_0 \langle t \rangle^{\overline{\nu}}} \sum_N \sum_{N' \approx N} \frac{1}{N'} \left\|A^{(\sigma + 1)}\widehat{\alpha\Phi}(t)\right\|_2 \left\| A\widehat{\rho}(t)_{N'}\right\|_2 \left\|A^{(\sigma + 1)}\mathcal{F}\{(1-|v|^2) \Phi\}(t)_{N} \right\|_2.
\end{align}
We finish the estimate with the same tricks as above (reducing the $1/2$ in the exponential to a $1/4$ so that we can trade $\langle t \rangle$ for $\langle t \rangle^{-1}$) to find
\begin{align}
\left|\mathcal{R}NL_{1,2}'(t) \right| \le & C \langle t \rangle^{-1} e^{-\frac{1}{4}\alpha_0 \langle t \rangle^{\overline{\nu}}}\left\| A\widehat{\rho}(t)\right\|_2\left\|A^{(\sigma + 1)}\widehat{\Phi}(t)\right\|_2^2 \nonumber\\
&+C \langle t \rangle^{-1} e^{-\frac{1}{4}\alpha_0 \langle t \rangle^{\overline{\nu}}}\left\| A\widehat{\rho}(t)\right\|_2\left\|A^{(\sigma + 1)}\widehat{\alpha \Phi}(t)\right\|_2^2
\end{align}

As above, we have exactly analogous estimates for $\mathcal{R}NL_{2,1}'$ and $\mathcal{R}NL_{2,2}'$:
\begin{align}
\left|\mathcal{R}NL_{2,1}'(t) \right| \le & C\langle t \rangle^{-1} e^{-\frac{1}{4}(1-c)\langle t \rangle^{\overline{\nu}}} \left\| A \widehat{\rho}(t)\right\|_2 \left\|A^{(\sigma + 1) }\widehat{\alpha \Phi}(t) \right\|_2^2\nonumber\\
&+C\langle t \rangle^{-1} e^{-\frac{1}{4}(1-c)\langle t \rangle^{\overline{\nu}}} \left\| A \widehat{\rho}(t)\right\|_2 \left\|A^{(\sigma + 1) }\widehat{\Phi}(t) \right\|_2^2,
\end{align}
\begin{align}
\left|\mathcal{R}NL_{2,2}'(t) \right| \le & C \langle t \rangle^{-1} e^{-\frac{1}{4}\alpha_0 \langle t \rangle^{\overline{\nu}}}\left\| A\widehat{\rho}(t)\right\|_2\left\|A^{(\sigma + 1)}\widehat{\Phi}(t)\right\|_2^2 \nonumber\\
&+C \langle t \rangle^{-1} e^{-\frac{1}{4}\alpha_0 \langle t \rangle^{\overline{\nu}}}\left\| A\widehat{\rho}(t)\right\|_2\left\|A^{(\sigma + 1)}\widehat{\alpha \Phi}(t)\right\|_2^2.
\end{align}

\subsection{$NL_3'$}

Having dealt with $NL_1'$ and $NL_2'$, we are left with the terms in $NL_3'$ where either the extra $\eta - kt/L$ simply does not occur or there are derivatives of $A$ which will give compensating factors (via Lemma \ref{est_on_DA}).  Fortunately, many of these terms come in groups which have similar estimates.  We will not belabor the details too much since most of these estimates are handled with tools developed above (and already used extensively throughout previous estimates).

For $NL_{3,1}'(t)$, we need only use Lemma \ref{product_lem} to obtain
\begin{align}
\left|NL_{3,1}'(t)\right| \le & C\left\|A^{(\sigma + 1)}\widehat{\alpha \Phi}(t)\right\|_2\nonumber\\
& \cdot \left(\left\|A^{(c\lambda,0)}\widehat{\Phi}(t) \right\|_2\left\|A^{(\sigma + 1)}\langle k \rangle^{-1}\widehat{\rho}(t) \right\|_2 + \left\|A^{(\sigma + 1)}\widehat{\Phi}(t) \right\|_2\left\|A^{(c\lambda,0)}\widehat{\rho}(t) \right\|_2 \right).
\end{align}
Note that we have retained a factor of $\langle k \rangle^{-1}$ coming from $\widehat{W}(k)k/L$ in one of the $\widehat{\rho}$ terms but dropped it in the other.  We have also already dropped the factor of $v(1-|v|^2)$ at the expense of a combinatorial factor.  For the first term, we only increase the estimate by replacing $A^{(c\lambda,0)}(t,k,\eta)$ by $A^{(\lambda,\sigma - \beta)}(t,k,\eta)$.  We can also reduce $\sigma + 1$ to $\sigma$ on the $\widehat{\rho}$ term at the expense of adding $\langle t \rangle$. For the second term, we use fact that $c \in (0,1)$ to introduce a decaying exponential factor.  The end result is
\begin{align}
\left|NL_{3,1}'(t)\right| \le & C\langle t \rangle\left\|A^{(\sigma + 1)}\widehat{\alpha \Phi}(t)\right\|_2\left\|A^{(\sigma -\beta)}\widehat{\Phi}(t) \right\|_2\left\|A\widehat{\rho}(t) \right\|_2 \nonumber\\
&+ C e^{-(1-c)\alpha_0\langle t \rangle^{\overline{\nu}}}\left\|A^{(\sigma + 1)}\widehat{\alpha \Phi}(t)\right\|_2\left\|A^{(\sigma + 1)}\widehat{\Phi}(t) \right\|_2\left\|A\widehat{\rho}(t) \right\|_2,
\end{align}
where, as always, $C$ depends only on the various parameters appearing in $A$.  As above, we write $\langle t \rangle = \langle t \rangle^2\langle t \rangle^{-1}$ and break up the first term as we did multiple times above.  This yields
\begin{align}
\left|NL_{3,1}'(t)\right| \le & C\langle t \rangle^4\left\|A^{(\sigma -\beta)}\widehat{\Phi}(t) \right\|_2\left\|A\widehat{\rho}(t) \right\|_2^2 \nonumber\\
&+ C \langle t \rangle^{-2}\left\|A^{(\sigma + 1)}\widehat{\alpha \Phi}(t)\right\|_2^2\left\|A^{(\sigma -\beta)}\widehat{\Phi}(t) \right\|_2\nonumber\\
&+ C e^{-(1-c)\alpha_0\langle t \rangle^{\overline{\nu}}}\left\|A^{(\sigma + 1)}\widehat{\alpha \Phi}(t)\right\|_2\left\|A^{(\sigma + 1)}\widehat{\Phi}(t) \right\|_2\left\|A\widehat{\rho}(t) \right\|_2.
\end{align}

For $NL_{3,2}'$ and $NL_{3,3}'$, we note that by Lemma \ref{est_on_DA}
\begin{equation}
\left| A^{(\sigma + 1)}\triangle_{\eta}A^{(\sigma + 1)}(t,k,\eta)\right| \le C\left(A^{(\sigma+\overline{\nu})}(t,k,\eta)\right)^2,
\end{equation}
and by simple estimates
\begin{equation}
\left|\eta - \frac{kt}{L}\right| \le C\langle t \rangle \langle k,\eta \rangle^{1/2}\langle k-\ell,\eta-\ell t/L \rangle^{1/2}.
\end{equation}
So, we have
\begin{align}
\left|NL_{3,2}'(t) \right| \le  C \langle t \rangle \sum_k \sum_{\ell \ne 0} \int & A^{(\sigma+\overline{\nu})}(t,k,\eta)^2 \left|\langle k, \eta \rangle^{1/2}\widehat{\Phi}(t,k,\eta)\right|\left|\widehat{W}(\ell)\widehat{\rho}(t,\ell)\frac{\ell}{L}\right|\nonumber\\
& \cdot \left|\left\langle k-\ell,\eta-\frac{\ell t}{L} \right\rangle^{1/2}\widehat{\alpha \Phi}\left(t,k-\ell,\eta-\frac{\ell t}{L}\right)\right| d\eta.
\end{align}
Another application of Lemma \ref{product_lem} gives
\begin{align}
&\left|NL_{3,2}'(t) \right|\nonumber\\
&\le  C \langle t \rangle \left\| A^{(\sigma + 1/2+ \overline{\nu})}\widehat{\Phi}(t) \right\|_2\nonumber\\
& \cdot \left( \left\|A^{(c\lambda,0)}\widehat{\rho}(t) \right\|_2\left\|A^{(\sigma + 1/2+ \overline{\nu})} \widehat{\alpha \Phi}(t)\right\|_2+\left\|A^{(\sigma + \overline{\nu})}\langle k \rangle^{-1}\widehat{\rho}(t) \right\|_2\left\|A^{(c\lambda,1/2)}\widehat{\alpha \Phi}(t) \right\|_2\right).
\end{align}
Noting that $1/2+ \overline{\nu}  < 1+ \overline{\nu}/2$ and making the same sort of estimates as above (also noticing that $NL_{3,3}'$ will be controlled similarly), we find
\begin{align}
&\left|NL_{3,2}'(t) \right| , \left|NL_{3,3}'(t) \right| \nonumber\\
&\le  C \langle t \rangle e^{-(1-c)\alpha_0\langle t \rangle^{\overline{\nu}}} \left\| A^{(\sigma + 1+ \overline{\nu}/2)}\widehat{\Phi}(t) \right\|_2\left\|A^{(\sigma + 1+ \overline{\nu}/2)} \widehat{\alpha \Phi}(t)\right\|_2\left\|A\widehat{\rho}(t) \right\|_2\nonumber\\
&+ C\langle t \rangle^{1+\overline{\nu}}\left\| A^{(\sigma + 1+ \overline{\nu}/2)}\widehat{\Phi}(t) \right\|_2\left\|A^{(\sigma - \beta)}\widehat{\alpha \Phi}(t) \right\|_2\left\|A\widehat{\rho}(t) \right\|_2\nonumber\\
&\le C \langle t \rangle e^{-(1-c)\alpha_0\langle t \rangle^{\overline{\nu}}} \left\| A^{(\sigma + 1+ \overline{\nu}/2)}\widehat{\Phi}(t) \right\|_2\left\|A^{(\sigma + 1+ \overline{\nu}/2)} \widehat{\alpha \Phi}(t)\right\|_2\left\|A\widehat{\rho}(t) \right\|_2\nonumber\\
& +C \langle t \rangle^{-2}\left\| A^{(\sigma + 1+ \overline{\nu}/2)}\widehat{\Phi}(t) \right\|_2^2\left\|A^{(\sigma - \beta)}\widehat{\alpha \Phi}(t) \right\|_2\nonumber\\
& +C \langle t \rangle^{4+2\overline{\nu}}\left\|A^{(\sigma - \beta)}\widehat{\alpha \Phi}(t) \right\|_2\left\|A\widehat{\rho}(t) \right\|_2^2.
\end{align}
Note, we can absorb the factor of $\langle t \rangle$ in the first term above at the expense of adding a factor of $1/2$ in the exponential.

$NL_{3,4}'(t)$ and $NL_{3,5}'(t)$ satisfy similar estimates.  Here, we use
\begin{equation}
\left|A^{(\sigma + 1)}\nabla_{\eta}A^{(\sigma + 1)}(t,k,\eta) \right| \le C A^{(\sigma + 1/2 + \overline{\nu}/2)}(t,k,\eta)^2
\end{equation}
to deduce an estimate of the form
\begin{align}
&\left|NL_{3,4}'(t) \right|, \left|NL_{3,5}'(t)\right| \nonumber\\
&\le C\langle t \rangle \left\|A^{(\sigma + 1 + \overline{\nu}/2)} \widehat{\Phi} \right\|_2\nonumber\\
&\cdot \left(\left\|A^{(c\lambda,1/2)} \widehat{\alpha \Phi}(t)\right\|_2\left\|A^{(\sigma + 1/2 + \overline{\nu}/2)}\langle k \rangle^{-1}\widehat{\rho}(t) \right\|_2+\left\|A^{(\sigma + 1 + \overline{\nu}/2)} \widehat{\alpha \Phi}(t)\right\|_2\left\|A^{(c\lambda,0)}\widehat{\rho}(t) \right\|_2 \right)\nonumber\\
&\le C\langle t \rangle e^{-(1-c)\alpha_0\langle t \rangle^{\overline{\nu}}}  \left\|A^{(\sigma + 1 + \overline{\nu}/2)} \widehat{\Phi} \right\|_2\left\|A^{(\sigma + 1 + \overline{\nu}/2)} \widehat{\alpha \Phi}(t)\right\|_2\left\|A\widehat{\rho}(t) \right\|_2\nonumber\\
&\qquad +C\langle t \rangle^{1+1/2+\overline{\nu}/2} \left\|A^{(\sigma + 1 + \overline{\nu}/2)} \widehat{\Phi} \right\|_2\left\|A^{(\sigma-\beta)} \widehat{\alpha \Phi}(t)\right\|_2 \left\|A\widehat{\rho}(t) \right\|_2.
\end{align}
Splitting up the term that grows with $t$ yields
\begin{align}
&\left|NL_{3,4}'(t) \right|, \left|NL_{3,5}'(t)\right| \nonumber\\
&\le C\langle t \rangle e^{-(1-c)\alpha_0\langle t \rangle^{\overline{\nu}}}  \left\|A^{(\sigma + 1 + \overline{\nu}/2)} \widehat{\Phi} \right\|_2\left\|A^{(\sigma + 1 + \overline{\nu}/2)} \widehat{\alpha \Phi}(t)\right\|_2\left\|A\widehat{\rho}(t) \right\|_2\nonumber\\
&\qquad +C\langle t \rangle^{-2} \left\|A^{(\sigma + 1 + \overline{\nu}/2)} \widehat{\Phi} \right\|_2^2\left\|A^{(\sigma-\beta)} \widehat{\alpha \Phi}(t)\right\|_2\nonumber\\
&\qquad +C\langle t \rangle^{5+\overline{\nu}} \left\|A^{(\sigma-\beta)} \widehat{\alpha \Phi}(t)\right\|_2\left\|A\widehat{\rho}(t) \right\|_2^2.
\end{align}

For $NL_{3,6}'$ and $NL_{3,7}'$ we use the estimate
\begin{align}
\Bigg|\left(\eta - \frac{kt}{L}\right)\cdot & \nabla_{\eta} \nabla_{\eta}A^{(\sigma + 1)}\left(t,k-\ell, \eta - \frac{\ell t}{L}\right) \Bigg|\nonumber\\
&\le C \langle t \rangle \langle k,\eta \rangle^{\overline{\nu}/2}A^{(\sigma + 1 +\overline{\nu}/2)}\left(t,k-\ell, \eta - \frac{\ell t}{L}\right)
\end{align}
(we have replaced $(3/2)\overline{\nu}$ by the larger $1+\overline{\nu}/2$) along with Lemma \ref{young_var_lem}[(i)] to conclude
\begin{align}
&\left| NL_{3,6}'(t) \right|, \left| NL_{3,7}'(t) \right| \nonumber\\
&\le C \langle t \rangle\left\|A^{(\sigma + 1 +\overline{\nu}/2)}\widehat{\Phi}(t) \right\|_2\left\|A^{(\sigma + 1 +\overline{\nu}/2)}\widehat{\alpha\Phi}(t)  \right\|_2\left\|\langle k \rangle^{\sigma} \widehat{\rho}(t) \right\|_2 \nonumber\\
& \le C \langle t \rangle e^{-\alpha_0\langle t \rangle^{\overline{\nu}}}\left\|A^{(\sigma + 1 +\overline{\nu}/2)}\widehat{\Phi}(t) \right\|_2\left\|A^{(\sigma + 1 +\overline{\nu}/2)}\widehat{\alpha\Phi}(t)  \right\|_2\left\|A\widehat{\rho}(t) \right\|_2
\end{align}

Similarly for $NL_{3,8}'$ - $NL_{3,11}'$, we have
\begin{align}
\Bigg|\left(\eta - \frac{kt}{L}\right)&\nabla_{\eta}A^{(\sigma + 1)}\left(t,k-\ell, \eta - \frac{\ell t}{L}\right) \Bigg|\nonumber\\
&\le C \langle t \rangle \langle k,\eta \rangle^{\overline{\nu}/2}A^{(\sigma + 1 +\overline{\nu}/2)}\left(t,k-\ell, \eta - \frac{\ell t}{L}\right)
\end{align}
from which we conclude
\begin{align}
&\left| NL_{3,8}'(t) \right|, \left| NL_{3,9}'(t) \right|, \left| NL_{3,10}'(t) \right|, \left| NL_{3,11}'(t) \right| \nonumber\\
&\le C \langle t \rangle\left\|A^{(\sigma + 1 +\overline{\nu}/2)}\widehat{\Phi}(t) \right\|_2\left\|A^{(\sigma + 1 +\overline{\nu}/2)}\widehat{\alpha\Phi}(t)  \right\|_2\left\|\langle k \rangle^{\sigma} \widehat{\rho}(t) \right\|_2 \nonumber\\
& \le C \langle t \rangle e^{-\alpha_0\langle t \rangle^{\overline{\nu}}}\left\|A^{(\sigma + 1 +\overline{\nu}/2)}\widehat{\Phi}(t) \right\|_2\left\|A^{(\sigma + 1 +\overline{\nu}/2)}\widehat{\alpha\Phi}(t)  \right\|_2\left\|A\widehat{\rho}(t) \right\|_2
\end{align}

This leaves us with $NL_{3,12}'$ and $NL_{3,13}'$.  In both of these, there is no additional factor of $\eta - kt/L$ to worry about.  As such, we can effectively ignore the derivative of $A$ in $NL_{3,12}$ since it only brings down a denominator of size greater than 1.  Appealing to Lemma \ref{young_var_lem} once more, and approximating $\|\langle k\rangle^{\sigma}\widehat{\rho}\|_2$ by $\|A\widehat{\rho} \|_2$ multiplied by a decaying exponential in $t$ yields
\begin{equation}
\left|NL_{3,12}'(t)\right|,\left|NL_{3,13}'(t)\right| \le Ce^{-\alpha_0\langle t \rangle^{\overline{\nu}}}\left\|A^{(\sigma + 1)}\widehat{\alpha \Phi}(t) \right\|_2\left\|A^{(\sigma + 1)} \widehat{\Phi}(t) \right\|_2\left\|A \widehat{\rho}(t)\right\|_2.
\end{equation}

\subsection{Final Estimate}

We now collect all of the estimates above.  We have already split many of the products into a sum of squares above when we needed to pair powers of $t$ with appropriate terms.  We will also split some of the remaining terms in this fashion.  In addition, we use Assumption \eqref{apriori_ass} to conclude that
$$(1-\varkappa^2)\left\| A^{(\sigma+m)} \widehat{\Phi}\right\|_2^2 \le \left\| A^{(\sigma+m)} \widehat{\Phi}\right\|_2^2 - \left\|\widehat{ v A^{(\sigma+m)} \Phi}\right\|_2^2   $$  for all values of $m$ appearing in the various computations above.  For simplicity, we write $\delta$ for the smallest coefficient appearing in all of the $\exp(-C\langle t\rangle^{\overline{\nu}})$ terms. Also, we drop $CK_3$ altogether since it is negative (and does not pair with anything else).  This gives an estimate of the form
\begin{align}
&\frac{d}{dt}\Bigg[\Big\|A^{(\sigma+1)} \widehat{ \Phi}(t) \Big\|^2_2 -\left\|\widehat{v A^{(\sigma+1)} \Phi}(t) \right\|^2_2+\left\|A^{(\sigma+1)}\widehat{\alpha \Phi}(t) \right\|^2_2\Bigg]\nonumber\\
& \le \left[2(1-\varkappa^2)\dot{\lambda}(t) + C e^{-\delta \langle t \rangle^{\overline{\nu}}}\left\|A\widehat{\rho}(t)\right\|_2 + C\frac{\left\|A^{(\sigma - \beta)}\widehat{\alpha \Phi}(t) \right\|_2}{\langle t \rangle^2} + \frac{Cb}{\langle t \rangle^2} \right]\left\| A^{(\sigma + 1 + \overline{\nu}/2)}\widehat{\Phi}(t)\right\|_2^2\nonumber\\
&+\left[2\dot{\lambda}(t) + C e^{-\delta \langle t \rangle^{\overline{\nu}}}\left\|A\widehat{\rho}(t)\right\|_2+ \frac{Cb}{\langle t \rangle^2}\right]\left\| A^{(\sigma + 1 + \overline{\nu}/2)}\widehat{\alpha \Phi}(t)\right\|_2^2 \nonumber\\
&+ \frac{C}{b}\langle t \rangle^4\left\|A\widehat{\rho}(t) \right\|_2^2 + C \left\|A^{(\sigma - \beta)}\widehat{\Phi}(t) \right\|_2\frac{\left\|A^{(\sigma +1)}\widehat{\alpha\Phi}(t) \right\|_2^2}{\langle t \rangle^2}\nonumber\\
&+C\left\|A^{(\sigma - \beta)}\widehat{\Phi}(t) \right\|_2\frac{\left\|A^{(\sigma +1)}\widehat{\Phi}(t) \right\|_2^2}{\langle t \rangle^2}+C\left\|A^{(\sigma - \beta)}\widehat{\alpha\Phi}(t) \right\|_2\frac{\left\|A^{(\sigma +1)}\widehat{\alpha\Phi}(t) \right\|_2^2}{\langle t \rangle^2} \nonumber\\
&+C\langle t \rangle^6 \left\|A^{(\sigma - \beta)}\widehat{\Phi}(t) \right\|_2\left\| A\widehat{\rho}(t)\right\|_2^2 + C\langle t \rangle^6 \left\|A^{(\sigma - \beta)}\widehat{\alpha\Phi}(t) \right\|_2\left\| A\widehat{\rho}(t)\right\|_2^2\nonumber\\
&+Ce^{-\delta \langle t \rangle^{\overline{\nu}}}\left\|A^{(\sigma+1)}\widehat{\Phi}(t)\right\|_2^2\left\|A\widehat{\rho}(t)\right\|_2+Ce^{-\delta \langle t \rangle^{\overline{\nu}}}\left\|A^{(\sigma+1)}\widehat{\alpha\Phi}(t)\right\|_2^2\left\|A\widehat{\rho}(t)\right\|_2\nonumber\\
&+Ce^{-\delta\langle t \rangle^{\overline{\nu}}}\left\|A^{(\sigma + 1)}\widehat{\Phi}(t)\right\|_2\left\|A^{(\sigma + 1)}\widehat{\alpha\Phi}(t)\right\|_2\left\|A\widehat{\rho}(t)\right\|_2,
\end{align}
for some $C$ depending on the various parameters appearing in the assumptions on the background and initial data.  Note we have used $\overline{\nu} < 1$ to absorb certain smaller terms.

For the first two terms, we note that $\dot{\lambda}(t)$ is negative and bounded for all time.  In contrast, the positive terms we have grouped with it are all decreasing in time.  Hence, if we choose $\epsilon$ and $b$ small enough, the first two terms will be negative for all time and can be dropped altogether.  Specifically, we require
\begin{eqnarray}
2C\sqrt{K_4}\langle t \rangle^{1/2}e^{-\delta \langle t \rangle^{\overline{\nu}}}\epsilon + \frac{2\sqrt{K_2}\epsilon}{\langle t \rangle^2} + \frac{Cb}{\langle t \rangle^2} &\le& 2(1-\varkappa^2)\left|\dot{\lambda}(t)\right|\\
2C\sqrt{K_4}\langle t \rangle^{1/2}e^{-\delta \langle t \rangle^{\overline{\nu}}}\epsilon + \frac{Cb}{\langle t \rangle^2} &\le & 2\left|\dot{\lambda}(t)\right|
\end{eqnarray}
Note that this requires choosing $\epsilon$ small with respect to $K_2$ and $K_4$, but $b$ can be chosen independently.  Making these choices, we then drop the first two terms and absorb $b^{-1}$ into the constant $C$ (presumably making it much larger).  Thus, we are left with (via \eqref{orig_bstrap_AI} and \eqref{orig_bstrap_AbI})
\begin{align}
&\frac{d}{dt}\Bigg[\Big\|A^{(\sigma+1)} \widehat{ \Phi}(t) \Big\|^2_2 -\left\|\widehat{v A^{(\sigma+1)} \Phi}(t) \right\|^2_2+\left\|A^{(\sigma+1)}\widehat{\alpha \Phi}(t) \right\|^2_2\Bigg]\nonumber\\
& \le C \langle t \rangle^4\left\|A\widehat{\rho}(t) \right\|_2^2 + 8C(1-\varkappa^2)^{-\frac{1}{2}} K_1\sqrt{K_2} \epsilon^3\langle t \rangle^{4} \nonumber\\
&+8C(1-\varkappa^2)^{-\frac{3}{2}}K_1\sqrt{K_2}\epsilon^3\langle t \rangle^{4}+8C K_1\sqrt{K_2}\epsilon^3 \langle t \rangle^{4} \nonumber\\
&+2C(1-\varkappa^2)^{-\frac{1}{2}}\sqrt{K_2}\epsilon\langle t \rangle^6 \left\| A\widehat{\rho}(t)\right\|_2^2 + 2C\sqrt{K_2}\epsilon \langle t \rangle^6 \left\| A\widehat{\rho}(t)\right\|_2^2\nonumber\\
&+4C(1-\varkappa^2)^{-1}K_1\epsilon^2 \langle t \rangle^{6}e^{-\delta \langle t \rangle^{\overline{\nu}}}\left\|A\widehat{\rho}(t)\right\|_2+4CK_1 \epsilon^2 \langle t \rangle^{6} e^{-\delta \langle t \rangle^{\overline{\nu}}}\left\|A\widehat{\rho}(t)\right\|_2\nonumber\\
&+4CK_1(1-\varkappa^2)^{-\frac{1}{2}}\epsilon^2 \langle t \rangle^{6} e^{-\delta\langle t \rangle^{\overline{\nu}}}\left\|A\widehat{\rho}(t)\right\|_2.
\end{align}
We now replace $t$ by $\tau$ and integrate the expression above from $0$ to $t$.  Using \eqref{orig_bstrap_rho} (and that our assumptions on the initial data give a bound at $t=0$), we find
\begin{align}
\Big\| A^{(\sigma+1)} &\widehat{ \Phi}(t) \Big\|^2_2  -\left\|\widehat{v A^{(\sigma+1)} \Phi}(t) \right\|^2_2 +\left\|A^{(\sigma+1)}\widehat{\alpha \Phi}(t) \right\|^2_2\nonumber\\
& \le 3\epsilon^2 + 4CK_3\epsilon^2\langle t \rangle^4 + 24C(1-\varkappa^2)^{-\frac{3}{2}}K_1\sqrt{K_2}\epsilon^3\langle t \rangle^{5}\nonumber\\
& \qquad + 16C(1-\varkappa^2)^{-\frac{1}{2}}K_3\sqrt{K_2}\epsilon^3\langle t \rangle^6+ 24C(1-\varkappa^2)^{-1}K_1\sqrt{K_3}\epsilon^3\langle t \rangle^{6}.
\end{align}

Thus, if we take
\begin{eqnarray}
K_1 &=& 3+4CK_3,\\
\epsilon &<& \frac{K_1(1-\varkappa^2)^{\frac{3}{2}}}{C} \left(24K_1\sqrt{K_2}+16K_3\sqrt{K_2}+24K_1\sqrt{K_3}\right)^{-1},
\end{eqnarray}
we establish \eqref{bstrap_AI}.

\section{Bootstrap Estimate \eqref{bstrap_AbI}}

The manipulations that allow us to establish \eqref{bstrap_AbI} are very similar to the ones in the previous section.  As such, it will suffice to point out the places where the analysis is slightly different from that above.  We first note
\begin{align}
\frac{1}{2}\frac{d}{dt}\Bigg[&\Big\|A^{(\sigma-\beta)} \widehat{ \Phi}(t) \Big\|^2_2 -\left\|\widehat{v A^{(\sigma-\beta)} \Phi}(t) \right\|^2_2+\left\|A^{(\sigma-\beta)}\widehat{\alpha \Phi}(t) \right\|^2_2\Bigg]\nonumber\\
&=\dot{\lambda}(t) \left(\left\|A^{(\sigma - \beta + \overline{\nu}/2)}\widehat{\Phi}(t)\right\|_2^2-\left\|\widehat{vA^{(\sigma - \beta + \overline{\nu}/2)}\Phi}(t)\right\|_2^2+\left\|A^{(\sigma - \beta + \overline{\nu}/2)}\widehat{\alpha\Phi}(t)\right\|_2^2\right)\nonumber\\
& + \frac{\overline{\nu}^2}{16\pi^2}\dot{\lambda}(t)\sum_k\int \left|A^{(\sigma - \beta - 1 +\overline{\nu}/2)}\widehat{\Phi}(t,k,\eta)\right|^2\frac{|\eta|^2}{\langle k,\eta \rangle^2}d\eta \nonumber\\
& + \Re \sum_k\int A^{(\sigma - \beta)}(t,k,\eta)^2 \overline{\widehat{\Phi}(t,k,\eta)} \partial_t \widehat{\Phi}(t,k,\eta) d\eta \nonumber\\
& + \Re \sum_k\int A^{(\sigma - \beta)}(t,k,\eta)^2 \overline{\widehat{\alpha \Phi}(t,k,\eta)} \partial_t \widehat{\alpha \Phi}(t,k,\eta) d\eta \nonumber\\
& - \Re \sum_k \int \overline{\frac{i}{2\pi}\nabla_{\eta}A^{(\sigma - \beta)}\widehat{\Phi}(t,k,\eta)} \cdot \frac{i}{2\pi}\nabla_{\eta}A^{(\sigma - \beta)}\partial_t \widehat{\Phi}(t,k,\eta)d\eta.
\end{align}
Since this is precisely analogous to the derivative of \eqref{bstrap_AI}, we obtain (after applying the same integral identities)
\begin{align}
\frac{1}{2}\frac{d}{dt}\Bigg[&\Big\|A^{(\sigma-\beta)} \widehat{ \Phi}(t) \Big\|^2_2 -\left\|\widehat{v A^{(\sigma-\beta)} \Phi}(t) \right\|^2_2+\left\|A^{(\sigma-\beta)}\widehat{\alpha \Phi}(t) \right\|^2_2\Bigg]\nonumber\\
&= \widetilde{CK}(t)+\widetilde{L}(t) + \widetilde{NL_1'}(t)+\widetilde{NL_2'}(t)+\widetilde{NL_3'}(t),
\end{align}
where the terms are precisely the same as the terms in the previous sections with $\sigma + 1$ replaced by $\sigma - \beta$.

For the linear terms, the same considerations as above now yield
\begin{align}
|\widetilde{L}(t)| &\le C \langle t \rangle^{-\beta}\left\|A\widehat{\rho}(t)\right\|_2\left(\left\|A^{(\sigma - \beta)}\widehat{\Phi}(t)\right\|_2+\left\|A^{(\sigma - \beta)}\widehat{\alpha\Phi}(t)\right\|_2\right)\nonumber\\
&\le Cb\langle t \rangle^{-\beta}\left(\left\|A^{(\sigma - \beta+\overline{\nu}/2)}\widehat{\Phi}(t)\right\|_2^2+\left\|A^{(\sigma - \beta+\overline{\nu}/2)}\widehat{\alpha\Phi}(t)\right\|_2^2\right)\nonumber\\
& \qquad + \frac{C}{b}\langle t \rangle^{-\beta}\left\|A\widehat{\rho}(t)\right\|_2^2,
\end{align}
where $b$ is a small, arbitrary parameter which will be chosen analogously to the previous section.  Note that we have replaced an exponentially decaying factor, $\exp(-(1-c)\langle t \rangle^{\nu})$, by the larger (up to a constant) $\langle t \rangle^{-\beta}$ .  Also, since $\beta > 1$, the integral of the last term above will be strictly decreasing in $t$.

As before, we break $\widetilde{NL_1'}(t)$ and $\widetilde{NL_2'}(t)$ in transport, reaction, and remainder terms.  The estimates for the transport and remainder portions are essentially unchanged (we need only replace $\sigma + 1$ with $\sigma - \beta$).  As for the reaction terms, we alter our analysis slightly.  We once again break up this term into two portions (e.g.  $\widetilde{RNL_{1,1}'}$ and $\widetilde{RNL_{1,2}'}$) where the first portion has the factor of $A^{(\sigma - \beta)}(t,k,\eta)$ from the difference.  The estimate for the second portion with $A^{(\sigma - \beta)}\left(t,k-\ell, \eta - \ell t/L\right)$ is precisely analogous to the previous case.  For the first portion, the Fourier localizations for this regime give us
\begin{equation}
A^{(\sigma - \beta)}(t,k,\eta) \le C e^{c\lambda(t)\langle k-\ell, \eta - \ell t/L \rangle^{\overline{\nu}}} A^{(\sigma - \beta)}\left(t,\ell,\frac{\ell t}{L}\right),
\end{equation}
where $c\in(0,1)$ (as per usual).  This gives
\begin{align}
\Big|\widetilde{RNL_{1,1}'}&(t)\Big| \nonumber\\
& \le  C\langle t \rangle \sum_N \sum_k\sum_{\ell \ne 0} \int \left|A^{(\sigma - \beta)}\mathcal{F}\{(1-|v|^2)\Phi\}(t,k,\eta)\right|\left|A^{(\sigma - \beta)}\widehat{\rho}(t,\ell)_N\right|\nonumber\\
& \qquad \cdot e^{c\lambda(t)\langle k-\ell, \eta - \ell t/L \rangle^{\overline{\nu}}}\!\left\langle k-\ell, \eta - \frac{\ell t}{L}\right\rangle \! \left|\widehat{\alpha \Phi}\left(t,k-\ell,\eta - \frac{\ell t}{L}\right)_{<N/8}\right|d\eta\nonumber\\
& \qquad + \ldots,
\end{align}
where the dots indicate the remaining term where the roles of $(1-|v|^2)\Phi$ and $\alpha\Phi$ are reversed.  Proceeding as before gives us a final estimate of the form
\begin{align}
\left|\widetilde{RNL_{1,1}'}(t)\right| \le  &C \langle t \rangle^{1-\beta}\left\|A^{(\sigma - \beta)}\widehat{\Phi}(t) \right\|_2\left\|A\widehat{\rho}\right\|_2^2\nonumber\\
&+C \langle t \rangle^{1-\beta}\left\|A^{(\sigma - \beta)}\widehat{\alpha\Phi}(t) \right\|_2\left\|A\widehat{\rho}\right\|_2^2\nonumber\\
&+C \langle t \rangle^{1-\beta}\left\|A^{(\sigma - \beta)}\widehat{\Phi}(t) \right\|_2\left\|A^{(\sigma - \beta)}\widehat{\alpha\Phi}(t) \right\|_2^2\nonumber\\
&+C \langle t \rangle^{1-\beta}\left\|A^{(\sigma - \beta)}\widehat{\alpha\Phi}(t) \right\|_2\left\|A^{(\sigma - \beta)}\widehat{\Phi}(t) \right\|_2^2.
\end{align}

We are now left with $\widetilde{NL_3'}(t)$.  For $\widetilde{NL_{3,1}'}(t)$, the analysis is changed only by the fact that we now have
\begin{equation}
\left\|A^{(\sigma - \beta)}\langle k\rangle^{-1} \widehat{\rho}(t)\right\|_2 \le C\langle t \rangle^{-\beta}\left\|A\widehat{\rho}(t)\right\|_2.
\end{equation}
For $\widetilde{NL_{3,2}'}(t)$ - $\widetilde{NL_{3,5}'}(t)$, we note that
\begin{eqnarray}
\left|A^{(\sigma - \beta)}\triangle_{\eta}A^{(\sigma - \beta)}(t,k,\eta)\right| &\le& C A^{(\sigma - \beta-1+\overline{\nu})}(t,k,\eta)^2,\\
\left|A^{(\sigma - \beta)}\nabla_{\eta}A^{(\sigma - \beta)}(t,k,\eta)\right| &\le& C A^{(\sigma - \beta-1/2+\overline{\nu}/2)}(t,k,\eta)^2.
\end{eqnarray}
We repeat the same computations for these bearing in mind that we pick up factors of $\langle t \rangle^{-\beta}$ when we estimate terms like $\| A^{(\sigma-\beta)}\widehat{\rho}(t)\|_2$.  All remaining terms are computed exactly the same.  Note that $\widetilde{NL_{3,6}'}(t)$ - $\widetilde{NL_{3,12}'}(t)$ all have exponentially decreasing factors in $t$.  Hence, we can absorb any positive powers of $\langle t \rangle$ by adding a factor of $1/2$ in the exponential (and increasing $C$, of course).  This was not crucial in the previous section since we can allow the $(\sigma+1)$-term to grow in $t$.  Here, we must be more careful.

Combining all of these observations, we find that
\begin{align}
\frac{d}{dt}\Bigg[&\Big\|A^{(\sigma-\beta)} \widehat{ \Phi}(t) \Big\|^2_2 -\left\|\widehat{v A^{(\sigma-\beta)} \Phi}(t) \right\|^2_2+\left\|A^{(\sigma-\beta)}\widehat{\alpha \Phi}(t) \right\|^2_2\Bigg]\nonumber\\
&\le \left\{2\left(1-\varkappa^2\right) \dot{\lambda}(t) + Ce^{-\frac{1}{2}(1-c)\alpha_0\langle t \rangle^{\overline{\nu}}}\left\|A \widehat{\rho}(t)\right\|_2 + \frac{C\left\|A^{(\sigma - \beta)}\widehat{\alpha\Phi}(t)\right\|}{\langle t \rangle^{\beta - 1}} + \frac{Cb}{\langle t \rangle^{\beta}} \right\}\nonumber\\
&\qquad \qquad \qquad \qquad \cdot \left\|A^{(\sigma - \beta + \overline{\nu}/2)}\widehat{\Phi}(t)\right\|_2^2\nonumber\\
&+ \left\{2\dot{\lambda}(t) + Ce^{-\frac{1}{2}(1-c)\alpha_0\langle t \rangle^{\overline{\nu}}}\left\|A \widehat{\rho}(t)\right\|_2 + \frac{Cb}{\langle t \rangle^{\beta}} \right\}\left\|A^{(\sigma - \beta + \overline{\nu}/2)}\widehat{\alpha\Phi}(t)\right\|_2^2\nonumber\\
&+\frac{C}{b\langle t \rangle^{\beta}}\left\|A\widehat{\rho}(t)\right\|_2^2 + C \langle t \rangle^{1-\beta}\left\|A^{(\sigma - \beta)}\widehat{\Phi}(t) \right\|_2\left\|A\widehat{\rho}(t)\right\|_2^2\nonumber\\
&+C \langle t \rangle^{1-\beta}\left\|A^{(\sigma - \beta)}\widehat{\alpha\Phi}(t) \right\|_2\left\|A\widehat{\rho}(t)\right\|_2^2 +C \langle t \rangle^{1-\beta}\left\|A^{(\sigma - \beta)}\widehat{\Phi}(t) \right\|_2\left\|A^{(\sigma - \beta)}\widehat{\alpha\Phi}(t) \right\|_2^2\nonumber\\
&+C \langle t \rangle^{1-\beta}\left\|A^{(\sigma - \beta)}\widehat{\alpha\Phi}(t) \right\|_2\left\|A^{(\sigma - \beta)}\widehat{\Phi}(t) \right\|_2^2\nonumber\\
&+C\langle t \rangle^{-1}\left(e^{-\frac{1}{4}\alpha_0\langle t \rangle^{\overline{\nu}}}+e^{-\frac{1}{4}(1-c)\langle t \rangle^{\overline{\nu}}}\right)\left\|A\widehat{\rho}(t)\right\|_2\left\|A^{(\sigma - \beta)}\widehat{\alpha \Phi}(t)\right\|_2^2\nonumber\\
&+C\langle t \rangle^{-1}\left(e^{-\frac{1}{4}\alpha_0\langle t \rangle^{\overline{\nu}}}+e^{-\frac{1}{4}(1-c)\langle t \rangle^{\overline{\nu}}}\right)\left\|A\widehat{\rho}(t)\right\|_2\left\|A^{(\sigma - \beta)}\widehat{\Phi}(t)\right\|_2^2,
\end{align}
where we have combined many terms with faster rates of decay into the terms displayed (as always, at the expense of possibly increasing the constant, $C$).  We let $\delta >0$ be a number slightly smaller than all coefficients appearing in the exponential factors above so that we may add a factor of $\langle t \rangle^{-\beta}$ in front of the last four terms (this might also require an increase in $C$).

As before, we must choose $\epsilon$ and $b$ small enough so that
\begin{eqnarray}
2C\sqrt{K_4}\langle t \rangle^{\frac{1}{2}} e^{-\delta \langle t \rangle^{\overline{\nu}}}\epsilon + 2C\sqrt{K_2}\langle t \rangle^{1-\beta}\epsilon + Cb\langle t \rangle^{-\beta} &\le& 2(1-\varkappa^2)\left|\dot{\lambda}(t)\right|,\\
2C\sqrt{K_4}\langle t \rangle^{\frac{1}{2}} e^{-\delta \langle t \rangle^{\overline{\nu}}}\epsilon + Cb\langle t \rangle^{-\beta} &\le& 2\left|\dot{\lambda}(t)\right|,
\end{eqnarray}
for all $t$.  This will ensure that the first two terms above are negative, and can therefore be dropped from the estimate.  Applying our bootstrap assumptions (and absorbing $b^{-1}$ into $C$) leaves us with
\begin{align}
\frac{d}{dt}\Bigg[&\Big\|A^{(\sigma-\beta)} \widehat{ \Phi}(t) \Big\|^2_2 -\left\|\widehat{v A^{(\sigma-\beta)} \Phi}(t) \right\|^2_2+\left\|A^{(\sigma-\beta)}\widehat{\alpha \Phi}(t) \right\|^2_2\Bigg]\nonumber\\
&\le C\langle t \rangle^{-\beta}\left\|A\widehat{\rho}(t)\right\|_2^2 + 4C \sqrt{K_2}(1-\varkappa^2)^{-1}\epsilon\langle t \rangle^{1-\beta}\left\|A\widehat{\rho}(t)\right\|_2^2\nonumber\\
& +16 C K_2 \sqrt{K_2}(1-\varkappa^2)^{-1}\epsilon^3 \langle t \rangle^{1-\beta} +16CK_2(1-\varkappa^2)^{-1}\epsilon^2e^{-\delta \langle t \rangle^{\overline{\nu}}}\left\|A\widehat{\rho}(t)\right\|_2.
\end{align}
Integrating this in $t$ (and recalling that our initial conditions give us an estimate at $t=0$), we find
\begin{align}
\Big\|A^{(\sigma-\beta)}& \widehat{ \Phi}(t) \Big\|^2_2 -\left\|\widehat{v A^{(\sigma-\beta)} \Phi}(t) \right\|^2_2+\left\|A^{(\sigma-\beta)}\widehat{\alpha \Phi}(t) \right\|^2_2\nonumber\\
\le &\left(3 + CK_3\right)\epsilon^2 \nonumber\\
&+C\left(4K_3\sqrt{K_2}+16K_2\sqrt{K_2}+16K_2\sqrt{K_3}\right)(1-\varkappa^2)^{-1} \epsilon^3
\end{align}
where we have once again enlarged $C$ so that the various integrals in $t$ are bounded for all time (note that all functions of $t$ above are integrable for $t \to \infty$).  So, we have established the theorem if we take
\begin{align}
K_2 &= 3 + CK_3,\\
\epsilon &< \frac{(1-\varkappa^2)K_2}{C}\left(4K_3\sqrt{K_2}+16K_2\sqrt{K_2}+16K_2\sqrt{K_3}\right)^{-1}.
\end{align}

\section{Remaining Estimates}

\addtocontents{toc}{\setcounter{tocdepth}{2}}

\subsection{Estimate \eqref{transport_control}}

To establish \eqref{transport_control}, we need to show that our bootstrap assumptions imply
\begin{align}
\sup_{\tau \ge 0} & \; e^{-(1-c_2)\lambda(0)\langle \tau \rangle^{\overline{\nu}}}\sum_{|m|\le 2}\sum_{k} \sup_{\omega \in \mathbb{Z}^3_{\ne 0} }\sup_{\zeta \in \mathbb{R}^3}\int_{-\infty}^{\infty}\left|A^{(\sigma + 1)}\widehat{v^m \alpha \Phi}\left(\tau,k, \frac{\omega}{|\omega|}s - \zeta\right)\right|^2ds \nonumber\\
 &\le C K_1 \epsilon^2,
\end{align}
for some positive constant $C$.  As usual, we can drop the sum over $m$ at the expense of picking up a combinatorial factor (from commuting the derivatives past the factor of $A$).  Applying the Trace Lemma \ref{trace_lem}, we have
\begin{align}
\sup_{\tau \ge 0} & \; e^{-(1-c_2)\lambda(0)\langle \tau \rangle^{\overline{\nu}}}\sum_{|m|\le 2}\sum_{k} \sup_{\omega \in \mathbb{Z}^3_{\ne 0} }\sup_{\zeta \in \mathbb{R}^3}\int_{-\infty}^{\infty}\left|A^{(\sigma + 1)}\widehat{v^m \alpha \Phi}\left(\tau,k, \frac{\omega}{|\omega|}s - \zeta\right)\right|^2ds \nonumber\\
&\le C \sup_{\tau \ge 0} \; e^{-(1-c_2)\lambda(0)\langle \tau \rangle^{\overline{\nu}}}\sum_{k} \sup_{\omega \in \mathbb{Z}^3_{\ne 0} }\sup_{\zeta \in \mathbb{R}^3}\int_{-\infty}^{\infty}\left|A^{(\sigma + 1)}\widehat{\alpha \Phi}\left(\tau,k, \frac{\omega}{|\omega|}s - \zeta\right)\right|^2ds\nonumber\\
& \le C \sup_{\tau \ge 0} \; e^{-(1-c_2)\lambda(0)\langle \tau \rangle^{\overline{\nu}}} \left\|A^{(\sigma + 1)}\widehat{\alpha\Phi}(t) \right\|_{H^2_{\eta}L^2_k}^2\nonumber\\
& \le C \sup_{\tau \ge 0} \; e^{-(1-c_2)\lambda(0)\langle \tau \rangle^{\overline{\nu}}} \left\|A^{(\sigma + 1)}\widehat{\alpha\Phi}(t) \right\|_{L^2_{\eta}L^2_k}^2,
\end{align}
where the last inequality is by the fact that derivatives in $\eta$ are equivalent to multiplication by $v$ (up to a multiple) on taking the inverse transform.  Hence, the $L^2$-norms are unaffected.  Applying bootstrap estimate \eqref{orig_bstrap_AI} yields
\begin{align}
\sup_{\tau \ge 0} & \; e^{-(1-c_2)\lambda(0)\langle \tau \rangle^{\overline{\nu}}}\sum_{|m|\le 2}\sum_{k} \sup_{\omega \in \mathbb{Z}^3_{\ne 0} }\sup_{\zeta \in \mathbb{R}^3}\int_{-\infty}^{\infty}\left|A^{(\sigma + 1)}\widehat{v^m \alpha \Phi}\left(\tau,k, \frac{\omega}{|\omega|}s - \zeta\right)\right|^2ds \nonumber\\
&\le 4C K_1 \sup_{\tau \ge 0} \; e^{-(1-c_2)\lambda(0)\langle \tau \rangle^{\overline{\nu}}}\langle \tau \rangle^{6}\epsilon^2.
\end{align}
Since the remaining function of $\tau$ is clearly bounded for all time, we have established the estimate for some appropriate value of $C$.

\subsection{Estimates \eqref{reaction_control} and \eqref{reaction_control_2} - Control of Plasma Echoes}

The crucial estimates \eqref{reaction_control} and \eqref{reaction_control_2} used in the proof of \eqref{bstrap_rho} are where an analysis of the phenomenon of plasma echoes occurs. This is the only place in the proof where a detailed expression for $\lambda(t)$ is important (most of what came before depended only on the fact that $\lambda(t)$ is decreasing and bounded). Our presentation follows the one given in \cite{BMM13}[\S 6] rather closely.  Recall the definition of the kernel $\overline{K}_{k,\ell}(t,\tau)$ given in \eqref{def_Kbar}.

\bigskip

\begin{lem}
Under the bootstrap hypotheses \eqref{orig_bstrap_AI} -- \eqref{orig_bstrap_rho}, there holds
\begin{equation}
\sup_{t\in [0,T]}\sup_k \int_0^t\sum_{\ell\ne 0} \overline{K}_{k,\ell}(t,\tau)d\tau \le C \sqrt{K_2}\epsilon,
\end{equation}
where $C$ depends on the various parameters appearing in the assumptions on the background and initial data.
\end{lem}

\bigskip

\textbf{Proof:}  Throughout, we will denote $c_2 \in(0,1)$ appearing in the exponential simply by $c$.  The remaining constants (including $c_1$) will be absorbed into a generic $C$.  First consider the case when $k=\ell$.  Since $\lambda(t)$ is decreasing, we have
\begin{eqnarray}
\int_0^t \overline{K}_{k,k}(t,\tau)d\tau &\le& C\int_0^t\frac{|k|(t-\tau)}{|k|^{\gamma}}e^{c\lambda(\tau)\langle k(t-\tau)/L \rangle^{\overline{\nu}}}\left|\widehat{\alpha\Phi}(\tau,0,k(t-\tau)/L)\right| d\tau\nonumber\\
&\le& C\int_0^t e^{-(1-c)\lambda(\tau)\langle k(t-\tau)/L \rangle^{\overline{\nu}}} \left(\sup_{\eta}e^{\lambda(\tau)\langle \eta \rangle^{\overline{\nu}}}|\eta|\left|\widehat{\alpha\Phi}(\tau,0,\eta)\right|\right)d\tau\nonumber\\
&\le& C\int_0^t e^{-(1-c)\lambda(\tau)\langle k(t-\tau)/L \rangle^{\overline{\nu}}} \left\|A^{(\sigma - \beta)}\widehat{\alpha \Phi}(\tau) \right\|_{H^2_{\eta}}d\tau.
\end{eqnarray}
where we have dropped the summation over $m$ (at the price of a combinatorial constant) since all such terms are controlled by the $m=0$ term.  The final inequality above is a consequence of the standard Sobolev embedding.  As we have noted above, the $H^M$-norm of $\widehat{\alpha\Phi}$ is controlled by the $L^2$-norm by the fact that $|v|\le 1$.  As such, \eqref{orig_bstrap_AbI} implies that
\begin{equation}
\int_0^t \overline{K}_{k,k}(t,\tau)d\tau \le C \sqrt{K_2}\epsilon,
\end{equation}
as the remaining integral is bounded independently of $t$.  This leaves us with the case $k\ne \ell$ where the plasma echoes become important.  For convenience, we will denote $$-\lambda(t,\tau) \equiv \lambda(t) - \lambda(\tau), $$ which is indeed negative for $t > \tau$.  We also use $\delta = (1-c)\alpha_0$.  Inserting the basic estimate $$\left|k(t-\tau)\right| \le L \langle \tau \rangle \left|k-\ell, \frac{kt-\ell\tau}{L} \right|,  $$ we have
\begin{align}
\mathds{1}_{k\ne\ell}\overline{K}_{k,\ell}(t,\tau) \le C & e^{-\lambda(t,\tau)\langle k,kt/L \rangle^{\overline{\nu}}} e^{c\lambda(\tau)\langle k-\ell, (kt-\ell\tau)/L \rangle^{\overline{\nu}}}\nonumber\\
&\qquad \cdot \frac{\langle \tau \rangle}{|\ell|^{\gamma}}\left|\widehat{\nabla\alpha\Phi}\left(\tau, k-\ell, \frac{kt-\ell\tau}{L}\right)\right|\mathds{1}_{k \ne \ell \ne 0}\nonumber\\
\le C & e^{-\lambda(t,\tau)\langle k,kt/L \rangle^{\overline{\nu}}}  \frac{\langle \tau \rangle}{|\ell|^{\gamma}} e^{-\delta \langle k-\ell, (kt-\ell\tau)/L \rangle^{\overline{\nu}}}\nonumber\\
&\cdot \left|e^{\lambda(\tau)\langle k-\ell, (kt-\ell\tau)/L \rangle^{\overline{\nu}}}\widehat{\nabla\alpha\Phi}\left(\tau, k-\ell, \frac{kt-\ell\tau}{L}\right)\right|\mathds{1}_{k \ne \ell \ne 0}.
\end{align}

Once again by Sobolev embedding,
\begin{align}
\Bigg|e^{\lambda(\tau)\langle k-\ell, (kt-\ell\tau)/L \rangle^{\overline{\nu}}}&\widehat{\nabla\alpha\Phi}\left(\tau, k-\ell, \frac{kt-\ell\tau}{L}\right)\Bigg|\mathds{1}_{k \ne \ell \ne 0}\nonumber\\
&\le C \sup_{\eta} e^{\lambda(\tau)\langle k-\ell, \eta \rangle^{\overline{\nu}}}\langle k-\ell, \eta \rangle\left|\widehat{\alpha\Phi}(\tau,k-\ell,\eta)\right|\nonumber\\
&\le C \left(\sum_k \sup_{\eta} e^{2\lambda(\tau)\langle k, \eta \rangle^{\overline{\nu}}}\langle k, \eta \rangle^2\left|\widehat{\alpha\Phi}(\tau,k,\eta)\right|^2 \right)^{\frac{1}{2}}\nonumber\\
&\le C \left\|A^{(\sigma - \beta)}\widehat{\alpha\Phi}(\tau) \right\|_{L^2_kH^2_{\eta}} \nonumber\\
&\le C\sqrt{K_2}\epsilon.
\end{align}
As such, we have
\begin{align}
\int_0^t\sum_{\ell\ne 0} & \overline{K}_{k,\ell}(t,\tau)\mathds{1}_{k\ne\ell}d\tau \nonumber\\
& \le C \sqrt{K_2}\epsilon \int_0^t\sum_{\ell\ne0}e^{-\lambda(t,\tau)\langle k,kt/L \rangle^{\overline{\nu}}}  \frac{\langle \tau \rangle}{|\ell|^{\gamma}} e^{-\delta \langle k-\ell, (kt-\ell\tau)/L \rangle^{\overline{\nu}}}\mathds{1}_{k\ne \ell}d\tau.
\end{align}
At this point, our analysis is precisely the same as \cite{BMM13}[Lemma 6.1], and so we only give the highlights (referring to this source for all details).  The goal is to show that the remaining summation and integral is bounded independently of $t$ and $k$.  The echo phenomenon refers to the fact that for a specific choice of wavevectors $k = (k_1,k_2,k_3)$ and $\ell = (\ell_1,\ell_2,\ell_3)$, the integrand above is rather sharply localized around times $\tau = tk_i/\ell_i$.  For small times, we have the basic estimate
\begin{equation}
\int_0^{\min\{1,t\}}e^{-\lambda(t,\tau)\langle k,kt/L \rangle^{\overline{\nu}}}  \frac{\langle \tau \rangle}{|\ell|^{\gamma}} e^{-\delta \langle k-\ell, (kt-\ell\tau)/L \rangle^{\overline{\nu}}}\mathds{1}_{k\ne \ell}d\tau \le \frac{Ce^{-C\delta\langle k - \ell \rangle^{\overline{\nu}}}}{\delta^{1/\overline{\nu}}|\ell|^{1+\gamma}}.
\end{equation}
Defining the \emph{resonant interval} as
\begin{equation}
I_R \equiv \left\{\tau \in [1,t] : \left|kt-\ell \tau\right|<\frac{t}{2}\right\},
\end{equation}
we can break up the integral into three portions
\begin{align}
\int_0^t& e^{-\lambda(t,\tau)\langle k,kt/L \rangle^{\overline{\nu}}}  \frac{\langle \tau \rangle}{|\ell|^{\gamma}} e^{-\delta \langle k-\ell, (kt-\ell\tau)/L \rangle^{\overline{\nu}}}\mathds{1}_{k\ne \ell}d\tau\nonumber\\
&= \left(\int_0^{\min\{1,t\}}+\int_{I_R} + \int_{[1,t]\setminus I_R}\right)e^{-\lambda(t,\tau)\langle k,kt/L \rangle^{\overline{\nu}}}  \frac{\langle \tau \rangle}{|\ell|^{\gamma}} e^{-\delta \langle k-\ell, (kt-\ell\tau)/L \rangle^{\overline{\nu}}}\mathds{1}_{k\ne \ell}d\tau\nonumber\\
&\le \frac{Ce^{-C\delta\langle k - \ell \rangle^{\overline{\nu}}}}{\delta^{1/\overline{\nu}}|\ell|^{1+\gamma}} + \mathcal{I}_R + \mathcal{I}_{NR}.
\end{align}
The non-resonant integral is easier to handle since we are bounded well away from the echo time:
\begin{equation}
\mathcal{I}_{NR} \le \frac{Ce^{-C\delta\langle k - \ell \rangle^{\overline{\nu}}}}{\delta^{2/\overline{\nu}}|\ell|^{1+\gamma}},
\end{equation}
which can be combined with the portion coming from the integral over small times.

The resonant integral is much trickier.  From the choice
\begin{equation}
\lambda(t) = \frac{1}{8}(\lambda_0 - \lambda')(1-t)_+ + \alpha_0 + \frac{1}{4}(\lambda_0-\lambda')\min\left\{1,\frac{1}{t^{a}}\right\},
\end{equation}
where
\begin{eqnarray}
\alpha_0 &=& \frac{1}{2}(\lambda_0 + \lambda')\\
a &=& \frac{(2+\gamma)\overline{\nu}-1}{1+\gamma},
\end{eqnarray}
we see that on $\tau \in [1,t]$
\begin{equation}
\lambda(t,\tau) = \delta'\left(\frac{t^{a} - \tau^{a}}{t^{a}\tau^{a}}\right),
\end{equation}
where $\delta' = \frac{1}{4}(\lambda_0-\lambda')$.  When $k \approx l$, the echo occurs at $\tau \approx t$ and hence $\lambda(t,\tau) \approx 0$.  This is where the subtlety lies in approximating the contribution from the plasma echoes.  Using the form of $\lambda(t)$ above, we find that
\begin{eqnarray}
\mathcal{I}_R &\le& C\frac{kt}{\delta^{1/\overline{\nu}}|\ell|^{2+\gamma}}\left(\frac{|\ell|^{\frac{1-a}{\overline{\nu}-a}}}{(a\delta')^{\frac{1}{\overline{\nu}-a}}kt}\right)e^{-C\delta\langle k - \ell \rangle^{\overline{\nu}}} \nonumber\\
&\le&\frac{Ce^{-C\delta \langle k-\ell \rangle^{\overline{\nu}}}}{\delta^{1/\overline{\nu}}(a\delta')^{\frac{1}{\overline{\nu}-a}}},
\end{eqnarray}
provided that $(2+\gamma)(\overline{\nu}-a)\ge 1-a$.  Taking equality in this expression gives rise to the form of $a$ above.  Since we need $a > 0$ for $\lambda(t)$ to be decreasing, we arrive at the condition
\begin{equation}
\overline{\nu} > (2+\gamma)^{-1}.
\end{equation}
Having these estimates, we finish the lemma by summing in $\ell$ and taking the supremum over $t$ and $k$.  Note that $\mathcal{I}_R$ is summable in either $k$ or $\ell$ but not both! $\blacksquare$

\bigskip

\begin{lem}
Under the bootstrap hypotheses \eqref{orig_bstrap_AI} -- \eqref{orig_bstrap_rho}, there holds
\begin{equation}
\sup_{\tau\in [0,T]}\sup_{\ell \ne 0} \sum_{k\ne 0} \int_{\tau}^T \overline{K}_{k,\ell}(t,\tau)dt\le C \sqrt{K_2}\epsilon,
\end{equation}
where $C$ depends on the various parameters appearing in the assumptions on the background and initial data.
\end{lem}

\bigskip

\textbf{Proof:}  The proof of this lemma is analogous to the proof of the lemma above.  We estimate the contribution coming from $k=\ell$ in the same way and end up with the same estimate.  By following the same procedures as above, we need to estimate
\begin{align}
\sum_{k\ne 0} & \int_{\tau}^T \overline{K}_{k,\ell}(t,\tau)\mathds{1}_{k\ne\ell}dt\nonumber\\
&\le C \sqrt{K_2}\epsilon \int_{\tau}^T e^{-\lambda(t,\tau)\langle k, kt/L\rangle^{\overline{\nu}}}e^{-\delta\langle k-\ell, (kt-\ell\tau)/L\rangle^{\overline{\nu}}}\frac{\langle \tau \rangle}{|\ell|^{\gamma}}\mathds{1}_{k\ne\ell}dt\nonumber\\
&= \mathcal{I}_{ST} +  \mathcal{I}_{R} + \mathcal{I}_{NR}
\end{align}
where the ``small time'' integral, $\mathcal{I}_{ST}$, is the contribution from $t \in [\tau, \max\{\tau,\min\{1,T\}\}]$.  The resonant contribution is over the interval
\begin{equation}
I_R = \left\{ t \in [\tau,T] : |kt - \ell\tau|\le \frac{\tau}{2} \right\},
\end{equation}
and the non-resonant portion is what is leftover.  As above, we have rather simple estimates for the small time and non-resonant portions:
\begin{eqnarray}
\mathcal{I}_{ST} &\le & \frac{Ce^{-C\delta\langle k-\ell \rangle^{\overline{\nu}}}}{\delta^{1/\overline{\nu}}|\ell|^{\gamma}|k|},\\
\mathcal{I}_{NR} &\le & \frac{Ce^{-C\delta\langle k-\ell \rangle^{\overline{\nu}}}}{\delta^{2/\overline{\nu}}|\ell|^{\gamma}|k|}.
\end{eqnarray}
The resonant integral yields
\begin{equation}
\mathcal{I}_{R} \le \frac{Ce^{-\frac{1}{2}C\delta\langle k-\ell \rangle^{\overline{\nu}}}}{\delta^{2/\overline{\nu}}(a\delta')^{\frac{1}{\overline{\nu}-a}}},
\end{equation}
where we again require $(2+\gamma)(\overline{\nu}-a)\ge 1-a$.  Combining these portions, summing in $k$, and taking the suprema in $\ell$ and $\tau$ gives the lemma. $\blacksquare$

\bigskip

\begin{lem}
Under the bootstrap hypotheses \eqref{orig_bstrap_AI} -- \eqref{orig_bstrap_rho}, there holds
\begin{equation}
\sup_{\tau\in [0,t]}\sup_{\ell \ne 0} \sum_{k\ne 0} \overline{K}_{k,\ell}(t,\tau) \le C \sqrt{K_2}\epsilon\langle t \rangle,
\end{equation}
where $C$ depends on the various parameters appearing in the assumptions on the background and initial data.
\end{lem}

\bigskip

\textbf{Proof:}  We proceed in the same way as above, but without the time integral
\begin{align}
\overline{K}_{k,\ell}(t,\tau) &\le C \sqrt{K_2}\epsilon e^{-\lambda(t,\tau)\langle k, kt/L\rangle^{\overline{\nu}}}e^{-\delta\langle k-\ell, (kt-\ell\tau)/L\rangle^{\overline{\nu}}}\frac{\langle \tau \rangle}{|\ell|^{\gamma}}\nonumber\\
&\le C \sqrt{K_2}\epsilon \langle \tau \rangle e^{-C\delta\langle k - \ell \rangle^{\overline{\nu}}}.
\end{align}
This last form is summable in $k$.  Taking the supremum in $\ell$ and $0 \le \tau \le t$ gives the lemma. $\blacksquare$

\section{Appendix: Relevancy of Assumption \eqref{ass_G_2} for Spherical, Decreasing Backgrounds}

We want to comment on Assumption \eqref{ass_G_2} - at least for a certain class of background data.  First, recall that for a complex number $z = r+is$, our convention on the Laplace Transform in $t$ is
\begin{eqnarray}
\mathcal{L}[f](z,k) &=& \int_0^{\infty}f(t,k)e^{-2 \pi zt}dt\\
&=& \int_{-\infty}^{\infty}H(t)f(t,k)e^{-2\pi rt}e^{-2 \pi i st}dt,
\end{eqnarray}
where $H$ is the Heaviside function.  The second form gives the relation between the Laplace Transform of $f$ and the convolution of $\widehat{f}^t$ (the Fourier Transform in the variable $t$) with the Fourier transform of a cut-off exponential function.  Our particular assumptions on $G_0$ show that the Laplace transform at least makes sense for $\Re(z) = r  \ge 0$, but the existence for any $z$ in the left half-plane seems questionable since we can never have exponential decay for $\widehat{\alpha G_0}$.  However, we will see that we can define the Laplace Transform of $L$ for all $z$ by analytic continuation -- at least for spherically symmetric backgrounds.

Abusing notation slightly, we will write $g_0(v) = g_0(|v|)$.  We have
\begin{equation}
\alpha G_0(v) = \frac{g_0'(|v|)}{1-|v|^2}\hat{v} \equiv \widetilde{G_0}(|v|)\hat{v}
\end{equation}
where $\hat{v}$ is the unit vector in the direction of $v$.  If we then take the Fourier Transform of this quantity and insert it into the kernel $L$ appearing in the evolution of $\widehat{\rho}$, we find
\begin{equation}
L(t,k) = -\frac{2L}{|k|}\widehat{W}(k)\int_0^1 \frac{2\pi\frac{|k|}{L}t\omega\cos\left(2\pi\frac{|k|}{L}t\omega\right)-\sin\left(2\pi\frac{|k|}{L}t\omega\right)}{t^2}\widetilde{G_0}(\omega)d\omega.
\end{equation}
Taking the Laplace Transform in $t$ of  $L$ (at least in the right half-plane) yields
\begin{equation}
\mathcal{L}[L](z,k) = -\frac{8\pi^2 |k|}{L}\widehat{W}(k)\int_0^{\infty}\int_0^1 f\left(2\pi\frac{|k|}{L}\omega t\right)\omega^2\widetilde{G_0}(\omega)e^{-2 \pi zt}d\omega dt,\label{linear2}
\end{equation}
where
\begin{equation}
f(at) = \frac{at\cos(at)-\sin(at)}{a^2t^2}.
\end{equation}
Note that $f$ is well-defined at $t=0$ (it vanishes like $-t$ near the origin) and decays like $t^{-2}$ for large $t$.  Hence, we can interchange the order of integration and compute the transform of $f$ at the very least for $\Re(z)\ge 0$.  Hence, we need to consider the Laplace Transform of $f(at)$ for $a = 2\pi\frac{|k|}{L}\omega \ge 0.$  Clearly for $r\ge0$ we have
\begin{equation}
\mathcal{L}[f](z=r+is) = \widehat{He^{-2\pi r \cdot}} \ast \widehat{f}(s)
\end{equation}
We have the well known fact that
\begin{eqnarray}
\widehat{He^{-2\pi r \cdot}} (s) &=& \left\{\begin{array}{cc} \frac{1}{2\pi(r+is)} & r > 0 \\ \frac{1}{2}\left(\frac{\textrm{P.V.}}{i\pi s}+\delta(s)\right)& r = 0\end{array}\right.,
\end{eqnarray}
in the sense of distributions.  Next, we note that
\begin{eqnarray}
\widehat{f}(s) &=& \int_{-\infty}^{\infty}\frac{at\cos(at)-\sin(at)}{a^2t^2}e^{-2\pi i s t}dt\nonumber\\
&=& \frac{2\pi i s}{a} \int_{-\infty}^{\infty}\frac{\sin(at)}{at}e^{-2\pi i s t}dt\nonumber\\
&=& \frac{2\pi^2 i s}{a^2} \widehat{\sinc}\left(\frac{\pi s}{a}\right),
\end{eqnarray}
where $\sinc(x) = \sin(\pi x)/(\pi x)$.  The Fourier Transform of sinc is well-known and given by
\begin{equation}
\widehat{\sinc}(s) = \left\{\begin{array}{cc}1 & |s| < \frac{1}{2}\\ \frac{1}{2} & s = \pm \frac{1}{2}\\ 0 & \textrm{ otherwise }\end{array}\right. ,
\end{equation}
which finally gives us
\begin{equation}
\widehat{f}(s) = \frac{2\pi^2 i s}{a^2} \cdot\left\{\begin{array}{cc}1 & |s| < \frac{a}{2\pi}\\ \frac{1}{2} & s = \pm \frac{a}{2\pi}\\ 0 & \textrm{ otherwise }\end{array}\right. .
\end{equation}
Using this to evaluate the Laplace Transform gives for $r>0$
\begin{equation}
\mathcal{L}[f](z=r+is) = \frac{1}{a}\left(\frac{z}{ia/2\pi}\arccoth\left(\frac{z}{ia/2\pi}\right)-1\right).
\end{equation}
Note that the since the principal domain of $\arccoth$ is $\mathbb{C} \setminus [-1,1] $, the formula above gives us the analytic continuation of our Laplace Transform to the complex plane minus the line segment $[-ia/2\pi, ia/2\pi]$ on the imaginary axis.

When $r=0$, we have three cases to consider.  When $|s| > a/2\pi$, we have
\begin{equation}
\mathcal{L}[f](z=is) =\frac{1}{a}\left(\frac{z}{ia/2\pi}\arccoth\left(\frac{z}{ia/2\pi}\right)-1\right),
\end{equation}
which (unsurprisingly) agrees with the analytic continuation found above.  When $|s| < a/2\pi$, we find
\begin{equation}
\mathcal{L}[f](z=is) =\frac{1}{a}\left(\frac{z}{ia/2\pi}\arctanh\left(\frac{z}{ia/2\pi}\right)-1\right)+\frac{\pi^2 z}{a^2}.
\end{equation}
This leaves $s = \pm a/2\pi$ where all of the formula above diverge.  This divergence is quite mild, however, since for any $\nu>0$ $$\lim_{x=1-}(1-x)^{\nu}\arctanh(x) =  \lim_{x=1+}(1-x)^{\nu}\arccoth(x) = 0, $$ (in fact, the divergence is precisely logarithmic).  Therefore, we have the existence of $\mathcal{L}[f](z)$ for all $z \in \mathbb{C}\setminus \{\pm ia/2\pi\}$, and the singularities at these exceptional points are merely logarithmic divergences.

We can now consider the Laplace Transform of the kernel $L$ given above.  For $\Re(z) \ne 0$, we have
\begin{eqnarray}
\mathcal{L}[L](z,k) &=& -4\pi\widehat{W}(k)\int_0^1\left(\frac{z}{i\frac{|k|}{L}\omega}\arccoth\left(\frac{z}{i\frac{|k|}{L}\omega}\right)-1\right) \omega \widetilde{G_0}(\omega)d\omega\nonumber\\
&=&-4\pi\widehat{W}(k)\int_0^1\left(\frac{z}{i\frac{|k|}{L}}\arctanh\left(\frac{i\frac{|k|}{L}\omega}{z}\right)-\omega\right) \widetilde{G_0}(\omega)d\omega,
\end{eqnarray}
which is well defined since this line integral never crosses the branch cut for $\arccoth$, and the integrand limits to $0$ as $\omega \to 0$.  For purely imaginary $z=is$, with $|s| \ge |k|/L$, we have
\begin{equation}
\mathcal{L}[L](z=is,k) = -4\pi \widehat{W}(k)\int_0^1 \left(\frac{Ls}{|k|}\arctanh\left(\frac{|k|\omega}{Ls}\right)-\omega\right)\widetilde{G_0}(\omega)d\omega
\end{equation}
which agrees with the integral for $\Re(z)\ne 0$.  Note that since $\widetilde{G_0} \in L^1([0,1])\cap L^2([0,1])$ and since the divergence of $\arctanh$ is so mild that it too is square integrable on $[0,1]$, this transform makes sense for the endpoints $s = \pm |k|/L$.

If $0< |s| < |k|/L$, we are forced to evaluate the integral over two disjoint intervals: $\omega \in \left[0,\frac{L|s|}{|k|}\right]$ and $\omega\in \left(\frac{L|s|}{|k|},1\right]$.  On the first portion, the integrand is exactly the same as above (with $z=is$):
\begin{equation}
 -4\pi\widehat{W}(k)\int_0^{\frac{L|s|}{|k|}}\left(\frac{Ls}{|k|}\arctanh\left(\frac{|k|\omega}{Ls}\right)-\omega\right) \widetilde{G_0}(\omega)d\omega.\nonumber\\
 \end{equation}
Again, this is integrable because $\arctanh$ has only a logarithmic divergence at the boundary.  Also note that the contribution from this portion goes to zero like $s^2$ as $s$ tends to zero.  On the second interval, the integrand changes to
\begin{equation}
-4\pi\widehat{W}(k)\int_{\frac{L|s|}{|k|}}^1\left(\frac{Ls}{|k|}\arccoth\left(\frac{|k|\omega}{Ls}\right)-\omega+\frac{i\pi Ls}{2|k|}\right)\widetilde{G_0}(\omega)d\omega.\nonumber
\end{equation}
Once again, $\arccoth$ has only a logarithmic divergence at the left-hand boundary (and so this integral is well defined).  We are left to consider the case of $s=0$.  If we return to the convolution integral, we can compute directly that
\begin{equation}
\mathcal{L}[L](0,k)= 4\pi\widehat{W}(k)\int_0^1\omega \widetilde{G_0}(\omega)d\omega.
\end{equation}
Hence, we see that $\mathcal{L}[L](z,k)$ exists and is finite for all $z \in \mathbb{C}$.

 In order to determine whether assumption \eqref{ass_G_2} is at all reasonable, we will consider a rather special case (though a case of some importance and/or interest).  Let us take  $$\widehat{W}(k) = \pm\frac{L^2}{|k|^2},  $$ which corresponds to $W$ being the kernel for the Laplacian.  Here, the $+1$ would give a Coulombic (repulsive) plasma while $-1$ would give a self-gravitating plasma.  For the background, let us consider the case where $\widetilde{G_0}(\omega)$ is strictly negative for all $\omega \in (0,1)$.  Recall that $\widetilde{G_0}(|v|) = g_0'(|v|)(1-|v|^2)^{-1}$ where $g_0$ is the actual background for our plasma.  So, assuming $\widetilde{G_0}$ is non-positive is equivalent to considering a background which is strictly decreasing (note that we must have $\widetilde{G_0}(0) = \widetilde{G_0}(1)=0$ for this to describe a smooth background).  This covers the most interesting case of the J\"uttner Distribution at temperature $T$:
\begin{equation}
g_{\textrm{J}}(|v|) = \frac{1}{4\pi k_B T K_2(1/k_B T)}e^{-\frac{1}{k_B T\sqrt{1-|v|^2}}}\mathds{1}_{[0,1)}(|v|) \label{Jdist},
\end{equation}
where $k_B$ is Boltzmann's constant and $K_2$ is the modified Bessel function of the second kind with index 2 (the indicator function is to emphasize that the support of this function is the unit ball in $\mathbb{R}^3$).  This describes the thermodynamic equilibrium distribution of momenta in a spatially uniform, relativistic ideal gas (analogous to the Maxwellian Distribution for non-relativistic gases).  We should note that this distribution appears to be Gevrey of order 3 (which corresponds to $\nu = 1/3$).  This is the critical class for Newtonian/Colombic interactions (where $\gamma=1$), and so it would appear that this is only an admissible background datum for interactions with $\gamma$ strictly greater than 1.  However, the J\"uttner Distribution is analytic except at the boundary $|v|=1$, and the Gevrey class it belongs to is entirely determined by its behavior at this boundary.  Hence, a very slight smoothing at the boundary should give us a background which is admissible for $\gamma =1$ and has essentially all the properties we detail below.

As we have already noted, $\mathcal{L}[L](|k|z,k)$ is well-defined for all $z\in \mathbb{C}$ and all non-zero $k \in \mathbb{Z}^3$ (our assumptions will ensure the zero-mode of our solutions is always zero and so of no interest to us).  To clarify the discussion, we list the various computations detailed above:
\begin{align}
&\mathcal{L}[L](|k|z=|k|r+i|k|s,k) = \nonumber\\
&\left\{\begin{array}{ll}  -4\pi\widehat{W}(k)\int_0^1\left(\frac{Lz}{i}\arctanh\left(\frac{i\omega}{Lz}\right)-\omega\right) \widetilde{G_0}(\omega)d\omega,& -iz \notin \left(-\frac{1}{L}, \frac{1}{L}\right)\\
 -4\pi\widehat{W}(k)\int_0^{L|s|}\left(Ls \: \arctanh\left(\frac{\omega}{Ls}\right)-\omega\right) \widetilde{G_0}(\omega)d\omega &\\
 \;\; -4\pi\widehat{W}(k)\int_{L|s|}^1\left(Ls \: \arccoth\left(\frac{\omega}{Ls}\right)-\omega+\frac{i\pi Ls}{2}\right)\widetilde{G_0}(\omega)d\omega,& z=is, \; 0<|s|<\frac{1}{L}\\
4\pi\widehat{W}(k)\int_0^1\omega \widetilde{G_0}(\omega)d\omega,& z = 0
\end{array}\right.
\end{align}
Note that this is expression is analytic in the disjoint half-planes $\Re(z) \ne 0$ (in fact, it is analytic in the complex plane minus the line segment $[-i/L,i/L]$ on the imaginary axis).  The imaginary part of $\mathcal{L}[L](|k|z,k)$ in this region is determined (up to a non-zero multiple) by integrating
\begin{equation}
-L\left(\frac{r}{2}\ln\left(\frac{r^2+\left(s+\frac{\omega}{L}\right)^2}{r^2+\left(s-\frac{\omega}{L}\right)^2}\right)+s\arctan\left(\frac{s-\frac{\omega}{L}}{r}\right)-s\arctan\left(\frac{s+\frac{\omega}{L}}{r}\right)\right)\nonumber
\end{equation}
against $\widetilde{G_0}$ over $\omega \in [0,1]$.  Note that this quantity will be zero when $s=0$ (and so, we will need to examine the behavior of the transform on the real axis).  Note that the derivative of this quantity in $\omega$ is
\begin{equation}
\frac{4sr\omega^2}{L^2}\left(r^2+\left(s+\frac{\omega}{L}\right)^2\right)^{-1}\left(r^2+\left(s-\frac{\omega}{L}\right)^2\right)^{-1}.\nonumber
\end{equation}
So when $s,r > 0$ or $s,r<0$, the integrand is increasing in $\omega$.  Since for $\omega = 0$ the term above gives zero, this portion of the integrand is strictly positive on the range of integration.  In the other quadrants, the integrand is strictly negative over the range of integration.  In all of these cases (i.e. $r,s\ne 0$), the imaginary part of $\mathcal{L}[L](|k|z,k)$ cannot be zero.  That leaves us with the real and imaginary axes.

On the real axis ($s=0$), we have
\begin{equation}
\mathcal{L}[L](|k|z=|k|r+i0,k) = \left\{ \begin{array}{ll} 4\pi\widehat{W}(k)\int_0^1\left(\omega - Lr \: \arctan\left(\frac{\omega}{Lr}\right)\right) \widetilde{G_0}(\omega)d\omega & r \ne 0 \\4\pi\widehat{W}(k)\int_0^1\omega \widetilde{G_0}(\omega)d\omega,& r = 0 \end{array}\right.\nonumber
\end{equation}
Note that this expression is continuous and even in $r$.  As is easily checked, $$\omega - Lr \: \arctan\left(\frac{\omega}{Lr}\right) $$ is strictly increasing in $\omega$ and is identically zero for $\omega=0$.  Hence, the integrand for $r \ne 0$ is strictly negative (recall that we assume $\widetilde{G_0}$ is strictly negative).  For a Coulombic plasma, $\widehat{W}(k)>0$ and we see that $\mathcal{L}[L](|k|z=|k|r+i0,k)$ is strictly negative (and so cannot possibly be equal to $+1$).  For the attractive case, $\widehat{W}(k)<0$, and we have $\mathcal{L}[L](|k|z=|k|r+i0,k) > 0$. We can also see that $\mathcal{L}[L](|k|z=|k|r+i0,k)$ will decrease as $|r|$ increases.  Hence, if
\begin{equation}
-4\pi\frac{L^2}{|k|^2}\int_0^1\omega \widetilde{G_0}(\omega)d\omega < 1\nonumber
\end{equation}
for a self-gravitating plasma, then the Laplace transform will not equal $1$ anywhere on the complex plane. Since $|k| \ge 1$, we will have no issues so long as
\begin{equation}
-4\pi L^2\int_0^1\omega \widetilde{G_0}(\omega)d\omega < 1.
\end{equation}
For a given background $\widetilde{G_0}$ (satisfying the hypotheses we are working under),  this gives an upper-bound on the allowable size of the torus:
\begin{equation}
 L < \left(-4\pi \int_0^1\omega \widetilde{G_0}(\omega)d\omega\right)^{-1/2}.\label{LmaxSG}
\end{equation}
Below, we plot the right-hand-side of \eqref{LmaxSG} for the J\"uttner Distribution as a function of $\theta = k_B T$.
\begin{figure}[!htbp]\centering
  \includegraphics[bb=75 202 542 584,clip=true,scale=0.65]{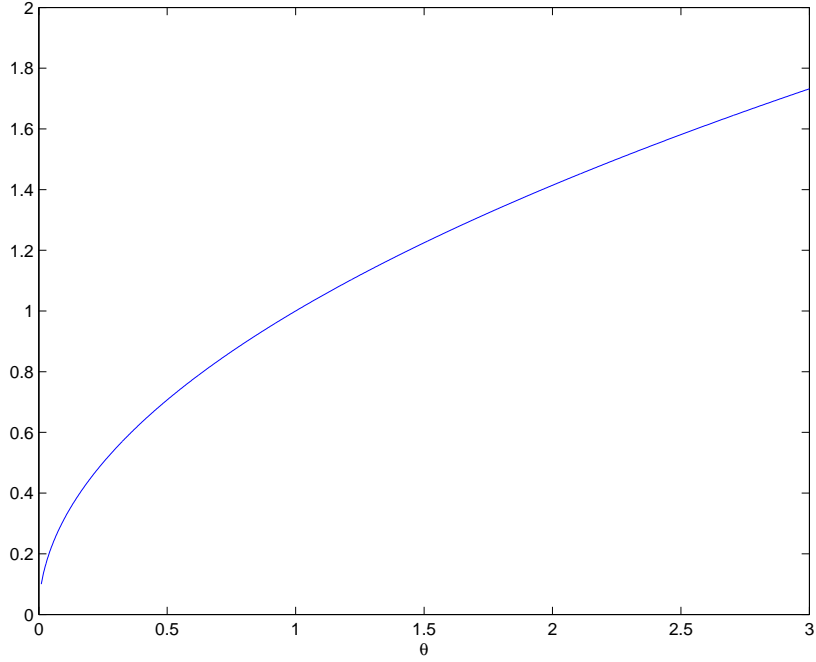}\\
  \caption{Plot of Maximum Torus Size for Self-Gravitating Plasmas vs. $\theta$ for the J\"uttner Distribution}
  \label{fig4}
\end{figure}
Note that as the temperature increases, the maximal size of the torus also increases.  Note that since we have taken the speed of light to be $1$, presumably we are working in some units like light-seconds.

That leaves us with the imaginary axis.  Note that for $0 < |s| < L^{-1}$, there will definitely be an imaginary component to $\mathcal{L}[L](i|k|s,k)$.  For $z=0$ and $z=i|k|s$ with $|s|\ge L^{-1}$, the transform will be purely real.  We have already dealt with the case $z=0$ above.  Hence, we concentrate on the remaining portion of the imaginary axis.  For $|s|\ge L^{-1}$ we have
\begin{equation}
\mathcal{L}[L](i|k|s,k) = 4\pi\widehat{W}(k)\int_0^{1}\left( \omega - Ls \:\arctanh\left(\frac{\omega}{Ls}\right)\right) \widetilde{G_0}(\omega)d\omega.
\end{equation}
This is even in $s$ and decreasing in $\omega$ for $s$ and $\omega$ in the ranges under consideration.  Since this expression is $0$ for $\omega = 0$, we see that it is negative on our range of integration.  Hence, the integral is a strictly positive quantity for $|s|\ge L^{-1}$.  Since for self-gravitating systems $\widehat{W}(k) < 0$, we see that such systems cannot have $\mathcal{L}[L](i|k|s,k)=1$.  For Coulombic systems on the other hand, we can have the transform equal to one on the imaginary axis. Since $$\omega - Ls \:\arctanh\left(\frac{\omega}{Ls}\right)  $$ is strictly increasing in $s$ and $\widetilde{G_0}$ is negative, we see that $\mathcal{L}[L](i|k|s,k)$ is decreasing on $|s| \ge L^{-1}$ for Coulombic systems.  Hence, if $$ 4\pi\frac{L^2}{|k|^2}\int_0^{1}\left( \omega - \arctanh\left(\omega\right)\right) \widetilde{G_0}(\omega)d\omega < 1,$$ then we will have no issues with the transform on the imaginary axis.  Thus, for Coulombic systems, the requirement that
\begin{equation}
4\pi L^2\int_0^{1}\left( \omega - \arctanh\left(\omega\right)\right) \widetilde{G_0}(\omega)d\omega < 1,
\end{equation}
would indicate that $\mathcal{L}[L](|k|z,k)$ is never equal to one on the entire complex plane.  This once again gives an upper bound on the size of the torus for a given background:
\begin{equation}
L < \left(4\pi \int_0^{1}\left( \omega - \arctanh\left(\omega\right)\right) \widetilde{G_0}(\omega)d\omega\right)^{-1/2}.\label{LmaxC}
\end{equation}
Below, we give a plot of the right hand side of \eqref{LmaxC} for the J\"uttner Distribution as a function of $\theta = k_B T$.
\begin{figure}[!htbp]\centering
  \includegraphics[bb=74 202 542 584,clip=true,scale=0.65]{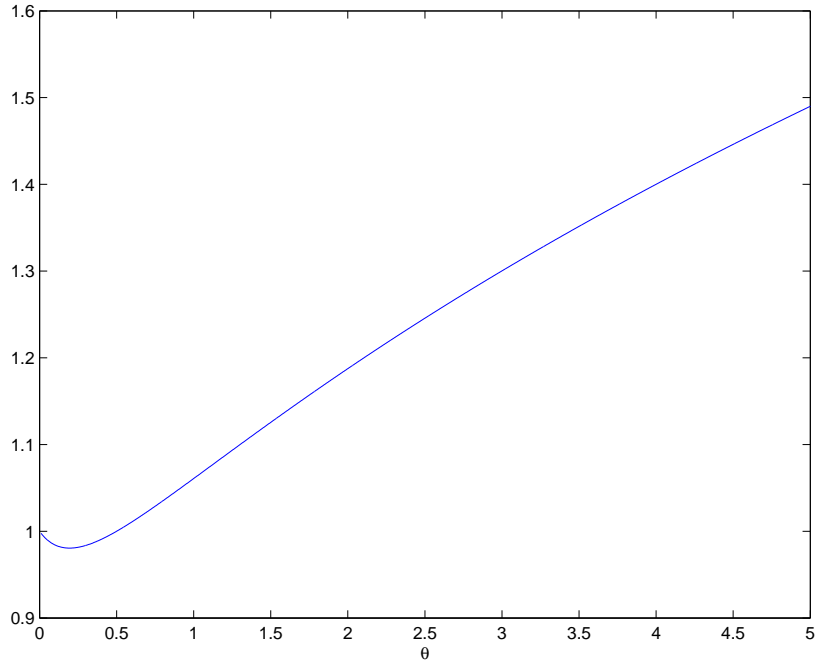}\\
  \caption{Plot of Maximum Torus Size for Coulombic Plasmas vs. $\theta$ for the J\"uttner Distribution}
  \label{fig5}
\end{figure}
Note that there is a critical size $L \approx 0.98$ below which Assumption \eqref{ass_G_2} is valid for all temperatures.  For larger plasmas, this assumption will only be valid for sufficiently large temperatures.

These results for the Coulombic plasma are exactly analogous to those found in \cite{Y14} (there is a discrepancy of $\sqrt{\pi}$ due to the different setup of the Poisson Equation in that paper).  The problematic modes for the self-gravitating system were also found in \cite{Y14}, but they could be effectively ignored since the behavior of a \emph{single} mode was under investigation (essentially, only the behavior of the transform on the imaginary axis came into play).  Since we are now enquiring about \emph{uniform} decay for all modes with spatial wavevectors $|k| \ge 1$, we need more breathing room around the imaginary axis.  Heuristic arguments in \cite{Y14} lead us to suspect that for plasmas exceeding these critical sizes, certain modes will not be rapidly damped out in the linearized system (and presumably not in the non-linear system either).  Note that these modes could still be slowly damped out (presumably at a rational rate in $t$) through dispersive effects.

\end{document}